\documentclass[12pt,dvipsnames,a4paper]{article}
\usepackage[top=80pt,bottom=85pt,left=85pt,right=85pt]{geometry}

\pdfoutput=1
\usepackage[utf8]{inputenc}
\usepackage[T1]{fontenc}
\usepackage[english]{babel}
\usepackage[dvipsnames]{xcolor}
\usepackage{booktabs}
\usepackage{array}
\usepackage{mathtools}
\usepackage{amsmath}
\usepackage{amssymb,amsthm}
\usepackage{mathrsfs}
\usepackage{cite}
\usepackage{microtype}
\usepackage[none]{hyphenat}
\usepackage{tikz}
\usetikzlibrary{calc,decorations.pathmorphing}
\definecolor{c0}{RGB}{20,20,20}
\definecolor{c1}{RGB}{200,60,40}
\definecolor{c2}{RGB}{40,90,190}
\definecolor{c3}{RGB}{20,140,90}
\definecolor{c4}{RGB}{130,50,160}
\tikzset{
  cc0/.style={c0,thick},
  cc1/.style={c1,thick,dashed},
  cc2/.style={c2,thick,decorate,decoration={snake,amplitude=.6pt,segment length=6pt}},
  cc3/.style={c3,thick,dash pattern=on 4pt off 2pt on 1pt off 2pt},
  cc4/.style={c4,very thick,densely dotted},
  bl/.style={circle,fill=black,inner sep=1.8pt},
  wh/.style={circle,draw=black,fill=white,inner sep=1.8pt}
}
\usetikzlibrary{calc,decorations.markings,arrows.meta,positioning,cd}
\usepackage[pdftex,colorlinks=true]{hyperref}
\hypersetup{urlcolor=MidnightBlue, citecolor=burntorange, linkcolor=darkmagenta}
\usepackage{float}

\numberwithin{equation}{section}

\definecolor{burntorange}{rgb}{0.8, 0.39, 0.0}
\definecolor{darkmagenta}{rgb}{0.55, 0.0, 0.55}

\newcommand{\Z}{\mathbb{Z}}
\newcommand{\Q}{\mathbb{Q}}
\newcommand{\R}{\mathbb{R}}
\newcommand{\C}{\mathbb{C}}
\newcommand{\CP}{\mathbb{CP}}
\newcommand{\CZ}{\mathcal{Z}}
\newcommand{\CD}{\mathcal{D}}
\newcommand{\CH}{\mathcal{H}}
\newcommand{\CE}{\mathcal{E}}
\newcommand{\Qd}{\mathcal{Q}_d}
\newcommand{\Dd}{\{0..d\}}   
\newcommand{\hG}{\widehat{\Gamma}}
\newcommand{\rs}[2]{#1|_{#2}}   
\newcommand{\hW}{\widehat{W}}
\newcommand{\lk}{\operatorname{lk}}
\newcommand{\Cone}{\operatorname{Cone}}
\newcommand{\sd}{\operatorname{sd}}
\newcommand{\kQ}{k_{\Q}}
\newcommand{\RP}{\mathbb{RP}}
\newcommand{\Sq}{\operatorname{Sq}}
\newcommand{\cupi}[1]{\cup_{#1}}

\newtheorem{theorem}{Theorem}
\newtheorem{proposition}[theorem]{Proposition}
\newtheorem{lemma}[theorem]{Lemma}
\newtheorem{corollary}[theorem]{Corollary}
\theoremstyle{remark}

\begin{document}

\begin{titlepage}

\phantom{wowiezowie}

\vspace{-1.6cm}

\begin{center}

{\Huge {\bf Multipartite Entanglement,}} \\

\vspace{0.3cm}

{\Huge {\bf TQFT Partition Functions}} \\

\vspace{0.3cm}

{\Huge {\bf and PL Bordism}}

\vspace{0.5cm}

{\Large Michele Del Zotto$^{\dagger\ddagger*}$, Abhijit Gadde$^{\star}$, and Pavel Putrov$^{\sharp}$ }\\

\vspace{0.6cm}

{\it
{\small

{\footnotesize$^\dagger$ Mathematics Institute, Uppsala University, \\ Box 480, SE-75106 Uppsala, Sweden}\\
\vspace{.15cm}
{\footnotesize$^\ddagger$ Department of Physics and Astronomy, Uppsala University,\\ Box 516, SE-75120 Uppsala, Sweden}\\
\vspace{.15cm}
{\footnotesize$^{*}$ Center for Geometry and Physics, Uppsala University,\\ Box 516, SE-75120 Uppsala, Sweden}\\
\vspace{.15cm}
{\footnotesize $^{\star}$ Department of Theoretical Physics \\ Tata Institute of Fundamental Research, Mumbai 400005
}\\
\vspace{.15cm}
{\footnotesize $^{\sharp}$ The Abdus Salam International Centre for Theoretical Physics, \\
 Strada Costiera 11, 34151 Trieste, Italy} \\
}}

\vskip 0.4cm
     	{\bf Abstract }
\vskip .05in
\end{center}

In a previous work we conjectured that the multi-invariants of the ground state of a gapped $(d-1)+1$-dimensional system, labeled by bipartite $(d+1)$-edge-colored graphs encoding closed oriented $d$-manifolds $M$, compute the partition functions $Z(M)$ of the low energy topological quantum field theory (TQFT), up to non-universal factors determined by the local structure of the triangulation. In this paper we determine which combinations of partition functions can be isolated from the non-universal factors. The answer is governed by oriented bordism. If $M$ is null-bordant, $Z(M)$ is obtained from an explicit, state-independent finite family of multi-invariants, up to a power of $Z(S^d)$ which is fixed to be $Z(S^d)^{\chi(M)/2}$ for even $d$ and can be removed for odd $d$. Conversely, the obstruction is measured by a quotient of the group of ``residues'' of colored graphs which receives a surjective map from the oriented bordism group, with kernel precisely the torsion subgroup as far as smooth manifolds are concerned. In $d=4$ the obstruction is the signature: $Z(M)/Z(S^4)^{\chi(M)/2}$ can be extracted if and only if $\sigma(M)=0$. This is due to the fact that the first Pontryagin class $p_1$ has a local expression and thus $e^{i\theta\sigma(M)}=e^{\frac{i}{3}\theta\int_M p_1}$ cannot be separated from the non-universal factors. Torsion invariants, in contrast, are not obstructed: for the ground state of the $4+1$-dimensional invertible phase with partition function $(-1)^{\int w_2w_3}$ we compute the multi-invariants exactly, and two graphs with identical local data, encoding the Dold manifold and $\CP^2\times S^1$, give opposite signs.

\vfill

\noindent\rule{4cm}{0.4pt}

{\footnotesize \noindent Emails: {\tt  michele.delzotto@math.uu.se, abhijit@theory.tifr.res.in, putrov@ictp.it}}

\eject

\end{titlepage}

{
  \hypersetup{linkcolor=black}
  \tableofcontents
}

\section{Introduction and summary}
\label{sec:intro}

It is by now well appreciated that the entanglement pattern of the ground state of a gapped quantum many-body system carries universal information about the topological phase it belongs to. The classic example is the topological entanglement entropy of Kitaev-Preskill and Levin-Wen \cite{Kitaev:2005dm,levin2006detecting}: a linear combination of entanglement entropies of the ground state of a $(2+1)$-dimensional gapped system, defined with respect to a decomposition of space into a few large regions, converges to the logarithm of the total quantum dimension of the low-energy topological order, that is to the logarithm of the partition function of the corresponding topological quantum field theory (TQFT) on the three-sphere. In \cite{DelZotto:2026dgp} we proposed a generalization of this statement, building on \cite{Gadde:2022cqi,Gadde:2023zzj,Gadde:2024taa,Gadde:2023zni,Gadde:2025csh,Gadde:2025ybn} and on the closely related work \cite{Sheffer:2025jtc,Sheffer:2025zyr,Sheffer:2026dgj}. The entanglement entropies are replaced there by \textit{multi-invariants}, which are local-unitary invariants of a state shared among several parties labeled by edge-colored graphs, and the parties are the regions of a decomposition of space modeled on the faces of a simplex. The key observation is that the graph labeling a multi-invariant can also be read as a recipe for gluing simplices into a closed manifold $M$ of dimension $d$, one higher than the spatial dimension of the system, and in the limit of large regions the multi-invariant is expected to compute the partition function $Z(M)$ of the low-energy TQFT, up to non-universal factors originating from the short-range entanglement of the ground state. We conjectured that these factors cancel in a suitable \textit{signal} of genuine multipartite entanglement \cite{Gadde:2026sig}, that is a combination of multi-invariants insensitive to lower-partite entanglement, resulting in a scheme that extracts the absolute value squared $|Z(M)|^2$ from the ground state. This was verified for all Levin-Wen string-net models \cite{Levin:2004mi} in $d=3$.

In the present paper we address the natural next question: can the partition function $Z(M)$ itself, including its phase and its normalization, be extracted in a similar manner? The motivation is that the phase of $Z(M)$ contains important physical information about the theory. In particular, the invertible TQFTs all have $|Z(M)|=1$ (up to the Euler counterterm) and thus can only be distinguished by the phase. The difficulty is that the non-universal factors in the multi-invariant are not known and depend on the microscopic details of the system, so one can only hope to cancel them against each other. We show that this is possible whenever a certain obstruction vanishes, assuming a refined version of the conjecture of \cite{DelZotto:2026dgp} which specifies the structure of the non-universal factors and which is satisfied by the Levin-Wen models. The obstruction is a purely combinatorial invariant of the graph labeling the multi-invariant, but the main result of this paper is that it is controlled by topology. Namely, a sufficient condition for it to vanish is that $M$ is null-bordant, i.e. that $M$ is the boundary of a compact oriented manifold of dimension $d+1$. In this case we construct explicitly, out of a triangulation of the bounding manifold, a signal which in the limit of large regions converges to $Z(M)$ divided by a power of the sphere partition function $Z(S^d)$. 

The null-bordance condition is not only sufficient but, in a certain sense, also close to necessary. We show that (i) the group of obstructions is a quotient of the oriented bordism group of $d$-manifolds, (ii) the two groups agree up to torsion, and (iii) for smooth manifolds the kernel is exactly the torsion subgroup. Statement (ii) follows from the existence of local combinatorial formulas for the rational Pontryagin numbers of a triangulated manifold \cite{LevittRourke1978,Gaifullin2004}. Statement (iii) follows from Wall's description of the torsion of the oriented bordism ring by explicit fiber bundles over lower-dimensional manifolds \cite{Wall1960}, the residues of which can be untwisted. In particular, for $d=4$ the obstruction is exactly the signature of $M$: the partition function of the complex projective plane cannot be isolated from the non-universal factors, while its absolute value squared can. There is a simple physical interpretation of this. Stacking the low-energy TQFT with an invertible topological phase, such as the gravitational theta-term in $d=4$ or the Euler counterterm in any even dimension, changes $Z(M)$ by a phase or a normalization which admits a local formula, and is therefore indistinguishable from an ultraviolet contribution by any scheme of the kind considered here. The ambiguity in the extracted partition function is precisely the freedom of stacking with such phases. Invertible phases associated with torsion bordism invariants of smooth manifolds behave differently and are detectable. We demonstrate this in an exactly solvable example: for the ground state of the non-trivial bosonic invertible phase in $4+1$ dimensions, with partition function $(-1)^{\int w_2w_3}$, we compute all the multi-invariants exactly and find that the two graphs of the Dold manifold and of $\CP^2\times S^1$ with identical residues have multi-invariants of the same magnitude and opposite sign.

For even $d$ the extra power of the sphere partition function is unavoidable and is fixed to be half the Euler characteristic of $M$. For odd $d$ the power can be made zero or one, and the sphere partition function itself is given by a separate signal, built from a $(d+2)$-cube graph, so that $Z(M)$ is determined completely. 

Since every closed oriented $3$-manifold is null-bordant, for $d=3$ one obtains by the construction above the full complex partition function of an arbitrary $3$-manifold. In particular this settles the question, raised in \cite{DelZotto:2026dgp}, of whether the phase of $Z(M)$ is accessible for $3$-manifolds that are not locally achiral in the sense of \cite{Sheffer:2026dgj}. The reason why the earlier scheme only saw $|Z(M)|^2$ is that a signal built from the graph of $M$ alone necessarily pairs $M$ with its orientation reversal. The additional data coming from the null-bordism is what evades this.

The paper is organized so that a physics-oriented reader may read only Section \ref{sec:main} and, possibly, Section \ref{sec:dold-example}. Section \ref{sec:main} contains the main statements. After introducing multi-invariants, graph-encoded manifolds and residues we formulate the conjecture that we assume throughout, in two forms, which are equivalent for $d\not\equiv 0\mod 4$, and in general when no partition function vanishes: Conjecture ELD (Effective Layer Decomposition), the refined version of the conjecture of \cite{DelZotto:2026dgp} which specifies the structure of the non-universal factors, and Conjecture TS (Topological Signal), which says that the signals built out of GEMs are topological. We then define the residue quotient, show that it is exactly the obstruction to the extraction of partition functions, describe its structure in terms of bordism (Theorem \ref{thm:structure}), and deduce the extraction scheme for null-bordant manifolds (Corollary \ref{cor:extract}). We also summarize what can be extracted in each dimension and work out a six-replica example for the lens space $L(3,1)$ explicitly.  Section \ref{sec:proofs} contains the proofs and the detailed versions of the statements, with the more technical piecewise-linear topology deferred to Appendix \ref{app:filling}. Appendix \ref{app:parity} contains a technical result on the power of $Z(S^d)$ in dimensions $d\equiv 1\mod 4$. Section \ref{sec:dold-example} contains the exactly solvable example in $d=5$, with the details in Appendix \ref{app:dold}. Appendix \ref{app:daughters} collects the substitution graphs of the $L(3,1)$ example, and Appendix \ref{app:glossary} is a glossary of the terminology of combinatorial topology used in the paper. In Section \ref{sec:discussion} we discuss the results and some future directions.

\section{Main statements}
\label{sec:main}

In this section we formulate the main conjectures and results of the paper in their basic form. Their strongest versions, together with the proofs and further details, are collected in Section \ref{sec:proofs}. We mostly follow the conventions of \cite{DelZotto:2026dgp}.

\subsection{Multi-invariants}
\label{sec:multi-invariants}

Consider a state $|\psi\rangle\in\CH=\CH^{(0)}\otimes\ldots\otimes\CH^{(d)}$ in the tensor product of $d+1$ Hilbert spaces. In the context of this paper, $|\psi\rangle$ is the ground state of a gapped local Hamiltonian $H$ on a $(d-1)$-dimensional space, which we take to be a lattice on a sphere $S^{d-1}$ (or, equivalently, on $\R^{d-1}$ with appropriate boundary conditions at infinity), decomposed into $d+1$ regions labeled by the elements of:
\begin{equation}
    \Dd:=\{0,1,\ldots,d\}.
\end{equation}
The regions are obtained by identifying $S^{d-1}$ with the boundary of a $d$-simplex, the $\alpha$-th region being the $\alpha$-th face. This is illustrated in Figure \ref{fig:regions}.

\begin{figure}[ht]
\centering
\begin{tikzpicture}[scale=0.85,font=\scriptsize,
    proj/.style={->,shorten >=2pt,gray},
    pt/.style={circle,fill=black,inner sep=1.0pt},
    inf/.style={circle,draw=black,fill=white,inner sep=1.2pt}]
  \node[gray] at (-0.85,0.75) {$d=2$};
  \begin{scope}
    \coordinate (A) at (0.65,0); \coordinate (B) at (2.35,0); \coordinate (C) at (1.50,1.45);
    \draw[c0,very thick] (A) -- (B); \draw[c1,very thick] (B) -- (C); \draw[c2,very thick] (C) -- (A);
    \node[pt] at (A) {}; \node[pt] at (B) {}; \node[pt] at (C) {};
    \node[inf] at (1.62,0) {}; \node[below,inner sep=3.5pt] at (1.62,-0.02) {$\infty$};
    \node[c0,below] at (0.98,0.02) {$0$}; \node[c1,right] at (2.05,0.78) {$1$}; \node[c2,left] at (0.95,0.78) {$2$};
  \end{scope}
  \draw[proj] (3.90,0.75) -- node[above,gray,inner sep=1.5pt]{$\cong$} (4.80,0.75);
  \begin{scope}[shift={(5.30,0.75)}]
    \draw[c0,very thick,<-] (-0.35,0) -- (1.05,0); \draw[c1,very thick] (1.05,0) -- (2.20,0);
    \draw[c2,very thick] (2.20,0) -- (3.35,0); \draw[c0,very thick,->] (3.35,0) -- (4.40,0);
    \node[pt] at (1.05,0) {}; \node[pt] at (2.20,0) {}; \node[pt] at (3.35,0) {};
    \node[c0,above] at (0.35,0.02) {$0$}; \node[c1,above] at (1.62,0.02) {$1$};
    \node[c2,above] at (2.78,0.02) {$2$}; \node[c0,above] at (3.90,0.02) {$0$};
    \node[gray] at (-0.62,-0.20) {$\infty$}; \node[gray] at (4.67,-0.20) {$\infty$};
    \node[gray,below] at (2.02,-0.45) {$\R^{1}$, $3$ regions};
  \end{scope}
  \node[gray] at (-0.85,-2.45) {$d=3$};
  \begin{scope}[shift={(0,-3.60)},scale=0.95]
    \coordinate (P) at (0,0.70); \coordinate (Q) at (1.6,0); \coordinate (R) at (3.2,1.10); \coordinate (S) at (1.5,2.50);
    \fill[c0,opacity=0.16] (P) -- (Q) -- (S) -- cycle; \fill[c1,opacity=0.16] (Q) -- (R) -- (S) -- cycle;
    \draw[gray,thick,dashed] (P) -- (R); \draw[thick] (P) -- (Q) -- (R) -- (S) -- cycle; \draw[thick] (Q) -- (S);
    \node[pt] at (P) {}; \node[pt] at (Q) {}; \node[pt] at (R) {}; \node[pt] at (S) {};
    \node[inf] at (1.05,1.50) {}; \node[above right,inner sep=1pt] at (1.05,1.50) {$\infty$};
    \node[c0] at (1.30,1.15) {$0$}; \node[c1] at (1.98,1.58) {$1$}; \node[c3] at (1.98,0.48) {$3$}; \node[c2] at (3.62,2.15) {$2$};
    \draw[c2,thin,-{Stealth[length=3pt]}] (3.46,2.00) -- (2.62,1.32);
  \end{scope}
  \draw[proj] (3.90,-2.45) -- node[above,gray,inner sep=1.5pt]{$\cong$} (4.80,-2.45);
  \begin{scope}[shift={(7.30,-2.35)}]
    \coordinate (Sp) at (0,1.30); \coordinate (Pp) at (-1.30,-0.85); \coordinate (Qp) at (1.30,-0.85); \coordinate (Rp) at (0,-0.13);
    \fill[c0,opacity=0.07] (-2.10,-1.42) rectangle (2.10,1.72);
    \fill[white] (Sp) -- (Pp) -- (Qp) -- cycle;
    \fill[c2,opacity=0.16] (Sp) -- (Pp) -- (Rp) -- cycle; \fill[c1,opacity=0.16] (Sp) -- (Qp) -- (Rp) -- cycle; \fill[c3,opacity=0.16] (Pp) -- (Qp) -- (Rp) -- cycle;
    \draw[thick] (Sp) -- (Pp) -- (Qp) -- cycle; \draw[thick] (Rp) -- (Sp); \draw[thick] (Rp) -- (Pp); \draw[thick] (Rp) -- (Qp);
    \node[pt] at (Sp) {}; \node[pt] at (Pp) {}; \node[pt] at (Qp) {}; \node[pt] at (Rp) {};
    \node[c2] at (-0.43,0.11) {$2$}; \node[c1] at (0.43,0.11) {$1$}; \node[c3] at (0,-0.61) {$3$}; \node[c0] at (-1.66,0.30) {$0$};
    \node[inf] at (-1.05,1.00) {}; \node[above left,inner sep=1.5pt] at (-1.05,1.00) {$\infty$};
    \node[gray,below] at (0,-1.52) {$\R^{2}$, $4$ regions};
  \end{scope}
\end{tikzpicture}
\caption{The decomposition of the $(d-1)$-dimensional space into $d+1$ regions, the faces of a $d$-simplex, for $d=2$ and $d=3$. Under the stereographic projection from a point $\infty$ in the interior of the face $0$, the sphere $S^{d-1}=\partial(\text{simplex})$ becomes $\R^{d-1}$ and the face $0$ becomes the unbounded region.}
\label{fig:regions}
\end{figure}
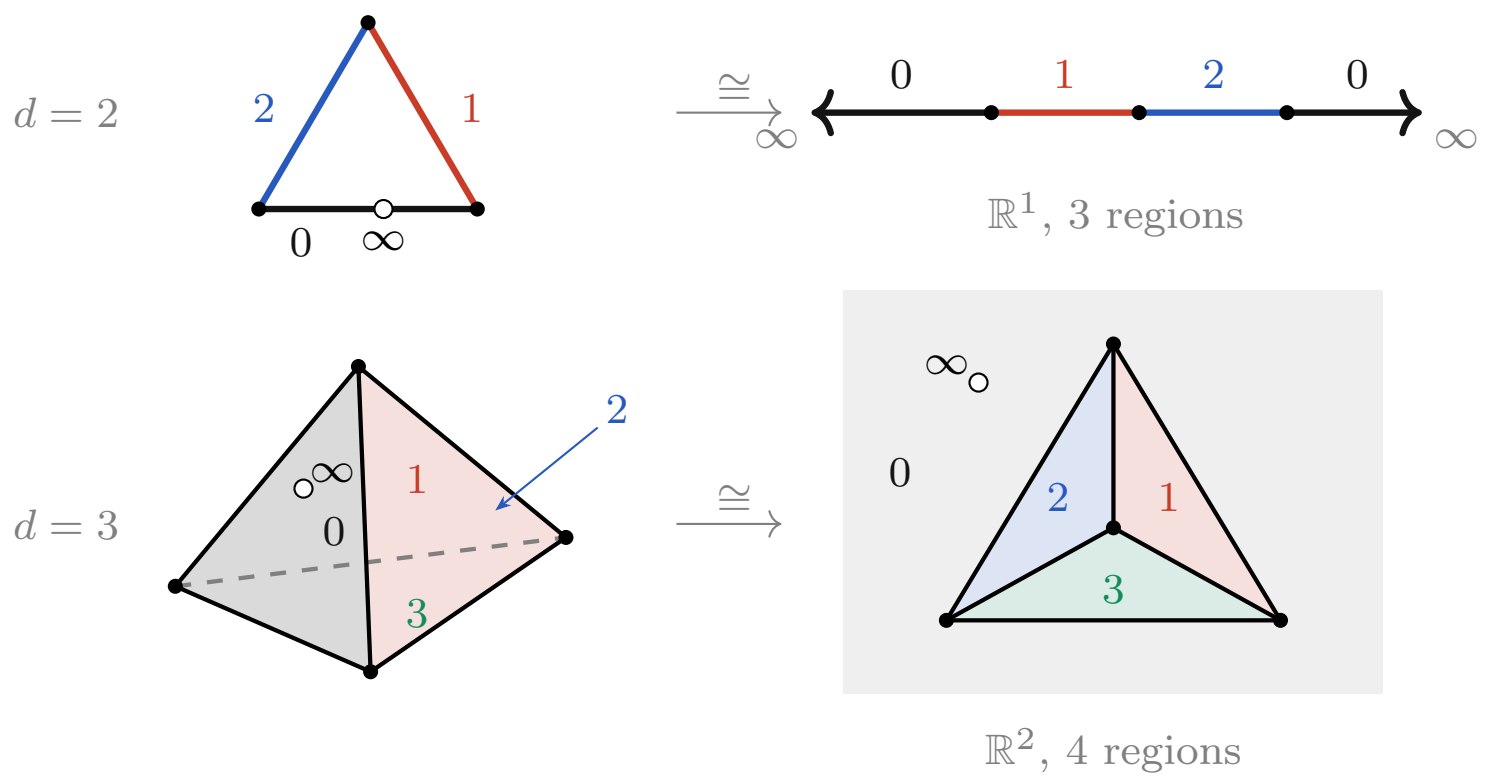 Each region is assumed to have linear size of order $L$, which is taken to be large compared to the correlation length and the lattice spacing, and to have a generic shape, with the boundary not aligned with the lattice directions or with other special directions of $H$. The genericity assumption excludes spurious contributions to the entanglement, insensitive to $L$ but unrelated to the low-energy TQFT, which are known to arise for special shapes of the regions already for the topological entanglement entropy \cite{Zou:2016dck,Williamson:2018zig}. The Hilbert space $\CH^{(\alpha)}$ is the Hilbert space of the degrees of freedom supported in the $\alpha$-th region.\footnote{As in \cite{levin2006detecting,DelZotto:2026dgp}, in gauge-theory-like lattice models this may require an extension of the physical Hilbert space, e.g. by relaxing the Gauss law constraints at the boundaries of the regions.}

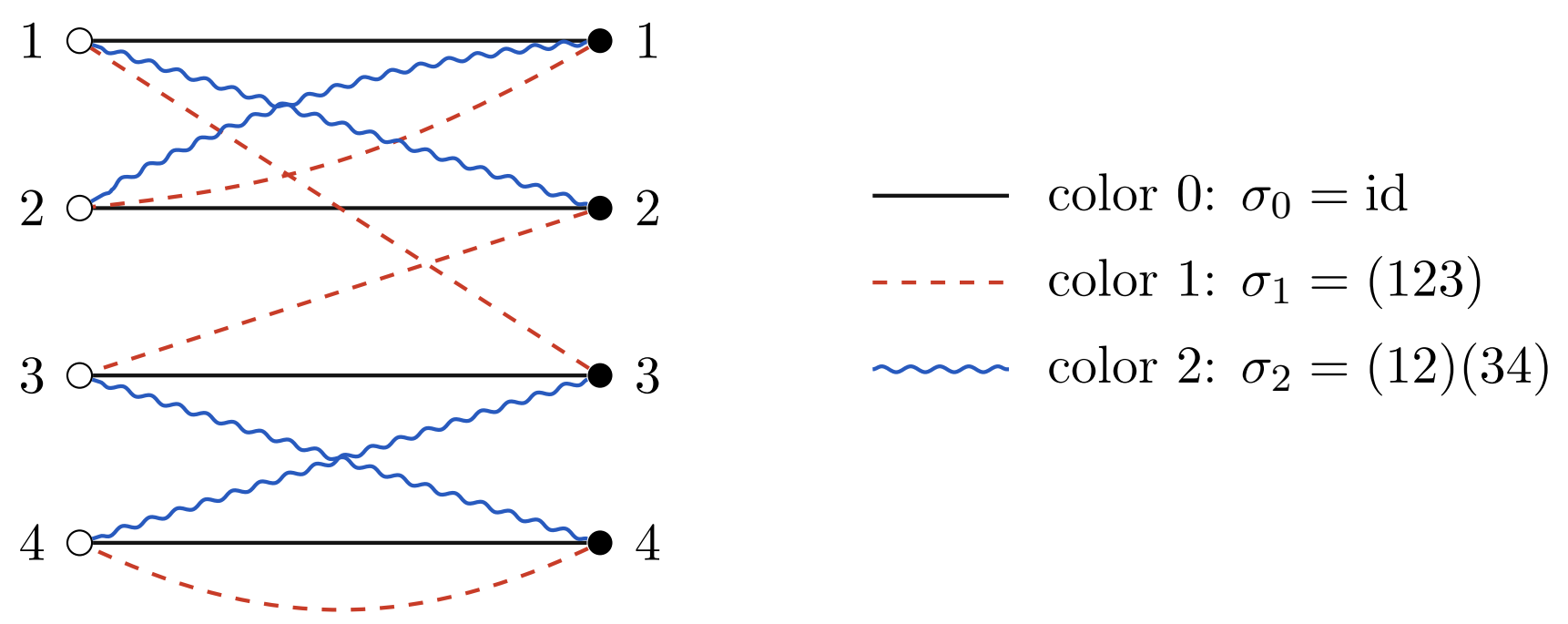
\begin{figure}[ht]
\centering
\begin{tikzpicture}[scale=0.9]
  \foreach \i in {1,...,4}{
    \node[wh] (W\i) at (0,-1.35*\i) {};
    \node[bl] (B\i) at (4.2,-1.35*\i) {};
    \node[left=3pt]  at (W\i) {\small $\i$};
    \node[right=3pt] at (B\i) {\small $\i$};
  }
  \foreach \i in {1,...,4}{ \draw[cc0] (B\i) -- (W\i); }
  \draw[cc1] (B1) to[bend left=12] (W2);
  \draw[cc1] (B2) -- (W3);
  \draw[cc1] (B3) -- (W1);
  \draw[cc1] (B4) to[bend left=25] (W4);
  \draw[cc2] (B1) to[bend right=12] (W2);
  \draw[cc2] (B2) -- (W1);
  \draw[cc2] (B3) -- (W4);
  \draw[cc2] (B4) -- (W3);
  \begin{scope}[shift={(6.4,-2.6)}]
    \draw[cc0] (0,0) -- (1.1,0);       \node[right] at (1.25,0) {\small color $0$: $\sigma_0=\mathrm{id}$};
    \draw[cc1] (0,-0.7) -- (1.1,-0.7); \node[right] at (1.25,-0.7) {\small color $1$: $\sigma_1=(123)$};
    \draw[cc2] (0,-1.4) -- (1.1,-1.4); \node[right] at (1.25,-1.4) {\small color $2$: $\sigma_2=(12)(34)$};
  \end{scope}
\end{tikzpicture}
\caption{The bipartite $3$-colored graph $\Gamma$ of the permutations $\sigma_0=\mathrm{id}$, $\sigma_1=(123)$, $\sigma_2=(12)(34)$ of $r=4$ replicas ($d=2$): the labels are the replica indices, the black vertices on the right are the ket replicas and the white vertices on the left the bra replicas, and an edge of color $\alpha$ joins $\bullet i$ to $\circ\sigma_\alpha(i)$. In the interpretation of Section \ref{sec:gem}, this graph encodes a two-sphere glued out of eight triangles.}
\label{fig:graph}
\end{figure}

Choose a \textit{replica number} $r\in\Z_{\geq 1}$ and permutations $\sigma_\alpha\in S_r$, $\alpha\in\Dd$. The corresponding \textit{multi-invariant} is
\begin{equation}
    \CZ(\sigma_0,\ldots,\sigma_d;|\psi\rangle):=\langle\psi|^{\otimes r}\,\sigma_0\otimes\ldots\otimes\sigma_d\,|\psi\rangle^{\otimes r},
    \label{multi-invariant-def}
\end{equation}
where $\sigma_\alpha$ acts by permuting the $r$ copies of $\CH^{(\alpha)}$. It is invariant under local unitary transformations $U_0\otimes\ldots\otimes U_d$, $U_\alpha\in U(\CH^{(\alpha)})$, and under the simultaneous multiplication of all $\sigma_\alpha$ by a fixed element of $S_r$ on the left or on the right. The invariant information is conveniently encoded in a bipartite $(d+1)$-edge-colored graph $\Gamma$: the black vertices $\bullet i$ correspond to the ket replicas, the white vertices $\circ i$ to the bra replicas, $i=1,\ldots,r$, and for each color $\alpha\in\Dd$ there is an edge of color $\alpha$ between $\bullet i$ and $\circ\sigma_\alpha(i)$. In this way $\Gamma$ is a $(d+1)$-regular graph (with multiple edges allowed) with a proper edge coloring by $\Dd$ (each vertex meets exactly one edge of each color, Appendix \ref{app:glossary}), and we write $\Gamma=\Gamma(\sigma_0,\ldots,\sigma_d)$ and $\CZ(\Gamma;|\psi\rangle)$ for (\ref{multi-invariant-def}). In (\ref{multi-invariant-def}) the index $\alpha$ of $\sigma_\alpha$ coincides with the position of the permutation in the argument, that is the color of the edges it produces. However, in general we treat the permutations without assigning them specific colors: the same element of $S_r$ may occupy several positions, as will be the case for the signals of Section \ref{sec:boolean}. The graphs are considered up to color-preserving isomorphism.
An example is shown in Figure \ref{fig:graph}.

For a fixed graph $\Gamma$ the multi-invariant is multiplicative with respect to the tensor product in the following sense. Let $\CH^{(\alpha)}=\CH^{(\alpha)}_1\otimes\CH^{(\alpha)}_2$ for each $\alpha\in\Dd$, and let $|\psi_i\rangle\in\bigotimes_{\alpha=0}^d\CH^{(\alpha)}_i$ for $i=1,2$. Then
\begin{equation}
    \CZ(\Gamma;|\psi_1\rangle\otimes|\psi_2\rangle)=\CZ(\Gamma;|\psi_1\rangle)\,\CZ(\Gamma;|\psi_2\rangle).
    \label{multiplicativity-states}
\end{equation}
Moreover, for a fixed state, multi-invariants are multiplicative with respect to the disjoint union of graphs,
\begin{equation}
    \CZ(\Gamma\sqcup\Gamma';|\psi\rangle)=\CZ(\Gamma;|\psi\rangle)\,\CZ(\Gamma';|\psi\rangle),
    \label{multiplicativity-graphs}
\end{equation}
which allows one to extend $\CZ$ multiplicatively to formal integer linear combinations of graphs, $\CZ(\sum_in_i\Gamma_i;|\psi\rangle):=\prod_i\CZ(\Gamma_i;|\psi\rangle)^{n_i}$, whenever the factors with negative exponents are non-zero. Accordingly, we will identify disjoint unions of graphs with sums in free abelian groups. Finally, exchanging black and white vertices, which we call the \textit{mirror} $\overline\Gamma$ of $\Gamma$, complex conjugates the multi-invariant.

\subsection{Graph-encoded manifolds}
\label{sec:gem}

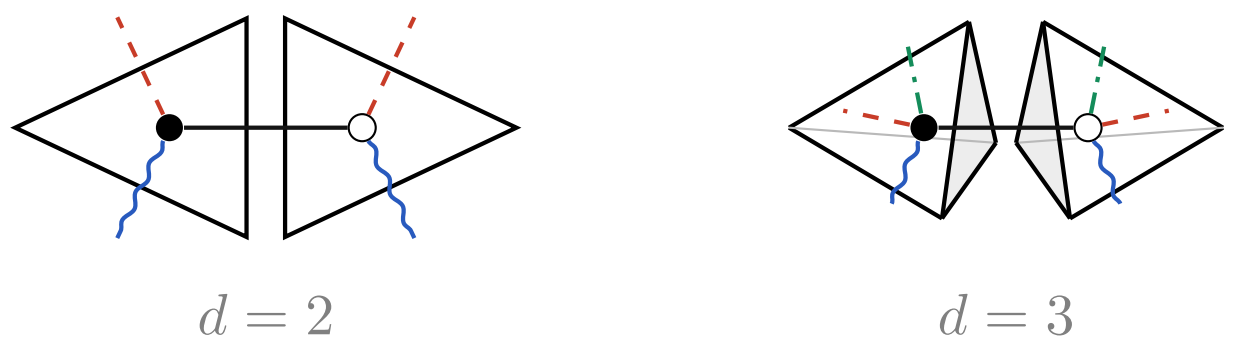
\begin{figure}[ht]
\centering
\begin{tikzpicture}[scale=0.85,baseline={(0,0)}]
  \draw[thick] (-0.15,0.85) -- (-0.15,-0.85) -- (-1.95,0) -- cycle;
  \draw[thick] ( 0.15,0.85) -- ( 0.15,-0.85) -- ( 1.95,0) -- cycle;
  \node[bl] (a) at (-0.75,0) {};
  \node[wh] (b) at (0.75,0) {};
  \draw[cc0] (a) -- (b);
  \draw[cc1] (a) -- ++(-0.406, 0.859);
  \draw[cc2] (a) -- ++(-0.406,-0.859);
  \draw[cc1] (b) -- ++( 0.406, 0.859);
  \draw[cc2] (b) -- ++( 0.406,-0.859);
  \node[gray,font=\small] at (0,-1.45) {$d = 2$};
\end{tikzpicture}
\hspace{1.5cm}
\begin{tikzpicture}[scale=0.70,baseline={(0,0)},
    vis/.style={thick}, sh/.style={thick}, hid/.style={gray!55,thin}, shf/.style={fill=black!7}]
  \fill[shf] (-0.350,1.000) -- (-0.605,-0.857) -- (-0.095,-0.143) -- cycle;
  \draw[vis] (-2.050,0.000) -- (-0.350,1.000);
  \draw[vis] (-2.050,0.000) -- (-0.605,-0.857);
  \draw[hid] (-2.050,0.000) -- (-0.095,-0.143);
  \draw[sh] (-0.350,1.000) -- (-0.605,-0.857);
  \draw[sh] (-0.605,-0.857) -- (-0.095,-0.143);
  \draw[sh] (-0.095,-0.143) -- (-0.350,1.000);
  \node[bl] (b) at (-0.775,0.000) {};
  \draw[cc1] (b) -- ++(-0.763,0.161);
  \draw[cc2] (b) -- ++(-0.305,-0.718);
  \draw[cc3] (b) -- ++(-0.152,0.765);
  \fill[shf] (0.350,1.000) -- (0.605,-0.857) -- (0.095,-0.143) -- cycle;
  \draw[vis] (2.050,0.000) -- (0.350,1.000);
  \draw[vis] (2.050,0.000) -- (0.605,-0.857);
  \draw[hid] (2.050,0.000) -- (0.095,-0.143);
  \draw[sh] (0.350,1.000) -- (0.605,-0.857);
  \draw[sh] (0.605,-0.857) -- (0.095,-0.143);
  \draw[sh] (0.095,-0.143) -- (0.350,1.000);
  \node[wh] (w) at (0.775,0.000) {};
  \draw[cc1] (w) -- ++(0.763,0.161);
  \draw[cc2] (w) -- ++(0.305,-0.718);
  \draw[cc3] (w) -- ++(0.152,0.765);
  \draw[cc0] (b) -- (w);
  \node[gray,font=\small] at (0,-1.76) {$d = 3$};
\end{tikzpicture}
\caption{The gluing rule: the vertices of $\Gamma$ are copies of a $d$-simplex ($\bullet$) and of its orientation reversal ($\circ$), and an edge of color $\alpha$ glues two simplices along their faces opposite to the vertex labeled $\alpha$. Shown are two triangles ($d=2$) and two tetrahedra ($d=3$) glued along the face of color $0$. The remaining edges at each vertex are dual to the remaining faces.}
\label{fig:gems}
\end{figure}

The same bipartite $(d+1)$-colored graph $\Gamma$ can be interpreted as a gluing pattern of simplices. Namely, associate to each black vertex a $d$-simplex with vertices labeled by the colors $\Dd$ and to each white vertex a copy of the same simplex with the opposite orientation. For each edge of color $\alpha$ between $\bullet i$ and $\circ i'$, glue the two simplices along their faces opposite to the vertex labeled $\alpha$, matching the labels (Figure \ref{fig:gems}). We denote the resulting topological space by $M_\Gamma$. The graph $\Gamma$ is called a \textit{$d$-GEM} (graph-encoded manifold) \cite{Pezzana1974,Ferri1976,FerriGagliardi1982,FerriGagliardiGrasselli1986,Lins1995,BandieriGagliardi2011arXiv} if $M_\Gamma$ is a closed piecewise-linear (PL) $d$-manifold.\footnote{There is a simple recursive characterization of $d$-GEMs in terms of the subgraphs of $\Gamma$, see Section \ref{sec:boolean}. It involves the recognition of PL spheres of dimension $d-1$, which is algorithmic for $d\leq 4$ \cite{Thompson1994}, open for $d=5$, and undecidable for $d\geq 6$ by a theorem of Novikov \cite{VolodinKuznetsovFomenko1974}.} We do not require $\Gamma$ to be connected: the connected components of $\Gamma$ encode the connected components of $M_\Gamma$, and a GEM of a disjoint union of manifolds is the disjoint union of GEMs. Since GEMs produce triangulations, PL manifolds are the natural setting for this paper, and we work with them throughout. Little is lost compared to the smooth setting: every smooth manifold admits a canonical PL structure \cite{Whitehead1940}, and by smoothing theory \cite{HirschMazur1974,KirbySiebenmann1977} every PL manifold of dimension at most $7$ admits a smooth structure, unique up to isotopy in dimension at most $6$. In particular, the natural map $\Omega^{\mathrm{SO}}_d\hookrightarrow\Omega^{\mathrm{SPL}}_d$ from the smooth to the PL oriented bordism group is surjective for $d\leq 7$. It is injective for all $d$, since the class of a smooth manifold is detected by its Pontryagin and Stiefel-Whitney numbers \cite{Wall1960}, which are PL bordism invariants \cite{Williamson1966,Kahn1972}. Therefore it is an isomorphism for $d\leq 7$, and it is an isomorphism after tensoring with $\Q$ for all $d$.\footnote{For an abelian group $G$, the vector space $G\otimes\Q$ is obtained by discarding the torsion elements of $G$ and allowing division by integers, e.g. $G\otimes\Q\cong\Q^r$ for $G\cong\Z^r\oplus T$ with $T$ finite. A homomorphism is an isomorphism after tensoring with $\Q$ if its kernel and cokernel consist of torsion elements, so that the two bordism groups can differ only in torsion invariants, such as the Stiefel-Whitney numbers, and not in the Pontryagin numbers.} The bipartite structure of $\Gamma$ is equivalent to the orientability of $M_\Gamma$, and we orient $M_\Gamma$ in such a way that the simplices corresponding to the black vertices, with the vertices ordered by increasing color, are positively oriented. With this convention $M_{\overline\Gamma}=\overline{M_\Gamma}$ is the orientation reversal. Every closed oriented PL $d$-manifold is $M_\Gamma$ for some bipartite $d$-GEM $\Gamma$ \cite{Pezzana1974,FerriGagliardiGrasselli1986}, and we say that $\Gamma$ is a GEM \textit{of} $M$ if $M_\Gamma\cong M$ as oriented PL manifolds. We call $\Gamma$ \textit{spherical} if $M_\Gamma\cong S^d$.\footnote{For $d\neq 4$ this is the same as $M_\Gamma$ being homotopy equivalent to $S^d$: every PL manifold homotopy equivalent to $S^d$ is PL homeomorphic to it, classically for $d\leq 2$, by \cite{Moise1952,Perelman:2006un,Perelman:2006up,Perelman:2003uq} for $d=3$, and by \cite{Smale1961,Stallings1960,Zeeman1961} for $d\geq 5$. For $d=4$ it is not known whether exotic PL $4$-spheres exist. A GEM of such a sphere would not be spherical in our sense, but every unitary TQFT would assign to it the value $Z(S^4)$ \cite{Reutter:2020bav}. Non-combinatorial triangulations of spheres, such as the double suspension of the Poincar\'e homology sphere, which is homeomorphic to $S^5$ \cite{Edwards1980,Cannon1979}, do not give GEMs at all, since some of their links are not PL spheres.} The simplest spherical GEM is the graph $\Gamma_{S^d}$ with one black and one white vertex connected by $d+1$ parallel edges of all colors. It realizes $S^d$ as two $d$-simplices glued along their common boundary, the equator $S^{d-1}$. The terminology of combinatorial topology used in this paper (simplicial cell complexes, PL structures, crystallizations, dipole moves) is summarized in Appendix \ref{app:glossary}.

Let $Z$ denote the partition function of the unitary TQFT describing the low energy limit of the gapped system with the ground state $|\psi\rangle$. We assume that the TQFT is defined on closed oriented manifolds without any additional structure (no framing-type anomalies, no spin structure dependence, etc.), and that its vacuum on $S^{d-1}$ is unique. Unitarity (reflection positivity) implies $Z(\overline M)=\overline{Z(M)}$ and
\begin{equation}
    Z(S^d)=\langle D^d|D^d\rangle >0,
    \label{Z-sphere-positive}
\end{equation}
where $|D^d\rangle\in\CH(S^{d-1})$ is the state produced by the $d$-dimensional ball.

As recalled in the introduction, the multi-invariant $\CZ(\Gamma;|\psi\rangle)$ of a $d$-GEM $\Gamma$ is expected to compute $Z(M_\Gamma)$ up to non-universal factors. The next subsection formulates what we assume about these factors. The main statements of the paper, collected in Section \ref{sec:results}, then describe which combinations of partition functions can be isolated from them.

\subsection{Residues and the two conjectures}
\label{sec:residues}

The non-universal factors are controlled by the subgraphs of $\Gamma$ with fewer colors, which we now introduce. For a subset $I\subseteq\Dd$ of colors, let $\rs{\Gamma}{I}$ (the \textit{restriction} of $\Gamma$ to the colors $I$) denote the subgraph of $\Gamma$ consisting of all the vertices and only the edges of colors in $I$, and let $\pi_0(\rs{\Gamma}{I})$ be the set of its connected components. Throughout, a hat denotes the complement in the set of colors, $\hat I:=\Dd\setminus I$ and $\hat\alpha:=\Dd\setminus\{\alpha\}$, so that $\rs{\Gamma}{\hat\alpha}$ is $\Gamma$ with the edges of color $\alpha$ deleted. If $\Gamma$ is a $d$-GEM, then for every proper subset $I\subsetneq\Dd$ each connected component of $\rs{\Gamma}{I}$ is a spherical $(|I|-1)$-GEM with the color set $I$: geometrically, it encodes the link (see Appendix \ref{app:glossary}) of a simplex of the cell decomposition of $M_\Gamma$ \cite{FerriGagliardiGrasselli1986,BandieriGagliardi2011arXiv}, see Section \ref{sec:boolean}. Here and below we allow $I=\emptyset$ with the following convention: $\rs{\Gamma}{\emptyset}$ consists of the $2r$ isolated vertices of $\Gamma$, which we group into the $r$ pairs $\{\bullet i,\circ i\}$ and regard as $r$ copies of the spherical $(-1)$-GEM, \textit{defined} to be a disjoint pair of a black and a white vertex encoding $S^{-1}=\emptyset$ (the isomorphism class of the pair does not depend on the grouping).

The relevance of these subgraphs is the following. By (\ref{multiplicativity-states}), if the state is a tensor product $|\psi\rangle=|\psi_1\rangle\otimes|\psi_2\rangle$ with respect to a decomposition $\CH^{(\alpha)}=\CH^{(\alpha)}_1\otimes\CH^{(\alpha)}_2$ of each party, then $\CZ(\Gamma;|\psi\rangle)=\CZ(\Gamma;|\psi_1\rangle)\CZ(\Gamma;|\psi_2\rangle)$. If, moreover, $|\psi_1\rangle$ is supported on a proper subset $I\subsetneq\Dd$ of the parties (meaning that it is a state in $\bigotimes_{\alpha\in I}\CH^{(\alpha)}_1$ times a fixed normalized product state in the remaining factors), then
\begin{equation}
    \CZ(\Gamma;|\psi_1\rangle)=\prod_{Q\in\pi_0(\rs{\Gamma}{I})}\CZ(Q;|\psi_1\rangle)\equiv \CZ(\rs{\Gamma}{I};|\psi_1\rangle).
    \label{lower-partite-factorization}
\end{equation}
That is, a state supported on the parties $I$ sees the graph $\Gamma$ only through $\rs{\Gamma}{I}$. A combination of multi-invariants which is trivial for any state that is a tensor product of states each of which is supported on a proper subset of the parties is called a \textit{signal} of genuine $(d+1)$-partite entanglement \cite{Gadde:2026sig}. The explicit family of signals used in this paper is described in Section \ref{sec:boolean}.

\begin{figure}[ht]
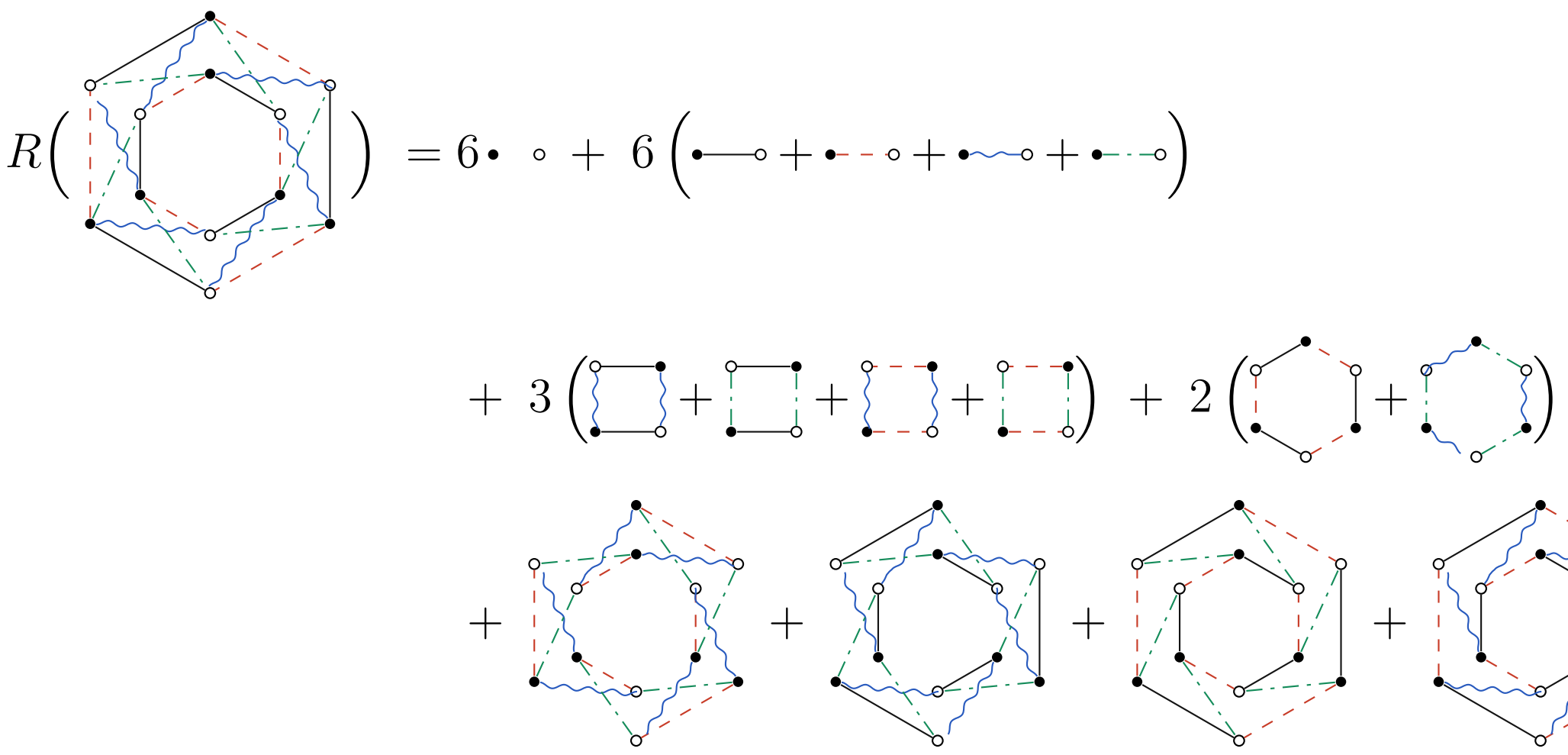

\centering
\newcommand{\lensN}{%
  \node[bl,inner sep=0.9pt] (b1) at ( 90:0.7) {}; \node[wh,inner sep=0.9pt] (w1) at ( 30:0.7) {};
  \node[bl,inner sep=0.9pt] (b2) at (330:0.7) {}; \node[wh,inner sep=0.9pt] (w2) at (270:0.7) {};
  \node[bl,inner sep=0.9pt] (b3) at (210:0.7) {}; \node[wh,inner sep=0.9pt] (w3) at (150:0.7) {};
  \node[bl,inner sep=0.9pt] (b4) at ( 90:1.2) {}; \node[wh,inner sep=0.9pt] (w5) at ( 30:1.2) {};
  \node[bl,inner sep=0.9pt] (b5) at (330:1.2) {}; \node[wh,inner sep=0.9pt] (w6) at (270:1.2) {};
  \node[bl,inner sep=0.9pt] (b6) at (210:1.2) {}; \node[wh,inner sep=0.9pt] (w4) at (150:1.2) {};}
\newcommand{\lensEa}{\draw[cc0,thin] (b1)--(w1) (b2)--(w2) (b3)--(w3) (b4)--(w4) (b5)--(w5) (b6)--(w6);}
\newcommand{\lensEb}{\draw[cc1,thin] (b1)--(w3) (b2)--(w1) (b3)--(w2) (b4)--(w5) (b5)--(w6) (b6)--(w4);}
\newcommand{\lensEc}{\draw[cc2,thin] (b1)--(w5) (b5)--(w1) (b2)--(w6) (b6)--(w2) (b3)--(w4) (b4)--(w3);}
\newcommand{\lensEd}{\draw[cc3,thin] (b1)--(w4) (b4)--(w1) (b2)--(w5) (b5)--(w2) (b3)--(w6) (b6)--(w3);}
\newcommand{\edgepic}[1]{\tikz[baseline=-0.6ex]{\node[bl,inner sep=0.9pt] (p) at (0,0) {}; \node[wh,inner sep=0.9pt] (q) at (0.55,0) {}; \draw[#1,thin] (p)--(q);}}
\newcommand{\sqpic}[2]{\tikz[baseline=-0.6ex]{\node[bl,inner sep=0.9pt] (v0) at (45:0.4) {}; \node[wh,inner sep=0.9pt] (v1) at (135:0.4) {}; \node[bl,inner sep=0.9pt] (v2) at (225:0.4) {}; \node[wh,inner sep=0.9pt] (v3) at (315:0.4) {}; \draw[#1,thin] (v0)--(v1) (v2)--(v3); \draw[#2,thin] (v1)--(v2) (v3)--(v0);}}
\newcommand{\hexpic}[2]{\tikz[baseline=-0.6ex]{\foreach \k in {0,...,5}{\pgfmathparse{int(mod(\k,2))}\ifnum\pgfmathresult=0 \node[bl,inner sep=0.9pt] (v\k) at (90+60*\k:0.5) {}; \else \node[wh,inner sep=0.9pt] (v\k) at (90+60*\k:0.5) {}; \fi} \draw[#1,thin] (v0)--(v1) (v2)--(v3) (v4)--(v5); \draw[#2,thin] (v1)--(v2) (v3)--(v4) (v5)--(v0);}}
\newcommand{\pairpic}{\tikz[baseline=-0.6ex]{\node[bl,inner sep=0.9pt] at (0,0) {}; \node[wh,inner sep=0.9pt] at (0.4,0) {};}}
\begin{align*}
R\Big(\;\tikz[baseline=-0.6ex]{\lensN\lensEa\lensEb\lensEc\lensEd}\;\Big)
 \;=\;& 6\,\pairpic\;+\;6\,\Big(\edgepic{cc0}+\edgepic{cc1}+\edgepic{cc2}+\edgepic{cc3}\Big)\\[1ex]
 &+\;3\,\Big(\sqpic{cc0}{cc2}+\sqpic{cc0}{cc3}+\sqpic{cc1}{cc2}+\sqpic{cc1}{cc3}\Big)\;+\;2\,\Big(\hexpic{cc0}{cc1}+\hexpic{cc2}{cc3}\Big)\\[1ex]
 &+\;\tikz[baseline=-0.6ex,scale=0.85]{\lensN\lensEb\lensEc\lensEd}\;+\;\tikz[baseline=-0.6ex,scale=0.85]{\lensN\lensEa\lensEc\lensEd}\;+\;\tikz[baseline=-0.6ex,scale=0.85]{\lensN\lensEa\lensEb\lensEd}\;+\;\tikz[baseline=-0.6ex,scale=0.85]{\lensN\lensEa\lensEb\lensEc}
\end{align*}
\caption{The residue of the six-replica GEM $\Gamma$ of the lens space $L(3,1)$ of Figure \ref{fig:lens}(a) (Section \ref{sec:example}), with the edge styles of that figure. The terms are, in order, the $I=\emptyset$ term ($6$ pairs), the six edges of each single color, the three $4$-cycles of each of the color pairs $\{0,2\},\{0,3\},\{1,2\},\{1,3\}$ and the two $6$-cycles of $\{0,1\}$ and $\{2,3\}$ (all GEMs of $S^1$), and the four connected $3$-colored subgraphs $\rs{\Gamma}{\hat\alpha}$, $\alpha=0,\ldots,3$, which are GEMs of $S^2$, the links of the four vertices of the triangulation of $L(3,1)$.}
\label{fig:residue}
\end{figure}

The conjecture of \cite{DelZotto:2026dgp} concerns a specific signal of genuine $(d+1)$-partite entanglement built out of the $d+1$ permutations of $\Gamma$, and states that in the limit $L\rightarrow\infty$ it converges to $|Z(M_\Gamma)|^2$ divided by a computable power of $Z(S^d)$. In this paper we assume a refined version of this conjecture, which describes the non-universal factors in a single multi-invariant. It implies a statement of the same kind for all signals built out of multi-invariants of GEMs. To formulate both, we first formalize the information about $\Gamma$ which is seen by the lower-partite parts of the state. Define the \textit{residue} of a $(d+1)$-colored graph $\Gamma$ (not necessarily a GEM) to be the formal sum of all its connected subgraphs obtained by retaining the edges of a proper subset of colors:
\begin{equation}
    R(\Gamma):=\sum_{I\subsetneq \Dd}\;\;\sum_{Q\in\pi_0(\rs{\Gamma}{I})}Q\;\;\in\;\;\Z\big[\text{$I$-colored graphs},\;I\subsetneq\Dd\big].
    \label{residue-def}
\end{equation}
Here, as always, graphs are considered up to color-preserving isomorphism, and $Q$ is regarded as a colored graph with the color set $I$ (in particular, the colors of the edges are remembered, so that the same abstract graph with different color sets gives different generators). In accordance with the convention above, for $I=\emptyset$ the inner sum is understood as $r\,(\bullet\sqcup\circ)$, one term for each pair $\{\bullet i,\circ i\}$. Let $\Z[d\text{-GEMs}]$ be the free abelian group generated by the isomorphism classes of connected $d$-GEMs, a disconnected GEM being identified with the sum of its components (Section \ref{sec:multi-invariants}), and let $\Z[\text{spherical }(<d)\text{-GEMs}]$ be the free abelian group generated by the connected spherical GEMs with color sets proper subsets of $\Dd$, including the $(-1)$-GEM $\bullet\sqcup\circ$. If $\Gamma$ is a GEM, all the $Q$'s are spherical GEMs of dimension $|I|-1$ (with the above convention for $I=\emptyset$), and $R$ is a homomorphism of abelian groups
\begin{equation}
\begin{array}{rrcl}
     R: & \Z[d\text{-GEMs}] & \longrightarrow & \Z[\text{spherical }(<d)\text{-GEMs}],  \\
     & \Gamma & \longmapsto & \sum_\gamma \#(\gamma,\Gamma)\,\gamma,
\end{array}
\label{residue-hom}
\end{equation}
where $\gamma$ runs over spherical GEMs with color sets proper subsets of $\Dd$ and $\#(\gamma,\Gamma)$ is the number of connected components of the subgraphs $\rs{\Gamma}{I}$ isomorphic to $\gamma$. Geometrically, $R(\Gamma)$ is the multiset of the (colored) links of all the simplices of dimension less than $d$ of the cell decomposition of $M_\Gamma$, together with the replica number $r$ recorded by the $I=\emptyset$ term. By (\ref{lower-partite-factorization}), $R(\Gamma)$ is exactly the information about $\Gamma$ that is seen by the lower-partite parts of the state. In particular, a product $\prod_i\CZ(\Gamma_i;|\psi\rangle)^{n_i}$ of multi-invariants of $d$-GEMs $\Gamma_i$ with integer exponents $n_i$ such that $\sum_in_iR(\Gamma_i)=0$ is a signal: for a state supported on the parties $I$ it is the product of the $\CZ(Q;|\psi\rangle)^{n_i}$ over the components $Q$ of the $\rs{\Gamma_i}{I}$, and every $Q$ occurs with total exponent zero. An example of a residue is shown in Figure \ref{fig:residue}.

The refined version of the conjecture of \cite{DelZotto:2026dgp} is what was actually established for the Levin-Wen models in Section 4 of that paper, and it explains the cancellation of the non-universal factors in the signal considered there. It says that, as far as the multi-invariants are concerned, the ground state behaves in the limit $L\rightarrow\infty$ as a product of effective layers, one for each proper subset of the parties, times a genuinely $(d+1)$-partite part which computes $Z(M_\Gamma)$.

\vspace{3ex}
\noindent\textbf{Conjecture ELD (Effective Layer Decomposition):} \textit{Let $\Gamma$ be a $d$-GEM. Then, in the limit $L\rightarrow\infty$,}
\begin{equation}
\begin{gathered}
    \CZ(\Gamma;|\psi\rangle)=\underbrace{U(\Gamma;|\psi\rangle)}_{\text{non-universal}}\;\times\;\big(Z(M_\Gamma)+O(e^{-\mathrm{const}\cdot L})\big),\\[4pt]
    U(\Gamma;|\psi\rangle)=\prod_{\gamma}A(\gamma;|\psi\rangle)^{\#(\gamma,\Gamma)},
\end{gathered}
    \label{conj-ELD}
\end{equation}
\textit{where $\gamma$ runs over colored graphs with color sets proper subsets $I\subsetneq\Dd$, connected for $I\neq\emptyset$ (i.e. spherical GEMs of dimension $|I|-1$), and the numbers $A(\gamma;|\psi\rangle)\in\C^*$ do not depend on $\Gamma$.}
\vspace{2ex}

\noindent
If $Z(M_\Gamma)\neq 0$, the error term can be written as a factor $1+O(e^{-\mathrm{const}\cdot L})$ multiplying $Z(M_\Gamma)$. If $Z(M_\Gamma)=0$, the conjecture says that the multi-invariant is exponentially small compared to the non-universal factor, which sets the scale of the error term. A priori there is no reason for it to vanish exactly.

The motivation for this form of the non-universal factors is the following. In the decomposition of $S^{d-1}$ into $d+1$ regions arranged as the faces of a simplex, there is no point common to all the regions. All the short-range correlations of the ground state therefore live on proper subsets $I\subsetneq\Dd$ of the parties, and by (\ref{lower-partite-factorization}) an effective factor of the state supported on the parties $I$ contributes a product over the connected components of $\rs{\Gamma}{I}$, which is of the form (\ref{conj-ELD}). The remaining, genuinely $(d+1)$-partite, part of the state is non-local and is what computes $Z(M_\Gamma)$. Factors of the form (\ref{conj-ELD}) are also expected to be exactly what a finite-depth quantum circuit applied to the state can change, see the paragraph on finite-depth circuits in Section \ref{sec:discussion}. The term $I=\emptyset$ accounts for the normalization of the state: under $|\psi\rangle\rightarrow\lambda|\psi\rangle$ the multi-invariant is multiplied by $|\lambda|^{2r}$, which is the factor $A(\bullet\sqcup\circ;|\psi\rangle)^{r}$ in (\ref{conj-ELD}), so that Conjecture ELD does not require $|\psi\rangle$ to be normalized. Note also that since multi-invariants are exactly multiplicative under disjoint unions (\ref{multiplicativity-graphs}), it is sufficient to assume Conjecture ELD for connected GEMs.

In terms of the residue, Conjecture ELD states that $U(\,\cdot\,;|\psi\rangle)$ is a homomorphism from $\Z[d\text{-GEMs}]$ to the multiplicative group $\C^*$ which factors through $R$. Namely, writing $A(\,\cdot\,;|\psi\rangle)$ also for the homomorphism $\Z[\text{spherical }(<d)\text{-GEMs}]\rightarrow\C^*$, $\gamma\mapsto A(\gamma;|\psi\rangle)$, one has $U(\Gamma;|\psi\rangle)=A(R(\Gamma);|\psi\rangle)$. This is the form in which the conjecture is used below.

Due to the inherent non-universality of the factor $U(\Gamma;|\psi\rangle)$, Conjecture ELD cannot be used directly to extract any universal information about the topological order. However, since $U(\Gamma;|\psi\rangle)=A(R(\Gamma);|\psi\rangle)$, by (\ref{multiplicativity-graphs}) the non-universal factors cancel in every product of multi-invariants of GEMs with vanishing total residue. This leads to the following statement, which involves only computable quantities. It is equivalent to Conjecture ELD for $d\not\equiv 0\mod 4$, and in general when the partition functions do not vanish (Proposition \ref{prop:equivalence}).

\vspace{3ex}
\noindent\textbf{Conjecture TS (Topological Signal):} \textit{Let $\Gamma_i$ be $d$-GEMs and $n_i$ integers such that $\sum_in_iR(\Gamma_i)=0$ and $Z(M_{\Gamma_i})\neq 0$ whenever $n_i<0$. Then, in the limit $L\rightarrow\infty$,}
\vspace{1ex}
\begin{equation}
    \boxed{\;
    \prod_i\CZ(\Gamma_i;|\psi\rangle)^{n_i}=\prod_iZ(M_{\Gamma_i})^{n_i}+O(e^{-\mathrm{const}\cdot L}).\;
    }
    \label{conj-TS}
\end{equation}
\vspace{1ex}

\noindent
The left-hand side of (\ref{conj-TS}) is a signal of genuine $(d+1)$-partite entanglement, by the remark following (\ref{residue-hom}), and the conjecture asserts that every signal of this form is topological: in the limit it computes the corresponding combination of partition functions of the low-energy TQFT, with no non-universal factor left. The condition on the negative exponents makes the right-hand side well defined. If all the partition functions in (\ref{conj-TS}) are non-zero, the error term can again be written as a factor $1+O(e^{-\mathrm{const}\cdot L})$. If the right-hand side vanishes, the signal is only claimed to be exponentially small. Both sides of (\ref{conj-TS}) are computable, the left-hand side from the ground state and the right-hand side from the data of the TQFT. The family of signals to which the conjecture applies is large. The signal of \cite{DelZotto:2026dgp} is of this form, so that Conjecture TS contains the conjecture of that paper. In this paper we construct signals of this form in which all but one of the $\Gamma_i$ are spherical and the remaining one enters with the exponent $1$. Such a signal exists provided a certain condition on the residue of the remaining GEM is satisfied (Section \ref{sec:results}), in particular whenever it encodes a null-bordant manifold, in which case it is built explicitly out of the substitution graphs of Section \ref{sec:boolean}.

\begin{proposition}
\label{prop:equivalence}
Conjecture ELD implies Conjecture TS. Conversely, Conjecture TS implies that (\ref{conj-ELD}) holds, with numbers $A(\gamma;|\psi\rangle)$ independent of $\Gamma$, for every $d$-GEM $\Gamma$ such that $NR(\Gamma)=\sum_in_iR(\Gamma_i)$ for some integer $N\geq 1$, $d$-GEMs $\Gamma_i$ with $Z(M_{\Gamma_i})\neq 0$ and integers $n_i$. This is the case if $Z(M_\Gamma)\neq 0$ or, by Theorem \ref{thm:structure} below, if $[M_\Gamma]\in\Omega^{\mathrm{SPL}}_d$ is a torsion class. In particular, for $d\not\equiv 0\mod 4$ the two conjectures are equivalent without any additional assumptions.
\end{proposition}

\begin{proof}
Assume Conjecture ELD, and let $\Gamma_i$ and $n_i$ be as in Conjecture TS. By (\ref{multiplicativity-graphs}) and (\ref{conj-ELD}),
\begin{equation}
    \prod_i\CZ(\Gamma_i;|\psi\rangle)^{n_i}=A\Big(\sum_in_iR(\Gamma_i);|\psi\rangle\Big)\prod_i\big(Z(M_{\Gamma_i})+O(e^{-\mathrm{const}\cdot L})\big)^{n_i},
    \label{TS-from-ELD}
\end{equation}
and the argument of $A$ vanishes by assumption. Since $Z(M_{\Gamma_i})\neq 0$ whenever $n_i<0$, the corresponding multi-invariants do not vanish for $L$ large enough, and the product on the right-hand side is $\prod_iZ(M_{\Gamma_i})^{n_i}+O(e^{-\mathrm{const}\cdot L})$. This gives (\ref{conj-TS}). The converse, which relies on Theorem \ref{thm:structure} for the torsion classes, is proven in Section \ref{sec:bordism}.
\end{proof}

\subsection{The residue quotient and the main results}
\label{sec:results}

Conjecture TS applies to combinations of GEMs with vanishing total residue. Combinations of partition functions of other manifolds can still be isolated by canceling the residue against spherical GEMs, for which the partition function is the positive number (\ref{Z-sphere-positive}). Let $\Z[\text{spherical }d\text{-GEMs}]\subseteq\Z[d\text{-GEMs}]$ be the subgroup generated by the spherical $d$-GEMs. We define the \textit{residue quotient}
\begin{equation}
    \Qd:=\frac{R(\Z[d\text{-GEMs}])}{R(\Z[\text{spherical }d\text{-GEMs}])},
    \label{Qd-def}
\end{equation}
and denote by $[\Gamma]\in\Qd$ the class of $R(\Gamma)$. By definition, $\sum_in_i[\Gamma_i]=0$ if and only if there is a formal linear combination $\eta=\sum_jm_j\Lambda_j$ of spherical $d$-GEMs with $R(\eta)=\sum_in_iR(\Gamma_i)$, that is, if and only if the residue of $\sum_in_i\Gamma_i$ can be ``canceled against spheres''. The residue quotient is the group of obstructions to such a cancellation. The following proposition says that, given the conjectures of Section \ref{sec:residues}, it is exactly the obstruction to the extraction of $\prod_iZ(M_{\Gamma_i})^{n_i}$.

\begin{proposition}
\label{prop:extract}
Assume Conjecture ELD, and let $\Gamma_i$ be $d$-GEMs and $n_i$ integers.
\begin{enumerate}
    \item[(i)] If $\sum_in_i[\Gamma_i]=0\in\Qd$ and $Z(M_{\Gamma_i})\neq 0$ whenever $n_i<0$, then for any spherical $d$-GEMs $\Lambda_j$ and integers $m_j$ with $\sum_jm_jR(\Lambda_j)=\sum_in_iR(\Gamma_i)$, in the limit $L\rightarrow\infty$,
    \begin{equation}
        \frac{\prod_i\CZ(\Gamma_i;|\psi\rangle)^{n_i}}{\prod_j\CZ(\Lambda_j;|\psi\rangle)^{m_j}}=\frac{\prod_iZ(M_{\Gamma_i})^{n_i}}{Z(S^d)^{\sum_jm_j}}+O(e^{-\mathrm{const}\cdot L}).
        \label{extract}
    \end{equation}
    \item[(ii)] Conversely, assume that $Z(M_{\Gamma_i})\neq 0$ for all $i$. If for some spherical $d$-GEMs $\Lambda_j$ and integers $m_j$ the left-hand side of (\ref{extract}) has a limit independent of the homomorphism $U$ in (\ref{conj-ELD}), then $\sum_jm_jR(\Lambda_j)=\sum_in_iR(\Gamma_i)$. In particular, $\sum_in_i[\Gamma_i]=0$.
    \item[(iii)] For even $d$, the identity $\sum_jm_jR(\Lambda_j)=\sum_in_iR(\Gamma_i)$ implies $2\sum_jm_j=\sum_in_i\chi(M_{\Gamma_i})$.
\end{enumerate}
\end{proposition}

\begin{proof}
Part (i) is Conjecture TS, which follows from Conjecture ELD by Proposition \ref{prop:equivalence}, for the collection consisting of the $\Gamma_i$ with the exponents $n_i$ and the $\Lambda_j$ with the exponents $-m_j$. This collection has vanishing total residue, and $Z(M_{\Lambda_j})=Z(S^d)>0$, so that the condition on the negative exponents is satisfied. For (ii), by (\ref{multiplicativity-graphs}) and (\ref{conj-ELD}) the left-hand side of (\ref{extract}) is $\prod_iZ(M_{\Gamma_i})^{n_i}\,Z(S^d)^{-\sum_jm_j}\,A\big(\sum_in_iR(\Gamma_i)-\sum_jm_jR(\Lambda_j);|\psi\rangle\big)\big(1+O(e^{-\mathrm{const}\cdot L})\big)$, where we used that all the partition functions involved are non-zero. If the argument of $A$ does not vanish, the limit depends on $U$, since $A$ ranges over all homomorphisms from the free abelian group $\Z[\text{spherical }(<d)\text{-GEMs}]$ to $\C^*$. Namely, there exists $A$ such that its value on this argument differs from $1$, the value of the trivial homomorphism. Finally, (iii) follows by applying the functional $\lambda_\chi$ of (\ref{chi-linear}) in Section \ref{sec:boolean} to the residue identity.
\end{proof}

We state the converse (ii) for completeness, but do not want to overemphasize it. It treats the homomorphism $U$ in (\ref{conj-ELD}) as arbitrary, and it is not obvious that all homomorphisms of this form are realized by ground states of local Hamiltonians. Part (i), however, is unconditional given the conjecture. Part (iii) says that for even $d$ the power of $Z(S^d)$ in (\ref{extract}) is forced, see Remark 2 below.

The structure of the residue quotient is described by the following theorem, which is proven in Sections \ref{sec:bordism} and \ref{sec:kernel}. We denote by $\Omega^{\mathrm{SPL}}_d$ the bordism group of closed oriented PL $d$-manifolds, so that $[M]=0\in\Omega^{\mathrm{SPL}}_d$ means that $M$ is the boundary of a compact oriented PL $(d+1)$-manifold. By Section \ref{sec:gem}, this is in particular the case if $M$ is the boundary of a compact oriented smooth manifold.

\begin{theorem}
\label{thm:structure}
The assignment $[M_\Gamma]\mapsto[\Gamma]$ defines a surjective homomorphism
\begin{equation}
    \Phi:\;\Omega^{\mathrm{SPL}}_d\longrightarrow\Qd
    \label{Phi-def}
\end{equation}
with the following properties.
\begin{enumerate}
    \item[(i)] $\Phi\otimes\Q$ is an isomorphism. In particular, the kernel of $\Phi$ consists of torsion elements, and $[\Gamma]=0\in\Qd$ implies that all rational Pontryagin numbers of $M_\Gamma$ vanish.
    \item[(ii)] $\Phi$ is an isomorphism for $d\leq 4$. In particular,
    \begin{equation}
        \Phi:\;\Omega^{\mathrm{SPL}}_4\cong\Z\xrightarrow{\;\;\cong\;\;}\mathcal{Q}_4,
    \end{equation}
    the isomorphism being given by the signature.
    \item[(iii)] $\Phi$ vanishes on the image of the torsion subgroup of $\Omega^{\mathrm{SO}}_d$ in $\Omega^{\mathrm{SPL}}_d$: if $M$ is a closed oriented smooth manifold with $N[M]=0\in\Omega^{\mathrm{SO}}_d$ for some $N\geq 1$, then $[\Gamma]=0\in \mathcal{Q}_d$ for every GEM $\Gamma$ of $M$. Consequently, the kernel of the composite $\Omega^{\mathrm{SO}}_d\hookrightarrow\Omega^{\mathrm{SPL}}_d\rightarrow\Qd$ is exactly the torsion subgroup of $\Omega^{\mathrm{SO}}_d$. In particular, $\Phi$ is not injective in general: for $d=5$, the generator of $\Omega^{\mathrm{SPL}}_5\cong\Z_2$, the Dold manifold $P(1,2)$, has a GEM with the same residue as a GEM of the null-bordant manifold $\CP^2\times S^1$, and $\mathcal{Q}_5=0$.
    \item[(iv)] The direct sum $\mathcal{Q}_*=\bigoplus_{d\geq 1}\Qd$ is a non-unital graded commutative ring, with the product induced by the product of manifolds, and $\Phi:\Omega^{\mathrm{SPL}}_*\rightarrow\mathcal{Q}_*$ is a homomorphism of graded rings.
\end{enumerate}
\end{theorem}

Part (iii) says that for smooth manifolds the kernel of $\Phi$ is exactly the torsion: the residue class of a GEM of a closed oriented smooth manifold $M$ vanishes if and only if all rational Pontryagin numbers of $M$ vanish, so that this class is precisely the collection of the rational Pontryagin numbers of $M$. The composite $\Omega^{\mathrm{SO}}_d\rightarrow\Qd$ is moreover surjective for $d\leq 7$, where every PL manifold is smoothable (Section \ref{sec:gem}), so that $\Qd\cong\Omega^{\mathrm{SO}}_d/\mathrm{torsion}$ for $d\leq 7$. Whether $\Qd\cong\Omega^{\mathrm{SPL}}_d/\mathrm{torsion}$ for all $d$ remains open: the proof of (iii) relies on Wall's description of the torsion of $\Omega^{\mathrm{SO}}_*$ by explicit manifolds \cite{Wall1960}, and in high dimensions $\Omega^{\mathrm{SPL}}_d$ contains torsion classes not represented by smooth manifolds \cite[Chapter 14]{MadsenMilgram1979}, to which the argument does not apply. Any torsion in $\Qd$ would have to come from such classes.

Theorem \ref{thm:structure} and Proposition \ref{prop:extract} together give the extraction scheme for null-bordant manifolds.

\begin{corollary}[null-bordism signal, NBS]
\label{cor:extract}
Assume Conjecture ELD, and let $\Gamma$ be a $d$-GEM with $[M_\Gamma]=0\in\Omega^{\mathrm{SPL}}_d$. Then there exist spherical $d$-GEMs $\Lambda_j$ and integers $m_j$, depending only on $\Gamma$ and not on $|\psi\rangle$, such that, in the limit $L\rightarrow\infty$,
\begin{equation}
    \boxed{\;
    \frac{Z(M_\Gamma)}{Z(S^d)^{N_\Gamma}}=\frac{\CZ(\Gamma;|\psi\rangle)}{\prod_j\CZ(\Lambda_j;|\psi\rangle)^{m_j}}+O(e^{-\mathrm{const}\cdot L}),\qquad N_\Gamma:=\sum_jm_j.\;
    }
    \label{extract-null}
\end{equation}
More generally, if $\Gamma_i$ are $d$-GEMs and $n_i$ integers with $\sum_in_i[M_{\Gamma_i}]=0\in\Omega^{\mathrm{SPL}}_d$ and $Z(M_{\Gamma_i})\neq 0$ whenever $n_i<0$, then (\ref{extract}) holds for suitable spherical $d$-GEMs $\Lambda_j$ and integers $m_j$.
\end{corollary}

\begin{proof}
By Theorem \ref{thm:structure}, $\Phi$ is a homomorphism, so that $\sum_in_i[\Gamma_i]=\Phi\big(\sum_in_i[M_{\Gamma_i}]\big)=0$, and Proposition \ref{prop:extract}(i) applies. The $\Lambda_j$ and $m_j$ form any combination of spherical GEMs with the same residue as $\sum_in_i\Gamma_i$, which is a condition on the graphs only.
\end{proof}

The part of Theorem \ref{thm:structure} used by the corollary, i.e. the fact that $\Phi$ is well defined, rests on two ingredients of a rather different nature. The first is an elementary combinatorial identity satisfied by the graphs obtained from \textit{any} $(d+2)$-colored graph by the replacement procedure of Remark 4 below (Lemma \ref{lem:boolean}), which is the algebraic reason for the cancellation of the non-universal factors in a signal. The second is a geometric construction from crystallization theory that produces, from a null-bordism $W$ of $M_\Gamma$, a $(d+2)$-colored graph for which all these graphs have a prescribed topology (Theorem \ref{thm:filling}). It is the second ingredient that requires the null-bordance assumption, and it is also what makes the graphs $\Lambda_j$ explicit.

\medskip\noindent\textbf{Remarks.}
\begin{enumerate}
    \item \textit{Beyond null-bordant manifolds.} The hypothesis $[M_\Gamma]=0\in\Omega^{\mathrm{SPL}}_d$ is not the weakest one under which (\ref{extract-null}) can be expected. By Proposition \ref{prop:extract} and Theorem \ref{thm:structure}(iii), Conjecture ELD implies (\ref{extract-null}) also when $[M_\Gamma]$ is a torsion class represented by a smooth manifold, i.e. when $M_\Gamma$ is PL-bordant to a smooth manifold $M$ with $N[M]=0\in\Omega^{\mathrm{SO}}_d$ for some $N\geq 1$. For $d\leq 7$ this simply means that $[M_\Gamma]$ is a torsion class, i.e. that all rational Pontryagin numbers of $M_\Gamma$ vanish. In this case, however, the graphs $\Lambda_j$ are only known to exist, and no construction comparable to the one of Remark 4 below is available, with the exception of $d=5$ (proof of Proposition \ref{prop:smooth-torsion}, used in Section \ref{sec:dold-example}). We therefore keep the null-bordance hypothesis in Corollary \ref{cor:extract}.

    \item \textit{The power of $Z(S^d)$.} For even $d$ one necessarily has
    \begin{equation}
        N_\Gamma=\frac{\chi(M_\Gamma)}{2},
    \end{equation}
    for \textit{any} choice of $(\Lambda_j,m_j)$ for which (\ref{extract-null}) holds independently of the non-universal factors (Proposition \ref{prop:extract}(iii)). This follows from the fact that the Euler characteristic $\chi(M_\Gamma)$ is a linear function of the same data of $\Gamma$ that enters the non-universal factors, see (\ref{chi-linear}) below, and $\chi(S^d)=2$. Physically, $Z(M)/Z(S^d)^{\chi(M)/2}$ is the combination insensitive to the Euler counterterm $Z(M)\rightarrow e^{\lambda\chi(M)}Z(M)$, which is the ambiguity in the normalization of the partition function of a TQFT in even dimensions. In particular, $Z(S^d)$ itself is not accessible for even $d$. For $d\equiv 3\mod 4$ the spherical GEMs $\Lambda_j$ can always be chosen (Proposition \ref{prop:exponent}) such that $N_\Gamma=0$. For $d\equiv 1\mod 4$ they can be chosen such that $N_\Gamma\in\{0,1\}$. In that case the parity of $N_\Gamma$ produced by the construction of Remark 4 below is fixed by $M_\Gamma$, and is given by its rational Kervaire semicharacteristic (Lemma \ref{lem:parity}).

    \item \textit{The sphere.} For odd $d$, $Z(S^d)^2$ is itself a ratio of multi-invariants of spherical GEMs. Namely (Proposition \ref{prop:exponent}), there exist spherical $d$-GEMs $\Lambda_j$ and integers $m_j$ with $\sum_jm_j=2$ such that
    \begin{equation}
        Z(S^d)^2=\prod_j\CZ(\Lambda_j;|\psi\rangle)^{m_j}\;\big(1+O(e^{-\mathrm{const}\cdot L})\big).
        \label{sphere-squared}
    \end{equation}
    This is Conjecture TS for the collection $(\Lambda_j,m_j)$, the total residue of which vanishes. Together with (\ref{Z-sphere-positive}) it determines $Z(S^d)$, and combining with (\ref{extract-null}), the partition function $Z(M_\Gamma)$ itself for any null-bordant $M_\Gamma$. In $d=3$, (\ref{sphere-squared}) computes $\CD^{-2}$ and can be considered as a multi-invariant analogue of the topological entanglement entropy of \cite{Kitaev:2005dm,levin2006detecting}. The graphs $\Lambda_j$ in (\ref{sphere-squared}) are the connected components of the graphs obtained from the $(d+2)$-cube graph, a GEM of $S^{d+1}$ (see Figure \ref{fig:hypercube}), by the replacement procedure of Remark 4 below. This construction generalizes the graph used in Appendix B of \cite{DelZotto:2026dgp} to normalize multi-invariants in $d=3$.

    \item \textit{The scheme.} There is a natural construction of $(\Lambda_j,m_j)$ starting from a compact oriented $(d+1)$-manifold $W$ with $\partial W=M_\Gamma$, see Section \ref{sec:daughters} and Appendix \ref{app:filling}. A suitable triangulation, colored by $d+2$ colors, of the closed pseudomanifold obtained from $W$ by coning off its boundary components, in which the links of the cone points are the boundary components of $W$ with the triangulations encoded by $\Gamma$, provides a $(d+2)$-tuple of permutations $(\sigma_0,\ldots,\sigma_d,\sigma_*)$ of a common set of replicas. The graphs $\Lambda_j$ are the connected components of the $2^{d+1}-1$ graphs obtained by replacing in the $(d+1)$-tuple $(\sigma_0,\ldots,\sigma_d)$ a non-empty subset of the entries by $\sigma_*$, together with the connected components of the graph of $(\sigma_0,\ldots,\sigma_d)$ other than $\Gamma$, if any. The numbers $m_j$ are equal to $+1$ or $-1$ according to whether the number of replaced entries is odd or even. The right-hand side of (\ref{extract-null}) is then a signal of genuine $(d+1)$-partite entanglement in the sense of \cite{Gadde:2026sig} (see Section \ref{sec:residues}), and $N_\Gamma$ is determined by the Euler characteristic of $W$ (Lemma \ref{lem:daughters}). In particular, for $d=3$ the power $N_\Gamma$ depends on the choice of the null-bordism $W$, and comparing two different choices (e.g. $W$ and $W\#\CP^2$) gives yet another way to extract $Z(S^3)$.
\end{enumerate}

\subsection{Summary by dimension}
\label{sec:table}

Combining Proposition \ref{prop:extract}, Theorem \ref{thm:structure} and Corollary \ref{cor:extract} with the remarks above, one obtains the following picture, summarized in Table \ref{tab:dims}.

For $d=2$ and $d=3$ the bordism groups (PL or smooth, which coincide for $d\leq 7$) vanish and every closed oriented manifold bounds. For $d=2$ the extractable quantity is $Z(M)/Z(S^2)^{\chi(M)/2}$, which is the standard combination independent of the Euler counterterm. For $d=3$ one obtains the full complex $Z(M)$ for every $M$, with $Z(S^3)$ itself provided by the $5$-cube signal. For $d=4$, $\Omega^{\mathrm{SPL}}_4\cong\Z$ is detected by the signature and coincides with $\mathcal{Q}_4$, so that $Z(M)/Z(S^4)^{\chi(M)/2}$ is extractable if and only if $\sigma(M)=0$, and more generally $\prod_iZ(M_i)^{n_i}/Z(S^4)^{\sum_in_i\chi(M_i)/2}$ is extractable if and only if $\sum_in_i\sigma(M_i)=0$. In particular, $|Z(M)|^2/Z(S^4)^{\chi(M)}$ is always extractable, in agreement with the fact that this is what the signals of \cite{DelZotto:2026dgp} produce. For $d=5$, although $\Omega^{\mathrm{SPL}}_5\cong\Z_2$, the residue quotient vanishes (Theorem \ref{thm:structure}(iii)) and $Z(M)$ is extractable for every closed oriented $5$-manifold. An exactly solvable example in which the sign $Z(P(1,2))=-1$ of the Dold manifold is extracted is worked out in Section \ref{sec:dold-example}. For $d=6,7$ the bordism groups vanish again. In general, the vanishing of all rational Pontryagin numbers of $\sum_in_i[M_i]$ is necessary for the extraction of $\prod_iZ(M_i)^{n_i}$, and it is also sufficient if all the $M_i$ are smooth (Theorem \ref{thm:structure}(iii)), or if $\Qd$ has no torsion. The rational Pontryagin numbers are non-trivial only for $d\equiv 0\mod 4$.

\begin{table}[ht]
\centering
\footnotesize
\begin{tabular}{@{}clll@{}}
\toprule
$d$ & $\Omega^{\mathrm{SPL}}_d$ & $\Qd$ & extractable (assuming Conjecture TS)\\
\midrule
$2$ & $0$ & $0$ & $Z(M)/Z(S^2)^{\chi(M)/2}$ for all $M$\\
$3$ & $0$ & $0$ & $Z(M)$ for all $M$, including $Z(S^3)$\\
$4$ & $\Z$ ($\sigma$) & $\Z$ & $Z(M)/Z(S^4)^{\chi(M)/2}$ iff $\sigma(M)=0$, $|Z(M)|^2/Z(S^4)^{\chi(M)}$ always\\
$5$ & $\Z_2$ ($w_2w_3$) & $0$ & $Z(M)$ for all $M$, including $Z(S^5)$\\
$6,7$ & $0$ & $0$ & $Z(M)/Z(S^6)^{\chi(M)/2}$, resp. $Z(M)$, for all $M$\\
$4k$ & $\Z^{p(k)}\oplus\text{tors}$ & $\supseteq\Z^{p(k)}$ & $\prod_iZ(M_i)^{n_i}/Z(S^d)^{\sum_in_i\chi(M_i)/2}$ only if $\sum_in_iP[M_i]=0$ for all $P$\\
\bottomrule
\end{tabular}
\caption{Partition functions extractable from multi-invariants of the ground state, by dimension. Here $P$ runs over the Pontryagin numbers of degree $d=4k$ and $p(k)$ is the number of partitions of $k$. For even $d$ the power of $Z(S^d)$ is forced, for odd $d$ it can be removed.}
\label{tab:dims}
\end{table}

\subsection{Example: six replicas for the lens space \texorpdfstring{$L(3,1)$}{L(3,1)}}
\label{sec:example}

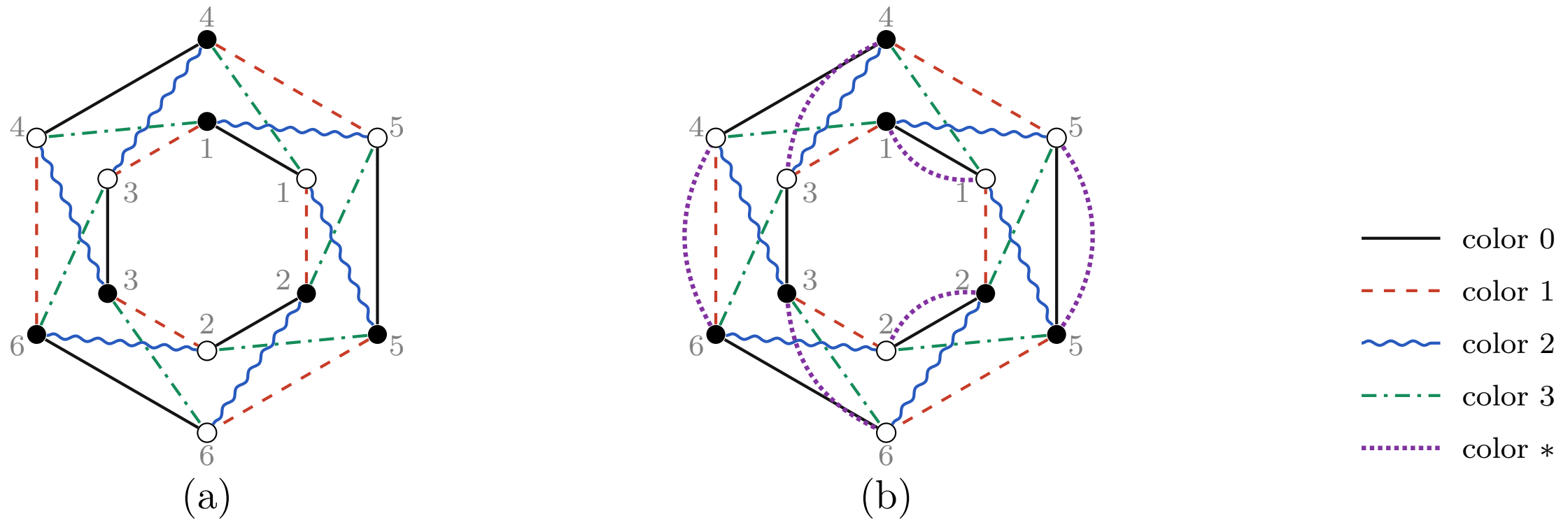
\begin{figure}[ht]
\centering
\newcommand{\lenshexagons}{%
  \def\rin{1.75}\def\rout{3.00}\def\lin{1.34}\def\lout{3.34}
  \node[bl] (b1) at ( 90:\rin) {}; \node[gray] at ( 90:\lin) {$1$};
  \node[wh] (w1) at ( 30:\rin) {}; \node[gray] at ( 30:\lin) {$1$};
  \node[bl] (b2) at (330:\rin) {}; \node[gray] at (330:\lin) {$2$};
  \node[wh] (w2) at (270:\rin) {}; \node[gray] at (270:\lin) {$2$};
  \node[bl] (b3) at (210:\rin) {}; \node[gray] at (210:\lin) {$3$};
  \node[wh] (w3) at (150:\rin) {}; \node[gray] at (150:\lin) {$3$};
  \node[bl] (b4) at ( 90:\rout) {}; \node[gray] at ( 90:\lout) {$4$};
  \node[wh] (w5) at ( 30:\rout) {}; \node[gray] at ( 30:\lout) {$5$};
  \node[bl] (b5) at (330:\rout) {}; \node[gray] at (330:\lout) {$5$};
  \node[wh] (w6) at (270:\rout) {}; \node[gray] at (270:\lout) {$6$};
  \node[bl] (b6) at (210:\rout) {}; \node[gray] at (210:\lout) {$6$};
  \node[wh] (w4) at (150:\rout) {}; \node[gray] at (150:\lout) {$4$};
  \draw[cc2] (b1) -- (w5); \draw[cc2] (b5) -- (w1); \draw[cc2] (b2) -- (w6);
  \draw[cc2] (b6) -- (w2); \draw[cc2] (b3) -- (w4); \draw[cc2] (b4) -- (w3);
  \draw[cc3] (b1) -- (w4); \draw[cc3] (b4) -- (w1); \draw[cc3] (b2) -- (w5);
  \draw[cc3] (b5) -- (w2); \draw[cc3] (b3) -- (w6); \draw[cc3] (b6) -- (w3);
  \draw[cc0] (b1) -- (w1); \draw[cc0] (b2) -- (w2); \draw[cc0] (b3) -- (w3);
  \draw[cc0] (b4) -- (w4); \draw[cc0] (b5) -- (w5); \draw[cc0] (b6) -- (w6);
  \draw[cc1] (b1) -- (w3); \draw[cc1] (b2) -- (w1); \draw[cc1] (b3) -- (w2);
  \draw[cc1] (b4) -- (w5); \draw[cc1] (b5) -- (w6); \draw[cc1] (b6) -- (w4);
}
\begin{tikzpicture}[scale=0.62,font=\scriptsize]
  \lenshexagons
  \node at (0,-4.0) {\small (a)};
\end{tikzpicture}
\hfill
\begin{tikzpicture}[scale=0.62,font=\scriptsize]
  \lenshexagons
  \draw[cc4] (b1) to[bend right=35] (w1); \draw[cc4] (b2) to[bend right=35] (w2); \draw[cc4] (b5) to[bend right=35] (w5);
  \draw[cc4] (b3) to[bend right=30] (w6); \draw[cc4] (b6) to[bend left=30] (w4); \draw[cc4] (b4) to[bend right=30] (w3);
  \node at (0,-4.0) {\small (b)};
\end{tikzpicture}
\hfill
\begin{tikzpicture}[scale=0.62,font=\scriptsize]
  \draw[cc0] (0,0) -- (1.2,0);           \node[right] at (1.3,0) {color $0$};
  \draw[cc1] (0,-0.8) -- (1.2,-0.8);     \node[right] at (1.3,-0.8) {color $1$};
  \draw[cc2] (0,-1.6) -- (1.2,-1.6);     \node[right] at (1.3,-1.6) {color $2$};
  \draw[cc3] (0,-2.4) -- (1.2,-2.4);     \node[right] at (1.3,-2.4) {color $3$};
  \draw[cc4] (0,-3.2) -- (1.2,-3.2);     \node[right] at (1.3,-3.2) {color $*$};
  \path (0,1.5) (0,-4.5);
\end{tikzpicture}
\caption{(a) The six-replica GEM $\Gamma$ of the lens space $L(3,1)$ defined by the permutations (\ref{L31-perms}). The labels are the replica indices. The two circles are the $\{0,1\}$-colored cycles of $\Gamma$. (b) The $5$-colored graph $\hG$ obtained by adjoining the permutation $\sigma_*$ of (\ref{L31-star}), so that $\Gamma=\rs{\hG}{\hat *}$.}
\label{fig:lens}
\end{figure}

The construction of Remark 4 in general increases the replica number, since the triangulation of the null-bordism $W$ has more $(d+1)$-simplices than the triangulation of $M_\Gamma$ has $d$-simplices. The following example shows that this is not always necessary. Consider the four permutations
\begin{equation}
    \sigma_0=\mathrm{id},\qquad
    \sigma_1=(132)(456),\qquad
    \sigma_2=(15)(26)(34),\qquad
    \sigma_3=(14)(25)(36),
    \label{L31-perms}
\end{equation}
of $r=6$ replicas.\footnote{In one-line notation, with the $i$-th entry being $\sigma_\alpha(i)$: $\sigma_0=(1,2,3,4,5,6)$, $\sigma_1=(3,1,2,5,6,4)$, $\sigma_2=(5,6,4,3,1,2)$, $\sigma_3=(4,5,6,1,2,3)$.} The corresponding graph $\Gamma$, shown in Figure \ref{fig:lens}(a), is the standard crystallization of the lens space $L(3,1)$ \cite{FerriGagliardi1982,Lins1995,BasakDatta2014}, which appeared in Appendix C of \cite{DelZotto:2026dgp}: the bicolored subgraphs $\rs{\Gamma}{I}$ have $2,3,3,3,3,2$ connected components for $I=\{0,1\},\{0,2\},\{0,3\},\{1,2\},\{1,3\},\{2,3\}$, and the four $3$-colored subgraphs $\rs{\Gamma}{\hat\alpha}$ are connected GEMs of $S^2$. Now adjoin the fifth permutation
\begin{equation}
    \sigma_*=(364),
    \label{L31-star}
\end{equation}
i.e. $\sigma_*=(1,2,6,3,5,4)$ in one-line notation. The resulting $5$-colored graph $\hG$ is shown in Figure \ref{fig:lens}(b). For $J\subseteq\{0..3\}$ let $H_J$ be the $4$-colored graph of the permutations $(\sigma_0,\ldots,\sigma_3)$ in which $\sigma_\alpha$ is replaced by $\sigma_*$ for every $\alpha\in J$, so that $H_\emptyset=\Gamma$. The sixteen graphs $H_J$ are drawn in Appendix \ref{app:daughters}. All the fifteen graphs $H_J$, $J\neq\emptyset$, turn out to be disjoint unions of GEMs of $S^3$, with the following numbers $c_J$ of connected components ($J$ written as the list of its elements):
\begin{center}
\begin{tabular}{c@{\,:\,}c@{\qquad}c@{\,:\,}c@{\qquad}c@{\,:\,}c@{\qquad}c@{\,:\,}c@{\qquad}c@{\,:\,}c}
$0$&1&$1$&1&$2$&1&$3$&1&$01$&1\\
$02$&1&$03$&1&$12$&2&$13$&2&$23$&1\\
$012$&3&$013$&3&$023$&2&$123$&4&$0123$&6
\end{tabular}
\end{center}
The fact that every component is indeed an $S^3$ was verified by reducing the graphs to $\Gamma_{S^3}$ by dipole moves \cite{FerriGagliardi1982,GrasselliMulazzani2020}. Corollary \ref{cor:extract} for $\Gamma$ then takes the form
\begin{equation}
    \frac{Z(L(3,1))}{Z(S^3)^2}=\lim_{L\rightarrow\infty}\prod_{J\subseteq\{0..3\}}\CZ(H_J;|\psi\rangle)^{(-1)^{|J|}},
    \label{L31-scheme}
\end{equation}
with the $\Lambda_j$ being the connected components of the $H_J$, $J\neq\emptyset$, with $m_j=-(-1)^{|J|}$, so that $N_\Gamma=-\sum_{J\neq\emptyset}(-1)^{|J|}c_J=4-8+12-6=2$. This is a ratio of sixteen six-replica multi-invariants, all of the same family $\{\sigma_0,\ldots,\sigma_3,\sigma_*\}$. The normalization $Z(S^3)^2$ is in turn given by the $5$-cube signal of Remark 3 (sixteen multi-invariants with $r=16$). The reason why this works is explained in Section \ref{sec:daughters}: the five permutations encode a colored triangulation of the pseudomanifold obtained from a null-bordism $W$ of $L(3,1)$ by coning off its boundary, as in Remark 4, with no additional vertices.

\section{Proofs and details}
\label{sec:proofs}

\begin{figure}[ht]
\centering
\begin{tikzpicture}[x=0.72cm,y=0.85cm,
  st/.style={draw,rounded corners=2pt,inner sep=2.5pt,font=\footnotesize,align=center,minimum height=1.9em},
  main/.style={st,thick,fill=black!6},
  conj/.style={st,dashed,fill=white},
  app/.style={st,fill=white,draw=gray},
  dep/.style={-{Stealth[length=5pt]},shorten >=1.5pt,shorten <=1.5pt}]
  \node[app] (L13) at (1.5,0) {Lemma \ref{lem:pseudo}\\[-2pt]\scriptsize pseudomanifolds};
  \node[app] (L15) at (12.1,0) {Lemma \ref{lem:dual}\\[-2pt]\scriptsize dual graph};
  \node[st] (T7) at (6.8,1.7) {Thm.\ \ref{thm:filling}\\[-2pt]\scriptsize colored filling};
  \node[st] (L8) at (0.6,3.1) {Lemma \ref{lem:daughters}\\[-2pt]\scriptsize spherical $H_J$, $N_\Gamma$};
  \node[st] (T11) at (13.35,3.2) {Thm.\ \ref{thm:bordism}\\[-2pt]\scriptsize $\Omega^{\mathrm{SPL}}_d\twoheadrightarrow\Qd$};
  \node[st] (P13) at (16.7,4) {Prop.\ \ref{prop:smooth-torsion}\\[-2pt]\scriptsize smooth torsion};
  \node[app] (L14) at (4.7,4.8) {Lemma \ref{lem:susp}\\[-2pt]\scriptsize suspension};
  \node[st] (L4) at (7.8,4.15) {Lemma \ref{lem:boolean}\\[-2pt]\scriptsize identity for $H_J$};
  \node[conj] (B) at (11,4.45) {Conj.\ ELD\\[-2pt]\scriptsize form of $U$};
  \node[st] (L10) at (0.5,6.9) {Lemma \ref{lem:parity}\\[-2pt]\scriptsize $\chi$ parity};
  \node[st] (P9) at (4.7,6.9) {Prop.\ \ref{prop:exponent}\\[-2pt]\scriptsize power of $Z(S^d)$};
  \node[st] (C5) at (8.2,6.7) {Cor.\ \ref{cor:S}\\[-2pt]\scriptsize $S(\Gamma)$};
  \node[main] (P1) at (12,6.45) {Prop.\ \ref{prop:equivalence}\\[-2pt]\scriptsize TS $\Leftrightarrow$ ELD};
  \node[st] (P12) at (16.5,6.8) {Prop.\ \ref{prop:kernel}\\[-2pt]\scriptsize local formulas};
  \node[main] (T1) at (3.3,8.75) {Cor.\ \ref{cor:extract} (NBS)\\[-2pt]\scriptsize null-bordant $M_\Gamma$};
  \node[main] (P2) at (12,7.85) {Prop.\ \ref{prop:extract}\\[-2pt]\scriptsize $\Qd$ criterion};
  \node[main] (T3) at (15.3,8.75) {Thm.\ \ref{thm:structure}\\[-2pt]\scriptsize structure of $\Qd$};
  \draw[dep] (L13) -- (L8);
  \draw[dep] (L13) -- (T7);
  \draw[dep] (L15) -- (T7);
  \draw[dep] (L15) -- (T11);
  \draw[dep] (L15) -- (P13);
  \draw[dep] (T7) -- (L8);
  \draw[dep] (L14) -- (L8);
  \draw[dep] (L14) -- (P9);
  \draw[dep] (L4) -- (L8);
  \draw[dep] (L4) -- (P9);
  \draw[dep] (L4) -- (C5);
  \draw[dep] (B) -- (C5);
  \draw[dep] (B) -- (P1);
  \draw[dep] (L8) -- (L10);
  \draw[dep] (L8) -- (P9);
  \draw[dep] (L8) -- (T1);
  \draw[dep] (L8) -- (T11);
  \draw[dep] (L10) -- (T1);
  \draw[dep] (P9) -- (T1);
  \draw[dep] (C5.north) -- (T1.south east);
  \draw[dep] (P1) -- (P2);
  \draw[dep] (P2) -- (T1);
  \draw[dep] (T3) -- (T1);
  \draw[dep] (T11) -- (P1);
  \draw[dep] (T11) -- (T3);
  \draw[dep] (T11) -- (P12);
  \draw[dep] (T11) -- (P13);
  \draw[dep] (P12) -- (T3);
  \draw[dep] (P12) -- (P13);
  \draw[dep] (P13.east) to[out=10,in=-10,looseness=0.6] (T3.east);
\end{tikzpicture}
\caption{Logical dependencies among the statements of the paper: an arrow from $X$ to $Y$ means that $X$ is used in the proof of $Y$. The main statements of Section \ref{sec:main} are shaded, and the lemmas stated and proven in Appendix \ref{app:filling} are drawn in gray. Conjecture ELD enters through Proposition \ref{prop:equivalence}, Proposition \ref{prop:extract} and Corollary \ref{cor:S}, and is inherited by Corollary \ref{cor:extract}. Everything else is unconditional. The arrows into Corollary \ref{cor:extract} from Proposition \ref{prop:exponent}, Lemma \ref{lem:parity}, Corollary \ref{cor:S} and Lemma \ref{lem:daughters} refer to the remarks following it, which are proven in Section \ref{sec:bordism}. In Proposition \ref{prop:equivalence}, Theorem \ref{thm:bordism} enters only the converse part, for the torsion classes. Proposition \ref{prop:ch} and Corollary \ref{cor:ch} of Section \ref{sec:dold-example} are not shown: they use the construction in the proof of Proposition \ref{prop:smooth-torsion}, Proposition \ref{prop:extract}, Corollary \ref{cor:S} and, for the closing remark of that section, Lemma \ref{lem:parity}.}
\label{fig:dependencies}
\end{figure}
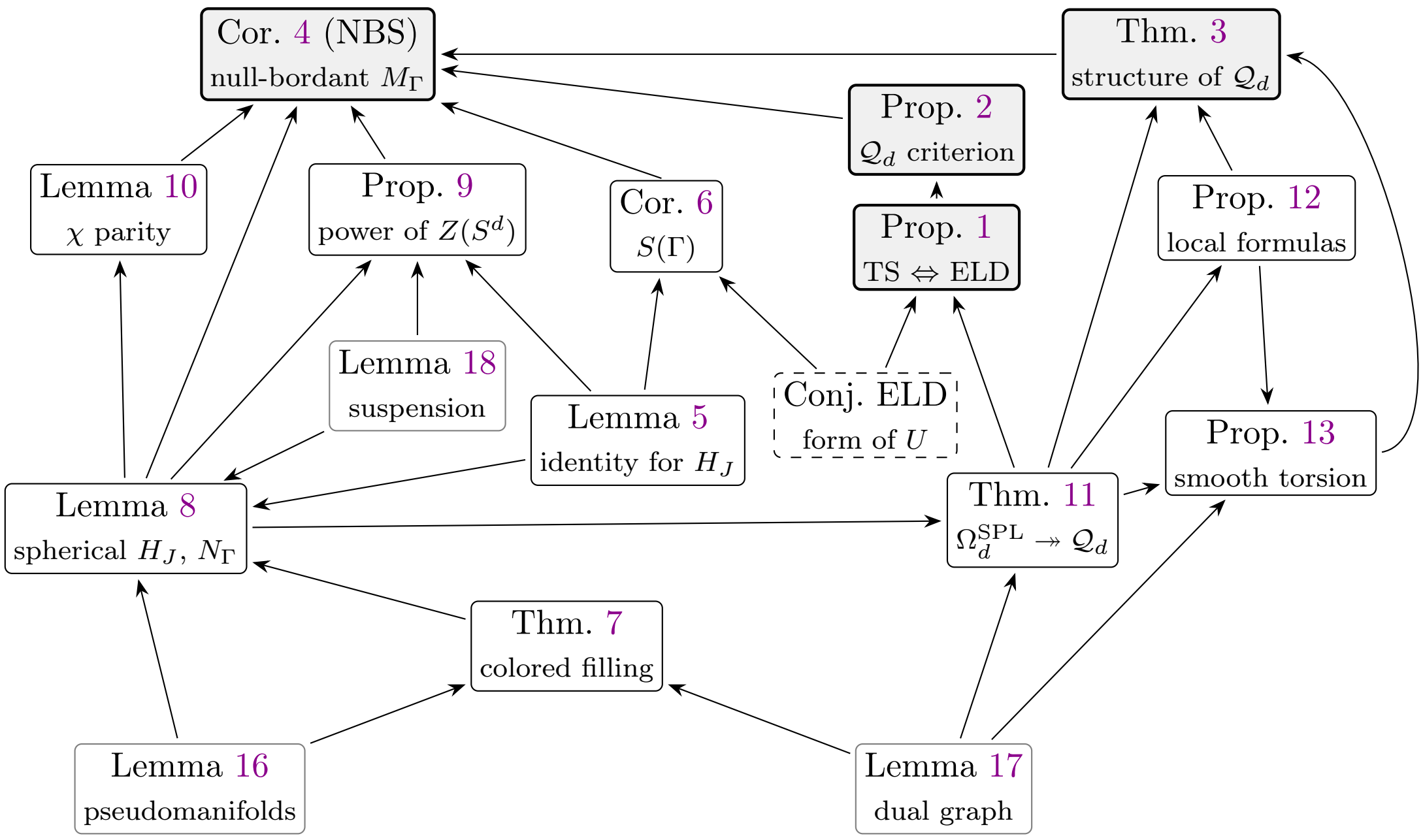

This section contains the proofs of the statements of Section \ref{sec:main}, in their detailed and, where possible, stronger form. The most technical part, the construction of colored triangulations with prescribed boundary, is deferred to Appendix \ref{app:filling}. The logical dependencies among the statements are summarized in Figure \ref{fig:dependencies}.

\subsection{Signals and substitution graphs}
\label{sec:boolean}

We start with the family of signals used in this paper. Given $d+2$ permutations $\sigma_0,\ldots,\sigma_d,\sigma_*\in S_r$, that is a bipartite $(d+2)$-colored graph $\hG$ with colors $\{0..d,*\}$, for each subset $J\subseteq\Dd$ let $H_J$ be the $(d+1)$-colored graph in which the permutation of color $\alpha$ is $\sigma_*$ if $\alpha\in J$ and $\sigma_\alpha$ otherwise:
\begin{equation}
    H_J:=\Gamma\big(\ldots,\underbrace{\sigma_\alpha}_{\alpha\notin J},\ldots,\underbrace{\sigma_*}_{\alpha\in J},\ldots\big).
    \label{H-def}
\end{equation}
We will refer to the graphs $H_J$ as the \textit{substitution graphs} of $\hG$. Note that $H_\emptyset=\rs{\hG}{\Dd}$ is obtained from $\hG$ by simply deleting the edges of color $*$, while for $|J|\geq 2$ the graph $H_J$ contains $|J|$ parallel copies of the color-$*$ edges of $\hG$, with different colors. Then
\begin{equation}
    \CZ(\sigma_0-\sigma_*,\sigma_1-\sigma_*,\ldots,\sigma_d-\sigma_*;|\psi\rangle):=\prod_{J\subseteq\Dd}\CZ(H_J;|\psi\rangle)^{(-1)^{|J|}}
    \label{signal-def}
\end{equation}
is a signal, as we will show momentarily. In the additive notation of \cite{Gadde:2026sig,DelZotto:2026dgp}, $\CE:=-\log\CZ$, it is the signal $\CE(\sigma_0-\sigma_*,\ldots,\sigma_d-\sigma_*;|\psi\rangle)$ obtained by taking the difference of two permutations in each argument. The multiplicative form (\ref{signal-def}) has the advantage of not requiring a choice of the branch of the logarithm, which is important since the multi-invariants of interest are in general complex.

We will also need some standard facts about the gluing of simplices described in Section \ref{sec:gem}. The graph $\Gamma$ produces a cell complex $K(\Gamma)$ with $|K(\Gamma)|=M_\Gamma$, the cells of which are $(|J|-1)$-simplices with vertices labeled by subsets $J\subseteq\Dd$. This is a \textit{simplicial cell complex}: two simplices can share more than one face. The cells of $K(\Gamma)$ with the set of vertex labels $J$ are in one-to-one correspondence with the connected components of $\rs{\Gamma}{\hat J}$, where, as in Section \ref{sec:residues}, $\hat J$ denotes the complement of $J$ in the set of colors of the graph. We denote their number by $c_J(\Gamma):=\#\pi_0(\rs{\Gamma}{\hat J})$. The link of the cell corresponding to a component $Q$ is $K(Q)$ \cite{FerriGagliardiGrasselli1986,GrasselliMulazzani2020}.  It follows that $\Gamma$ is a $d$-GEM if and only if every connected component of every $\rs{\Gamma}{\hat\alpha}$, $\alpha\in\Dd$, is a GEM of the $(d-1)$-sphere \cite{FerriGagliardiGrasselli1986,BandieriGagliardi2011arXiv}. Applying this recursively, all the connected components of $\rs{\Gamma}{I}$, $\emptyset\neq I\subsetneq\Dd$ of a $d$-GEM $\Gamma$ are spherical $(|I|-1)$-GEMs, as claimed in Section \ref{sec:residues}. 

Next we remark on the structure of the residue, which also clarifies the form of Conjecture ELD. The terms of $R(\Gamma)$ with $|I|=d$, the links of the vertices, determine all the others. Indeed, a connected component of $\rs{\Gamma}{I}$ with $|I|<d$ is a connected component of $(\rs{\Gamma}{\hat\alpha})|_I=\rs{\Gamma}{I}$ for each of the $d+1-|I|$ colors $\alpha\notin I$, and conversely, so that for every connected $I$-colored $\gamma$ with $|I|<d$
\begin{equation}
    \#(\gamma,\Gamma)=\frac{1}{d+1-|I|}\sum_{\gamma'}\#(\gamma,\gamma')\,\#(\gamma',\Gamma),
    \label{vertex-links-determine}
\end{equation}
where $\gamma'$ runs over the spherical $(d-1)$-GEMs with color set containing $I$ (for $I=\emptyset$ this says that $2r(d+1)$ is the total number of vertices of the $d+1$ graphs $\rs{\Gamma}{\hat\alpha}$, $\alpha\in\Dd$, i.e. of all the vertex links). Therefore two GEMs, or two formal combinations of GEMs, have the same residue if and only if they have the same multiset of colored vertex links, and the residue quotient $\Qd$ would not change if $R$ were defined to retain the vertex links only. Correspondingly, in Conjecture ELD one may assume without loss of generality that $A(\gamma;|\psi\rangle)=1$ unless $|I|=d$: choosing for each $\gamma$ with $|I|<d$ a $(d+1-|I|)$-th root $a_\gamma$ of $A(\gamma;|\psi\rangle)$ and setting $A'(\gamma';|\psi\rangle):=A(\gamma';|\psi\rangle)\prod_{\gamma}a_\gamma^{\#(\gamma,\gamma')}$ for the $(d-1)$-GEMs $\gamma'$, one has $\prod_{\gamma'}A'(\gamma';|\psi\rangle)^{\#(\gamma',\Gamma)}=U(\Gamma;|\psi\rangle)$ for all $\Gamma$ by (\ref{vertex-links-determine}), the choice of the roots being immaterial since $a_\gamma$ appears to a power divisible by $d+1-|I|$. In this form Conjecture ELD states that the non-universal factor is a product over the vertices of the triangulation of a quantity depending only on the colored link of the vertex, i.e. it is a \textit{local formula} in the sense of Section \ref{sec:kernel}, where only vertex links will enter for the same reason. We nevertheless keep the finer form (\ref{conj-ELD}), in which the factors are labeled by the actual support $I$ of the corresponding lower-partite part of the state, cf.\ (\ref{lower-partite-factorization}): it is the form in which the non-universal factors arise physically, and in which finer statements can be made. For instance, in the Levin-Wen models the factors with $|I|=3$ are trivial while those with $|I|\leq 2$, including the vertex term with $I=\emptyset$, are not \cite{DelZotto:2026dgp}, a statement that would become obscured by the absorption above.

Two simple quantities are linear functions of the residue: the replica number $r$, which is the coefficient of the $I=\emptyset$ term, and the Euler characteristic of $M_\Gamma$,
\begin{equation}
    \chi(M_\Gamma)=\sum_{\emptyset\neq J\subseteq \Dd}(-1)^{|J|-1}c_J(\Gamma),
    \label{chi-linear}
\end{equation}
where $c_J(\Gamma)$ is the number of $(|J|-1)$-dimensional cells of $K(\Gamma)$ (in particular $c_{\Dd}(\Gamma)=2r$). In other words, there is a $\Z$-linear functional $\lambda_\chi$ on $\Z[\text{spherical }(<d)\text{-GEMs}]$ such that $\chi(M_\Gamma)=\lambda_\chi(R(\Gamma))$ for every $d$-GEM $\Gamma$.

\begin{lemma}
\label{lem:boolean}
Let $\hG$ be a bipartite $(d+2)$-colored graph with colors $\{0..d,*\}$, not necessarily a GEM, and let $H_J$, $J\subseteq\Dd$, be its substitution graphs (\ref{H-def}). For every proper subset $I\subsetneq\Dd$ one has $\sum_J(-1)^{|J|}\,\rs{H_J}{I}=0$ in $\Z[\text{$I$-colored graphs}]$. Consequently,
\begin{equation}
    \sum_{J\subseteq\Dd}(-1)^{|J|}R(H_J)=0,
    \label{boolean-identity}
\end{equation}
and the product (\ref{signal-def}) equals $1$ on any state that is a tensor product of states supported on proper subsets of the parties.
\end{lemma}

\begin{proof}
Choose a color $\alpha\in\Dd\setminus I$ and pair each $J$ that does not contain $\alpha$ with $J\cup\{\alpha\}$. The graphs $H_J$ and $H_{J\cup\{\alpha\}}$ differ only in the edges of color $\alpha$, hence $\rs{H_J}{I}=\rs{H_{J\cup\{\alpha\}}}{I}$, while the two terms enter (\ref{boolean-identity}) with opposite signs. The statement about the product (\ref{signal-def}) follows from the same pairing together with (\ref{lower-partite-factorization}): a factor of the state supported on the parties $I$ contributes to $\CZ(H_J;\cdot)$ only through $\rs{H_J}{I}$.
\end{proof}
Alternatively, the last statement of the lemma follows from the general statement of \cite{Gadde:2026sig} that, for any permutations $\sigma_k$ and any coefficients $a^{(\alpha)}_k$ such that $\sum_ka^{(\alpha)}_k=0$ for every $\alpha\in\Dd$,
\begin{equation}
    \CE\Big(\sum_{k_0}a^{(0)}_{k_0}\sigma_{k_0},\ldots,\sum_{k_d}a^{(d)}_{k_d}\sigma_{k_d};|\psi\rangle\Big):=\sum_{k_0,\ldots,k_d}a^{(0)}_{k_0}\cdots a^{(d)}_{k_d}\,\CE(\sigma_{k_0},\ldots,\sigma_{k_d};|\psi\rangle)
\end{equation}
vanishes on any state that is a tensor product of states supported on proper subsets of the parties, that is, it is a \textit{signal}. The product (\ref{signal-def}) is the case of two permutations, $\sigma_\alpha$ and $\sigma_*$, in each argument.

Note that no assumption on $\hG$ is needed. The identity (\ref{boolean-identity}) is the algebraic reason, mentioned after Corollary \ref{cor:extract}, for the cancellation of the non-universal factors, and the role of the geometry below is only to choose $\hG$ in such a way that the topology of all the substitution graphs is under control. Namely, we have the following

\begin{corollary}
\label{cor:S}
Assume Conjecture ELD. Suppose that $H_\emptyset=\Gamma\sqcup T$, where $\Gamma$ is a (not necessarily connected) $d$-GEM and $T$ is a disjoint union of $t$ spherical $d$-GEMs, and that every $H_J$ with $J\neq\emptyset$ is a disjoint union of $c_J$ spherical $d$-GEMs (thus $c_J=\#\pi_0(H_J)=c_J(\hG)$, since parallel edges do not affect connectivity). Then
\begin{equation}
    S(\Gamma):=-T-\sum_{J\neq\emptyset}(-1)^{|J|}H_J\;\in\;\Z[\text{spherical }d\text{-GEMs}],\qquad R(S(\Gamma))=R(\Gamma),
    \label{S-def}
\end{equation}
the sum of the coefficients of $S(\Gamma)$ is
\begin{equation}
    N_\Gamma=-\sum_{J}(-1)^{|J|}c_J,\qquad c_\emptyset:=t,
    \label{N-Gamma}
\end{equation}
and
\begin{equation}
    \frac{\CZ(\Gamma;|\psi\rangle)}{\CZ(S(\Gamma);|\psi\rangle)}=\prod_J\CZ(H_J;|\psi\rangle)^{(-1)^{|J|}}\;\xrightarrow{\;L\rightarrow\infty\;}\;\frac{Z(M_\Gamma)}{Z(S^d)^{N_\Gamma}}.
    \label{extract-S}
\end{equation}
\end{corollary}

\begin{proof}
Equation (\ref{S-def}) is (\ref{boolean-identity}) rearranged, and the coefficient sum is $-t-\sum_{J\neq\emptyset}(-1)^{|J|}c_J$. For (\ref{extract-S}), by (\ref{multiplicativity-graphs}) and (\ref{conj-ELD}) the left-hand side is $U(R(\Gamma))\big(Z(M_\Gamma)+O(e^{-\mathrm{const}\cdot L})\big)/\big(U(R(S(\Gamma)))\big(Z(S^d)^{N_\Gamma}+O(e^{-\mathrm{const}\cdot L})\big)\big)$, where we used that $U$ factors through $R$ and is a homomorphism. Since $R(S(\Gamma))=R(\Gamma)$ and $Z(S^d)>0$, it tends to $Z(M_\Gamma)/Z(S^d)^{N_\Gamma}$.
\end{proof}

\subsection{Spherical substitution graphs from a null-bordism}
\label{sec:daughters}

We now describe the geometric input. Let $\hG$ be a bipartite $(d+2)$-colored graph with colors $\{0..d,*\}$. We call $\hG$ \textit{admissible} if every connected component of every $\rs{\hG}{\hat\alpha}$, $\alpha\in\{0..d,*\}$, is a $d$-GEM of some closed manifold. In this case $\hW:=K(\hG)$ is a closed oriented $(d+1)$-dimensional pseudomanifold, the only possible singular points of which are the vertices. Specifically, the link of the vertex corresponding to a connected component $Q$ of $\rs{\hG}{\hat\alpha}$ is $K(Q)$, and the vertex is singular if $M_Q$ is not a sphere (Lemma \ref{lem:pseudo}(i),(ii)). The following theorem, which is essentially a classical result of crystallization theory \cite{CavicchioliGagliardi1980,Gagliardi1983}, is proven in Appendix \ref{app:filling}.

\begin{theorem}
\label{thm:filling}
Let $W$ be a compact oriented PL $(d+1)$-manifold and let $\Gamma$ be a bipartite GEM of $\partial W$ (as oriented manifolds), not necessarily connected. Then there exists a bipartite $(d+2)$-colored graph $\hG$ with colors $\{0..d,*\}$ such that
\begin{enumerate}
    \item[(i)] $\rs{\hG}{\hat *}=\Gamma\sqcup T$, where $T$ is a disjoint union of spherical $d$-GEMs.
    \item[(ii)] $\rs{\hG}{\hat\alpha}$ is a disjoint union of spherical $d$-GEMs for every color $\alpha\in\Dd$.
    \item[(iii)] $K(\hG)$ is a triangulation of the closed pseudomanifold obtained from $W$ by coning off each connected component of $\partial W$, the cone points being vertices of color $*$ with links the components of $K(\Gamma)$.
\end{enumerate}
In particular $\hG$ is admissible. Conversely, (i) and (ii) alone imply that $K(\hG)$ is a triangulation of this kind of the coning off of some compact oriented $(d+1)$-manifold with boundary $M_\Gamma$, namely $\hW:=K(\hG)$ with open regular neighborhoods of the vertices of color $*$ with links the components of $K(\Gamma)$ removed, the other vertices of color $*$ being interior manifold points by (i), (ii) and Lemma \ref{lem:pseudo}.
\end{theorem}

\begin{lemma}
\label{lem:daughters}
Let $\hG$ be a bipartite $(d+2)$-colored graph with colors $\{0..d,*\}$ satisfying (i) and (ii) of Theorem \ref{thm:filling} for some $d$-GEM $\Gamma$, for instance the graph provided by the theorem, and let $H_J$, $J\subseteq\Dd$, be its substitution graphs (\ref{H-def}). Then $H_\emptyset=\Gamma\sqcup T$, while every $H_J$ with $J\neq\emptyset$ is a disjoint union of spherical $d$-GEMs, and the numbers of components are
\begin{equation}
    c_J:=\#\pi_0(H_J)=c_J(\hG)\quad (J\neq \emptyset),\qquad t:=\#\pi_0(T)=c_{\{*\}}(\hG)-\#\pi_0(\Gamma),
    \label{c-x}
\end{equation}
i.e. $c_J$ is the number of cells of $\hW=K(\hG)$ with the set of vertex colors $J$, and $t+\#\pi_0(\Gamma)$ is the number of vertices of color $*$. Consequently,
\begin{equation}
    R(\Gamma)=-R(T)-\sum_{J\neq\emptyset}(-1)^{|J|}R(H_J)\;\in\;R(\Z[\text{spherical }d\text{-GEMs}]),
    \label{bordism-relation}
\end{equation}
so that $[\Gamma]=0$ in $\Qd$, and the coefficient sum $N_\Gamma$ of the combination $S(\Gamma)$ of (\ref{S-def}), which is the power of $Z(S^d)$ in (\ref{extract-S}), is
\begin{equation}
    N_\Gamma=\chi(W)-t\,\big(1-(-1)^d\big),
    \label{N-chi}
\end{equation}
where $W$ is the compact oriented $(d+1)$-manifold with $\partial W=M_\Gamma$ of which $\hW$ is the coning off, as in Theorem \ref{thm:filling}(iii) and the remark after it, i.e. $\hW$ with open regular neighborhoods of the cone points removed (for the graph provided by the theorem, the given manifold $W$). In particular, for even $d$, $N_\Gamma=\chi(W)=\chi(M_\Gamma)/2$ independently of the choice of $W$.
\end{lemma}

\begin{proof}
Let $J\neq\emptyset$ and $|J|=s$. Deleting $s-1$ of the $s$ parallel copies of the $\sigma_*$-edges in $H_J$ and renaming the remaining one $*$ gives $\rs{\hG}{\hat J\cup\{*\}}$, the connected components of which are spherical $(d+1-s)$-GEMs: for $s=1$ this is Theorem \ref{thm:filling}(ii), and for $s\geq 2$ the graph $\rs{\hG}{\hat J\cup\{*\}}$ is the restriction of $\rs{\hG}{\hat\alpha}$, for any $\alpha\in J$, to a proper subset of its colors, so that its components are proper-color subgraphs of spherical $d$-GEMs, hence spherical GEMs (Section \ref{sec:residues}). Restoring the deleted copies one by one is an $(s-1)$-fold suspension (Lemma \ref{lem:susp}), which gives spherical $d$-GEMs. Parallel edges do not change connectivity, and by Lemma \ref{lem:pseudo}(i) the connected components of $\rs{\hG}{\hat J\cup\{*\}}$ correspond to the cells of $\hW$ with the set of vertex colors $J$, which gives (\ref{c-x}). For $J=\emptyset$ the components are the links of the vertices of color $*$: the cone points over the components of $K(\Gamma)$, and $t$ other vertices. The relation (\ref{bordism-relation}) is Lemma \ref{lem:boolean}.

For (\ref{N-chi}), by (\ref{c-x}),
\begin{equation}
    N_\Gamma=-t-\sum_{\emptyset\neq J\subseteq\Dd}(-1)^{|J|}c_J=\chi(W^\circ)-t,
\end{equation}
where $W^\circ$ is the subcomplex of $\hW$ formed by the cells with no vertex of color $*$, the $(|J|-1)$-dimensional cells of which are counted by the $c_J$. Every cell of $\hW$ containing a vertex $v$ of color $*$ is the join (Appendix \ref{app:glossary}) of $v$ with a cell of $W^\circ$, so that $\hW$ with its vertices of color $*$ removed deformation retracts onto $W^\circ$, by the radial projection away from $v$ in each such cell. On the other hand, $\hW$ with its vertices of color $*$ removed is homotopy equivalent to $W$ with its $t$ interior vertices of color $*$ removed, since a punctured regular neighborhood of a vertex deformation retracts onto its boundary. Removing a point from the interior of a $(d+1)$-manifold changes its Euler characteristic by $(-1)^d$, which gives (\ref{N-chi}). For even $d$ the manifold $W$ is odd-dimensional and $\chi(W)=\frac12\chi(\partial W)$.
\end{proof}

Corollary \ref{cor:S} identifies the right-hand side of (\ref{extract-null}) with the signal (\ref{signal-def}) of the $d+2$ permutations defining $\hG$,
\begin{equation}
    \frac{Z(M_\Gamma)}{Z(S^d)^{N_\Gamma}}=\lim_{L\rightarrow\infty}\CZ(\sigma_0-\sigma_*,\ldots,\sigma_d-\sigma_*;|\psi\rangle),
    \label{conj-A-signal}
\end{equation}
which is the precise form of Remark 4 after Corollary \ref{cor:extract}. In the example of Section \ref{sec:example}, the graph $\hG$ of the five permutations (\ref{L31-perms}), (\ref{L31-star}), shown in Figure \ref{fig:lens}(b), satisfies (i) and (ii) of Theorem \ref{thm:filling} with $T=\emptyset$: $\rs{\hG}{\hat *}=\Gamma$ and $\rs{\hG}{\hat\alpha}$, $\alpha\in\{0..3\}$, are GEMs of $S^3$, so that $\hW$ is a $4$-dimensional pseudomanifold with a single singular point, the link of which is $L(3,1)$. It is not of the form constructed in the proof of the theorem, the star of the singular point being a quotient of the cone over $K(\Gamma)$ rather than the cone itself, which is why no additional replicas are needed. Removing a neighborhood of this point one obtains a $4$-manifold $W$ with $\partial W=L(3,1)$ and $t=0$, and (\ref{N-chi}) gives $N_\Gamma=\chi(W)=2$, in agreement with the count of Section \ref{sec:example}. The complex $W^\circ$ of the proof of Lemma \ref{lem:daughters}, a deformation retract of $W$, has $4$ vertices, $8$ edges, $12$ triangles and $6$ tetrahedra, and its mod $2$ Betti numbers are $(1,0,1,0)$, so that $W$ has the homology of the disc bundle over $S^2$ of Euler number $3$, another $4$-manifold bounded by $L(3,1)$.

\subsection{The power of the sphere partition function}
\label{sec:exponent}

\begin{figure}[ht]
\centering
\begin{tikzpicture}[scale=0.95]
  \begin{scope}[shift={(0,0.9)}]
    \coordinate (a00) at (0,0); \coordinate (a10) at (1.5,0); \coordinate (a01) at (0,1.5); \coordinate (a11) at (1.5,1.5);
    \draw[cc0] (a00)--(a10) (a01)--(a11);
    \draw[cc4] (a00)--(a01) (a10)--(a11);
    \node[bl] at (a00) {}; \node[wh] at (a10) {}; \node[wh] at (a01) {}; \node[bl] at (a11) {};
    \node at (0.75,-1.55) {(a) $d=0$};
  \end{scope}
  \begin{scope}[shift={(3.6,0.55)}]
    \foreach \x in {0,1} \foreach \y in {0,1} \foreach \z in {0,1} { \coordinate (b\x\y\z) at (1.7*\x+0.85*\z,1.7*\y+0.85*\z); }
    \foreach \y in {0,1} \foreach \z in {0,1} { \draw[cc0] (b0\y\z)--(b1\y\z); }
    \foreach \x in {0,1} \foreach \z in {0,1} { \draw[cc1] (b\x0\z)--(b\x1\z); }
    \foreach \x in {0,1} \foreach \y in {0,1} { \draw[cc4] (b\x\y0)--(b\x\y1); }
    \foreach \x in {0,1} \foreach \y in {0,1} \foreach \z in {0,1} {
      \pgfmathtruncatemacro{\p}{mod(\x+\y+\z,2)}
      \ifnum\p=0 \node[bl] at (b\x\y\z) {}; \else \node[wh] at (b\x\y\z) {}; \fi }
    \node at (1.3,-1.2) {(b) $d=1$};
  \end{scope}
  \begin{scope}[shift={(10.0,1.75)}]
    \foreach \w in {0,1} {
      \pgfmathsetmacro{\s}{0.4+0.6*\w}
      \foreach \x in {0,1} \foreach \y in {0,1} \foreach \z in {0,1} {
        \coordinate (c\x\y\z\w) at ({\s*((2*\x-1)*1.5+(2*\z-1)*0.75)},{\s*((2*\y-1)*1.5+(2*\z-1)*0.45)}); } }
    \foreach \w in {0,1} {
      \foreach \y in {0,1} \foreach \z in {0,1} { \draw[cc0] (c0\y\z\w)--(c1\y\z\w); }
      \foreach \x in {0,1} \foreach \z in {0,1} { \draw[cc1] (c\x0\z\w)--(c\x1\z\w); }
      \foreach \x in {0,1} \foreach \y in {0,1} { \draw[cc2] (c\x\y0\w)--(c\x\y1\w); } }
    \foreach \x in {0,1} \foreach \y in {0,1} \foreach \z in {0,1} { \draw[cc4] (c\x\y\z0)--(c\x\y\z1); }
    \foreach \w in {0,1} \foreach \x in {0,1} \foreach \y in {0,1} \foreach \z in {0,1} {
      \pgfmathtruncatemacro{\p}{mod(\x+\y+\z+\w,2)}
      \ifnum\p=0 \node[bl] at (c\x\y\z\w) {}; \else \node[wh] at (c\x\y\z\w) {}; \fi }
    \node at (0,-2.4) {(c) $d=2$};
  \end{scope}
\end{tikzpicture}
\caption{The $(d+2)$-cube graphs $\hG_d$ of Proposition \ref{prop:exponent}(ii) for $d=0,1,2$: (a) the square, (b) the $3$-cube, (c) the $4$-cube, drawn as two nested $3$-cubes. The vertices are the points of $\{0,1\}^{d+2}$, black for an even and white for an odd sum of coordinates, and the edges of color $\alpha\in\{0..d,*\}$ join the vertices differing in the $\alpha$-th coordinate, with the edge styles of Figure \ref{fig:lens}: color $0$ solid, $1$ dashed, $2$ wavy, $*$ dotted. Each $\hG_d$ is a GEM of $S^{d+1}$, obtained from $\hG_{d-1}$ by doubling along the last color. Deleting the edges of color $*$ gives $H_\emptyset$, two copies of the $(d+1)$-cube graph, a GEM of $S^d$. For $d=3$ the graph has $32$ vertices.}
\label{fig:hypercube}
\end{figure}
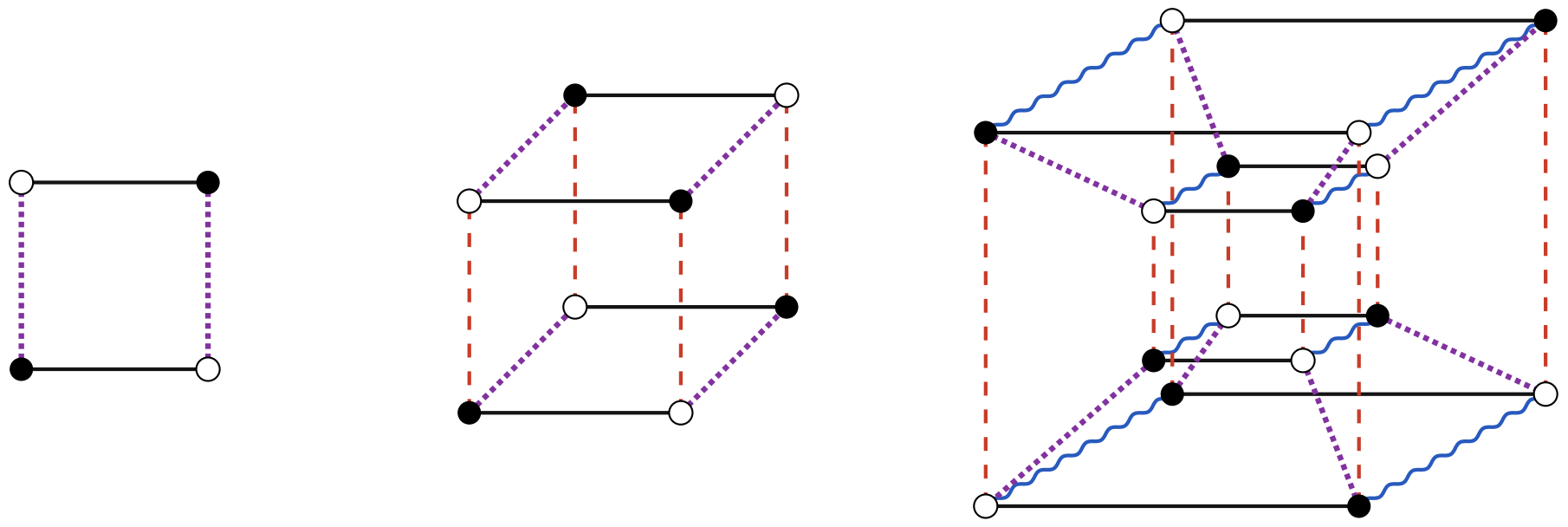

\begin{proposition}
\label{prop:exponent}
Let $\Gamma$ be a $d$-GEM and $\eta=\sum_jm_j\Lambda_j$ a formal combination of spherical $d$-GEMs with $R(\eta)=R(\Gamma)$. Put $N:=\sum_jm_j$.
\begin{enumerate}
    \item[(i)] If $d$ is even, then $N=\chi(M_\Gamma)/2$.
    \item[(ii)] If $d$ is odd, there exists a combination $\zeta$ of spherical $d$-GEMs with $R(\zeta)=0$ and coefficient sum $2$. Explicitly, let $\hG$ be the $(d+2)$-cube graph, i.e. the graph with vertices $\{0,1\}^{d+2}$ and edges of color $\alpha$ connecting the vertices differing in the $\alpha$-th coordinate, $\alpha\in\{0..d,*\}$ (Figure \ref{fig:hypercube}). Then all its substitution graphs are disjoint unions of spherical $d$-GEMs, with $2^{|J|}$ components for $J\neq\emptyset$ and $2$ components for $J=\emptyset$, and $\zeta=\sum_J(-1)^{|J|}H_J$. Consequently, $N$ can be shifted by an arbitrary even integer.
    \item[(iii)] If $d\equiv 3\mod 4$, there also exists a combination $\zeta'$ of spherical $d$-GEMs with $R(\zeta')=0$ and coefficient sum $1$. Consequently $N$ can be made zero.
\end{enumerate}
\end{proposition}

Conjecture TS applied to the relation $\zeta$ of (ii), the total residue of which vanishes, gives the $(d+2)$-cube signal
\begin{equation}
    \prod_J\CZ(H_J;|\psi\rangle)^{(-1)^{|J|}}\;\xrightarrow{\;L\rightarrow\infty\;}\;Z(S^d)^{1+(-1)^{d+1}}=\left\{\begin{array}{ll}Z(S^d)^2, & d\text{ odd},\\ 1, & d\text{ even}.\end{array}\right.
    \label{cube-signal}
\end{equation}

\begin{proof}
(i) Applying the functional $\lambda_\chi$ of (\ref{chi-linear}) to $R(\Gamma)=R(\eta)$ gives $\chi(M_\Gamma)=\sum_jm_j\chi(S^d)=2N$. (In particular, adding to $\eta$ any $\zeta$ with $R(\zeta)=0$ does not change $N$ for even $d$.)

(ii) Write $\hG_d$ for the $(d+2)$-cube graph. We show by induction on $d$ that $\hG_d$ is a GEM of $S^{d+1}$, that all its substitution graphs are disjoint unions of spherical $d$-GEMs, and that the coefficient sum $N_d:=\sum_{J\subseteq\Dd}(-1)^{|J|}\#\pi_0(H_J)$ equals $1+(-1)^{d+1}$. We use two elementary operations on a colored graph $\Lambda$, both of which replace $M_\Lambda$ by its suspension $\Sigma M_\Lambda$ and hence take GEMs of $S^m$ to GEMs of $S^{m+1}$. The first is \textit{doubling}: two copies of $\Lambda$ with opposite bipartitions, joined by edges of a new color between corresponding vertices. The two vertices of the new color of the resulting complex have $K(\Lambda)$ as their links, every top-dimensional cell contains exactly one of them, and the cells containing neither form a single copy of $K(\Lambda)$. Hence the complex is the union of two cones over $K(\Lambda)$ glued along $K(\Lambda)$, i.e. the suspension. The second is the addition of a parallel copy of the edges of one color with a new color, Lemma \ref{lem:susp}.

Now $\hG_{d+1}$ is the doubling of $\hG_d$ with the new color $d+1$, so it is a GEM of $\Sigma S^{d+1}=S^{d+2}$, and the $\sigma_*$-edges of $\hG_{d+1}$ preserve the new coordinate. The subsets $J'\subseteq\{0..d+1\}$ are of three kinds. (a) If $d+1\notin J'$, then $H_{J'}$ is the doubling of the substitution graph $H_{J'}$ of $\hG_d$: its components are the suspensions of the components of the latter, hence GEMs of $S^{d+1}$, and their number is unchanged. These terms contribute $N_d$. (b) If $J'=J\cup\{d+1\}$ with $\emptyset\neq J\subseteq\Dd$, then $H_{J'}$ is two disjoint copies of the substitution graph $H_J$ of $\hG_d$ with one more parallel copy of the $\sigma_*$-edges: its components are again suspensions, hence GEMs of $S^{d+1}$, there are twice as many of them as for $H_J$, and they enter with the sign $-(-1)^{|J|}$. These terms contribute $-2\big(N_d-\#\pi_0(H_\emptyset)\big)=-2(N_d-2)$. (c) If $J'=\{d+1\}$, then $H_{J'}$ is two disjoint copies of $\hG_d$ itself, a connected GEM of $S^{d+1}$, and contributes $-2$. Altogether
\begin{equation}
    N_{d+1}=N_d-2(N_d-2)-2=2-N_d.
\end{equation}
For $d=0$, $\hG_0$ is a square with the colors $0,*$, the doubling of a single edge of color $0$ (the GEM of $S^0$), hence a GEM of $S^1$. The graphs $H_\emptyset$ and $H_{\{0\}}$ each consist of two edges, i.e. of two GEMs of $S^0$, so that $N_0=2-2=0$. Therefore $N_d=1+(-1)^{d+1}$. The same induction gives $\#\pi_0(H_J)=2^{|J|}$ for $J\neq\emptyset$ and $\#\pi_0(H_\emptyset)=2$. The relation $R(\zeta)=0$ is Lemma \ref{lem:boolean}.

(iii) Let $d+1=4k$. Apply Lemma \ref{lem:daughters} to the two-vertex spherical GEM $\Gamma_{S^d}$ and the null-bordism $W=D^{4k}\#\CP^{2k}$ of $S^d$. It provides a relation $R(\Gamma_{S^d})+R(T)+\sum_{J\neq\emptyset}(-1)^{|J|}R(H_J)=0$ among spherical GEMs with coefficient sum $1-N_{\Gamma_{S^d}}=1-\chi(W)+2t$, by (\ref{N-chi}). Since $\chi(W)=\chi(D^{4k})+\chi(\CP^{2k})-2=2k$ is even, the sum is odd, and subtracting a suitable multiple of $\zeta$ from (ii) gives $\zeta'$.
\end{proof}

The graphs $\hG_d$ for small $d$ are shown in Figure \ref{fig:hypercube}. For $d=3$ the substitution graph $H_\emptyset$ consists of two copies of the $16$-vertex $4$-cube graph used in Appendix B of \cite{DelZotto:2026dgp} to normalize the multi-invariants, and $\CD^{-2}=Z(S^3)^2$ is what (\ref{cube-signal}) computes. The relation in (ii) has coefficient sum $2$ rather than $1$, and for $d\equiv 1\mod 4$ this cannot be improved by constructions of the same kind: the parity of the power $N_\Gamma$ of Lemma \ref{lem:daughters} is then fixed by $M_\Gamma$, both for the fillings and for the relations of Lemma \ref{lem:boolean}.

\begin{lemma}
\label{lem:parity}
Let $d\equiv 1\mod 4$, let $\Gamma$ be a $d$-GEM, not necessarily connected and possibly empty, and let $W$ be a compact oriented $(d+1)$-manifold, not necessarily connected, with $\partial W=M_\Gamma$. Then the power $N_\Gamma=\chi(W)-2t$ produced by Lemma \ref{lem:daughters} satisfies
\begin{equation}
    N_\Gamma\equiv\chi(W)\equiv \kQ(M_\Gamma)\mod 2,\qquad \kQ(M):=\sum_{i=0}^{(d-1)/2}\dim H_i(M;\Q)\mod 2,
    \label{chi-semichar}
\end{equation}
where $\kQ(M)\in\Z_2$ is the rational Kervaire semicharacteristic \cite{LusztigMilnorPeterson1969}. In particular, the parity of $N_\Gamma$ does not depend on $W$.
\end{lemma}

The proof is given in Appendix \ref{app:parity}. As a corollary, the relations (\ref{boolean-identity}) of Lemma \ref{lem:boolean} obtained from graphs $\hG$ all the substitution graphs of which are disjoint unions of spherical $d$-GEMs have even coefficient sum. Indeed, $X:=|K(\hG)|$ is then a closed $(d+1)$-manifold and $\hG$ satisfies (i) and (ii) of Theorem \ref{thm:filling} with $W=X$ and $\Gamma=\emptyset$, so that the coefficient sum of the relation is, up to sign, $N_\emptyset=\chi(X)-2t$ by (\ref{N-chi}). Applying the lemma to $\Gamma=\emptyset$ and $W=X$ shows that $\chi(X)$ is even, as is well known. Consequently, every combination $\eta$ with $R(\eta)=R(\Gamma)$ obtained from a filling of $M_\Gamma$ and from such relations has coefficient sum $N\equiv\kQ(M_\Gamma)\mod 2$, and the construction of Proposition \ref{prop:exponent}(iii) has no analogue for $d\equiv 1\mod 4$. Whether for $d\equiv 1\mod 4$, $d\geq 5$, a relation $R(\zeta')=0$ with odd coefficient sum exists by some other mechanism is an open question. For $d=1$ such relations trivially exist, since the residue of a cycle of length $2n$ is $n$ times the residue of the cycle of length $2$. If they do not exist for $d=5$, then $N_\Gamma\mod 2$ is an invariant of $\Gamma$, equal to $\kQ(M_\Gamma)$ for null-bordant $M_\Gamma$. In any case, by (\ref{cube-signal}) and (\ref{Z-sphere-positive}) this does not affect the extraction of $Z(M_\Gamma)$.

\subsection{The bordism theorem}
\label{sec:bordism}

Recall that the oriented PL bordism group $\Omega^{\mathrm{SPL}}_d$ is the quotient of the free abelian group generated by the closed oriented PL $d$-manifolds (up to orientation preserving PL homeomorphism) by the relations $M\sqcup M'=M+M'$ and $\partial W=0$ for every compact oriented PL $(d+1)$-manifold $W$. The relation $\overline M=-M$ is a consequence, since $M\sqcup\overline M=\partial(M\times[0,1])$.

\begin{theorem}
\label{thm:bordism}
The assignment $M\mapsto[\Gamma]$, where $\Gamma$ is any GEM of $M$, defines a surjective homomorphism
\begin{equation}
\begin{array}{rrcl}
     \Phi: & \Omega^{\mathrm{SPL}}_d & \longrightarrow & \Qd,  \\
     & [M]=[M_\Gamma] & \longmapsto & [\Gamma].
\end{array}
\end{equation}
Moreover, $\mathcal{Q}_*:=\bigoplus_{d\geq 1}\Qd$ is a graded commutative ring with the product
\begin{equation}
    [\Gamma]\cdot[\Gamma']:=[\Gamma\times\Gamma'],
    \label{Q-product}
\end{equation}
where $\Gamma\times\Gamma'$ is the GEM of $M_\Gamma\times M_{\Gamma'}$ dual to the barycentric subdivision of $K(\Gamma)\times K(\Gamma')$ with vertices colored by dimension, and $\Phi:\Omega^{\mathrm{SPL}}_*\rightarrow\mathcal{Q}_*$ is a homomorphism of graded rings.
\end{theorem}

\begin{figure}[htp]
\centering
\begin{tikzpicture}[scale=1.3]
\input{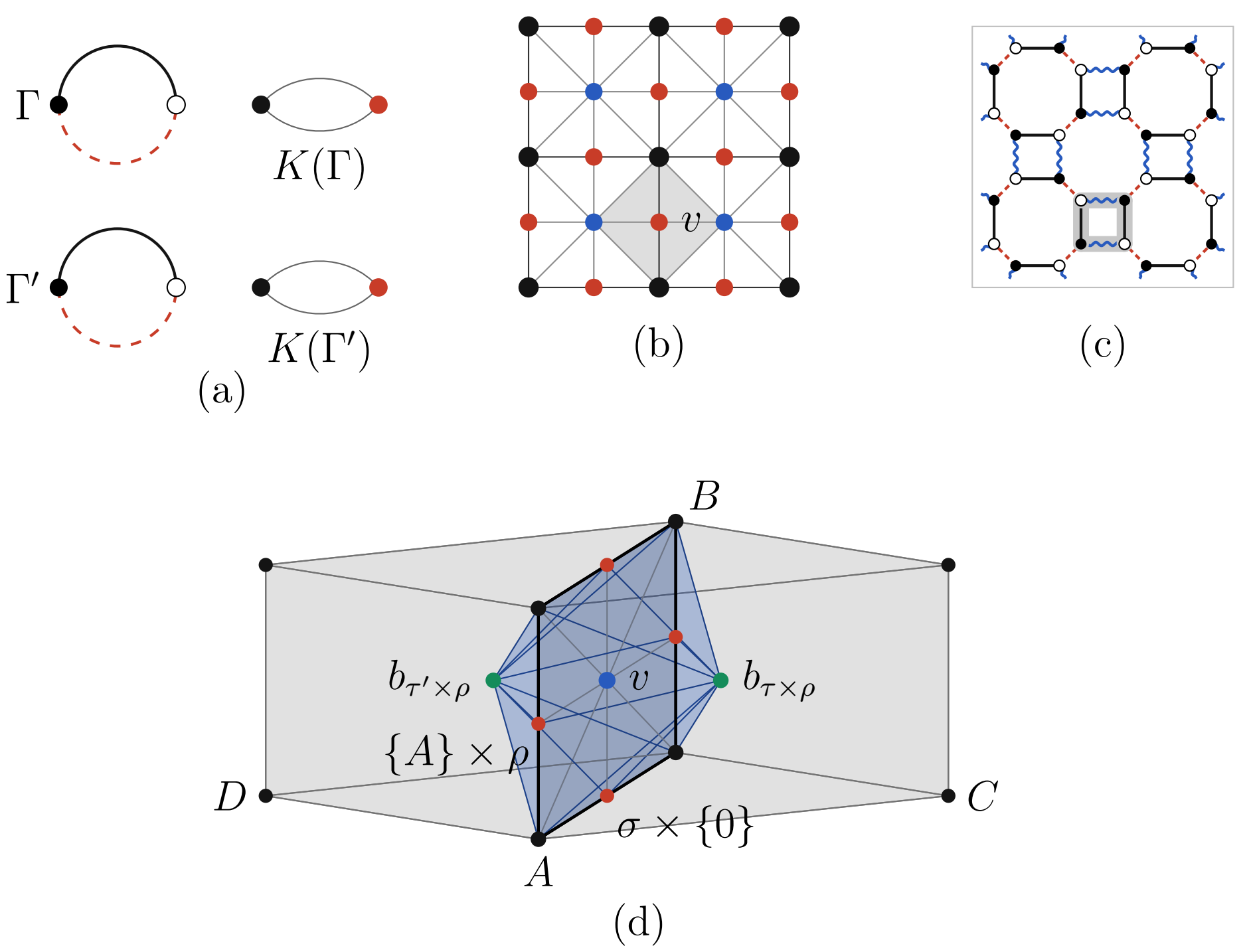}
\end{tikzpicture}\\[2ex]
\begin{tikzpicture}[scale=1.0]
  \fill[black,fill opacity=0.06] (0.000,0.000) -- (1.368,0.864) -- (4.084,0.432) -- cycle;
  \fill[black,fill opacity=0.06] (0.000,2.300) -- (1.368,3.164) -- (4.084,2.732) -- cycle;
  \fill[black,fill opacity=0.06] (0.000,0.000) -- (1.368,0.864) -- (1.368,3.164) -- (0.000,2.300) -- cycle;
  \fill[black,fill opacity=0.06] (1.368,0.864) -- (4.084,0.432) -- (4.084,2.732) -- (1.368,3.164) -- cycle;
  \fill[black,fill opacity=0.06] (4.084,0.432) -- (0.000,0.000) -- (0.000,2.300) -- (4.084,2.732) -- cycle;
  \draw[black!55,thin] (0.000,0.000) -- (1.368,0.864) -- (4.084,0.432) -- cycle;
  \draw[black!55,thin] (0.000,2.300) -- (1.368,3.164) -- (4.084,2.732) -- cycle;
  \draw[black!55,thin] (0.000,0.000) -- (1.368,0.864) -- (1.368,3.164) -- (0.000,2.300) -- cycle;
  \draw[black!55,thin] (1.368,0.864) -- (4.084,0.432) -- (4.084,2.732) -- (1.368,3.164) -- cycle;
  \draw[black!55,thin] (4.084,0.432) -- (0.000,0.000) -- (0.000,2.300) -- (4.084,2.732) -- cycle;
  \fill[c2,fill opacity=0.16] (0.000,0.000) -- (0.684,0.432) -- (1.817,1.582) -- cycle;
  \draw[c2!70!black,thin] (0.000,0.000) -- (1.817,1.582); 
  \fill[c2,fill opacity=0.16] (0.684,0.432) -- (1.368,0.864) -- (1.817,1.582) -- cycle;
  \draw[c2!70!black,thin] (0.684,0.432) -- (1.817,1.582); 
  \fill[c2,fill opacity=0.16] (1.368,0.864) -- (1.368,2.014) -- (1.817,1.582) -- cycle;
  \draw[c2!70!black,thin] (1.368,0.864) -- (1.817,1.582); 
  \fill[c2,fill opacity=0.16] (1.368,2.014) -- (1.368,3.164) -- (1.817,1.582) -- cycle;
  \draw[c2!70!black,thin] (1.368,2.014) -- (1.817,1.582); 
  \fill[c2,fill opacity=0.16] (1.368,3.164) -- (0.684,2.732) -- (1.817,1.582) -- cycle;
  \draw[c2!70!black,thin] (1.368,3.164) -- (1.817,1.582); 
  \fill[c2,fill opacity=0.16] (0.684,2.732) -- (0.000,2.300) -- (1.817,1.582) -- cycle;
  \draw[c2!70!black,thin] (0.684,2.732) -- (1.817,1.582); 
  \fill[c2,fill opacity=0.16] (0.000,2.300) -- (0.000,1.150) -- (1.817,1.582) -- cycle;
  \draw[c2!70!black,thin] (0.000,2.300) -- (1.817,1.582); 
  \fill[c2,fill opacity=0.16] (0.000,1.150) -- (0.000,0.000) -- (1.817,1.582) -- cycle;
  \draw[c2!70!black,thin] (0.000,1.150) -- (1.817,1.582); 
  \draw[black,thick] (0.000,0.000) -- (1.368,0.864) -- (1.368,3.164) -- (0.000,2.300) -- cycle;
  \draw[black!45,thin] (0.684,1.582) -- (0.000,0.000);
  \draw[black!45,thin] (0.684,1.582) -- (0.684,0.432);
  \draw[black!45,thin] (0.684,1.582) -- (1.368,0.864);
  \draw[black!45,thin] (0.684,1.582) -- (1.368,2.014);
  \draw[black!45,thin] (0.684,1.582) -- (1.368,3.164);
  \draw[black!45,thin] (0.684,1.582) -- (0.684,2.732);
  \draw[black!45,thin] (0.684,1.582) -- (0.000,2.300);
  \draw[black!45,thin] (0.684,1.582) -- (0.000,1.150);
  \fill[black,fill opacity=0.06] (0.000,0.000) -- (1.368,0.864) -- (-2.716,0.432) -- cycle;
  \fill[black,fill opacity=0.06] (0.000,2.300) -- (1.368,3.164) -- (-2.716,2.732) -- cycle;
  \fill[black,fill opacity=0.06] (0.000,0.000) -- (1.368,0.864) -- (1.368,3.164) -- (0.000,2.300) -- cycle;
  \fill[black,fill opacity=0.06] (1.368,0.864) -- (-2.716,0.432) -- (-2.716,2.732) -- (1.368,3.164) -- cycle;
  \fill[black,fill opacity=0.06] (-2.716,0.432) -- (0.000,0.000) -- (0.000,2.300) -- (-2.716,2.732) -- cycle;
  \draw[black!55,thin] (0.000,0.000) -- (1.368,0.864) -- (-2.716,0.432) -- cycle;
  \draw[black!55,thin] (0.000,2.300) -- (1.368,3.164) -- (-2.716,2.732) -- cycle;
  \draw[black!55,thin] (0.000,0.000) -- (1.368,0.864) -- (1.368,3.164) -- (0.000,2.300) -- cycle;
  \draw[black!55,thin] (1.368,0.864) -- (-2.716,0.432) -- (-2.716,2.732) -- (1.368,3.164) -- cycle;
  \draw[black!55,thin] (-2.716,0.432) -- (0.000,0.000) -- (0.000,2.300) -- (-2.716,2.732) -- cycle;
  \fill[c2,fill opacity=0.16] (0.000,0.000) -- (0.684,0.432) -- (-0.449,1.582) -- cycle;
  \draw[c2!70!black,thin] (0.000,0.000) -- (-0.449,1.582); 
  \fill[c2,fill opacity=0.16] (0.684,0.432) -- (1.368,0.864) -- (-0.449,1.582) -- cycle;
  \draw[c2!70!black,thin] (0.684,0.432) -- (-0.449,1.582); 
  \fill[c2,fill opacity=0.16] (1.368,0.864) -- (1.368,2.014) -- (-0.449,1.582) -- cycle;
  \draw[c2!70!black,thin] (1.368,0.864) -- (-0.449,1.582); 
  \fill[c2,fill opacity=0.16] (1.368,2.014) -- (1.368,3.164) -- (-0.449,1.582) -- cycle;
  \draw[c2!70!black,thin] (1.368,2.014) -- (-0.449,1.582); 
  \fill[c2,fill opacity=0.16] (1.368,3.164) -- (0.684,2.732) -- (-0.449,1.582) -- cycle;
  \draw[c2!70!black,thin] (1.368,3.164) -- (-0.449,1.582); 
  \fill[c2,fill opacity=0.16] (0.684,2.732) -- (0.000,2.300) -- (-0.449,1.582) -- cycle;
  \draw[c2!70!black,thin] (0.684,2.732) -- (-0.449,1.582); 
  \fill[c2,fill opacity=0.16] (0.000,2.300) -- (0.000,1.150) -- (-0.449,1.582) -- cycle;
  \draw[c2!70!black,thin] (0.000,2.300) -- (-0.449,1.582); 
  \fill[c2,fill opacity=0.16] (0.000,1.150) -- (0.000,0.000) -- (-0.449,1.582) -- cycle;
  \draw[c2!70!black,thin] (0.000,1.150) -- (-0.449,1.582); 
  \draw[black,thick] (0.000,0.000) -- (1.368,0.864) -- (1.368,3.164) -- (0.000,2.300) -- cycle;
  \fill[c0] (0.000,0.000) circle (2.2pt);
  \fill[c0] (1.368,0.864) circle (2.2pt);
  \fill[c0] (1.368,3.164) circle (2.2pt);
  \fill[c0] (0.000,2.300) circle (2.2pt);
  \fill[c1] (0.684,0.432) circle (2pt);
  \fill[c1] (1.368,2.014) circle (2pt);
  \fill[c1] (0.684,2.732) circle (2pt);
  \fill[c1] (0.000,1.150) circle (2pt);
  \fill[c2] (0.684,1.582) circle (2.4pt);
  \fill[c3] (1.817,1.582) circle (2.2pt);
  \fill[c3] (-0.449,1.582) circle (2.2pt);
  \fill[c0] (4.084,0.432) circle (2pt);
  \fill[c0] (4.084,2.732) circle (2pt);
  \fill[c0] (-2.716,0.432) circle (2pt);
  \fill[c0] (-2.716,2.732) circle (2pt);
  \node[right=2pt] at (0.684,1.582) {$v$};
  \node[below right=-2pt] at (0.684,0.432) {$\sigma\times\{0\}$};
  \node[below left=-2pt] at (0.000,1.150) {$\{A\}\times\rho$};
  \node[right=2pt] at (1.817,1.582) {$b_{\tau\times\rho}$};
  \node[left=2pt] at (-0.449,1.582) {$b_{\tau'\times\rho}$};
  \node[right=1pt] at (4.084,0.432) {$C$}; \node[left=1pt] at (-2.716,0.432) {$D$};
  \node[below=1pt] at (0.000,0.000) {$A$}; \node[above right=-1pt] at (1.368,3.164) {$B$};
  \node at (1.0,-0.85) {(d)};
\end{tikzpicture}
\caption{The product of two GEMs in the proof of Theorem \ref{thm:bordism}. (a)--(c): the case $d=d'=1$. (a) $\Gamma$ and $\Gamma'$ are two-vertex GEMs of $S^1$, and $K(\Gamma)$, $K(\Gamma')$ are $2$-gons. (b) The product $K(\Gamma)\times K(\Gamma')$ is a torus tiled by four squares, the opposite sides of the drawing being identified, shown with its barycentric subdivision. The vertices are colored by the dimension of the cell they subdivide: the corners have color $0$, the midpoints of the edges color $1$ and the centers of the squares color $2$. The vertex $v$ subdivides the product of a vertex $w$ of one factor with an edge $e$ of the other, and its link is the join of $\partial(w\times e)$, two points, with $\lk w$, two points corresponding to the two squares containing $w\times e$ and represented by their centers, i.e. a $4$-cycle, the boundary of the shaded star. (c) The dual graph $\Gamma\times\Gamma'$, a $2$-GEM of the torus, with the edges crossing the identified sides drawn as stubs. The $4$-cycle dual to the link of $v$ is highlighted. (d) A local picture for $d=2$, $d'=1$: two triangles $\tau=ABC$ and $\tau'=ABD$ of $K(\Gamma)$ sharing the edge $\sigma=AB$, times an edge $\rho$ of $K(\Gamma')$, give two prisms sharing the rectangle $\sigma\times\rho$ (subdivided, with its corners of color $0$, edge midpoints of color $1$ and center $v$ of color $2$). The link of $v$ is the join of the subdivided boundary of the rectangle, an octagon, with $\lk\sigma*\lk\rho=\lk\sigma=\{C,D\}$. The points $C$ and $D$ correspond to the two prisms $\tau\times\rho$ and $\tau'\times\rho$ containing the rectangle and are represented by their barycenters $b_{\tau\times\rho}$ and $b_{\tau'\times\rho}$, of color $3$. The result is the bipyramid drawn in blue.}
\label{fig:product}
\end{figure}

\begin{proof}
Consider the homomorphism $\widetilde\Phi:\Z[d\text{-GEMs}]\rightarrow\Qd$, $\Gamma\mapsto[\Gamma]$, on the free abelian group generated by the connected $d$-GEMs. As usual, a disconnected GEM is identified with the sum of its components, so that $\widetilde\Phi(\Gamma\sqcup\Gamma')=\widetilde\Phi(\Gamma)+\widetilde\Phi(\Gamma')$. By Lemma \ref{lem:daughters}, $\widetilde\Phi(\Gamma)=0$ whenever $M_\Gamma$ is the boundary of a compact oriented PL manifold. In particular, $\widetilde\Phi$ does not depend on the choice of the GEM of a given manifold: if $M_\Gamma\cong M_{\Gamma'}\cong M$ as oriented manifolds, then both $\Gamma\sqcup\overline{\Gamma'}$ and $\Gamma'\sqcup\overline{\Gamma'}$ are GEMs of $M\sqcup\overline M=\partial(M\times[0,1])$, so that $\widetilde\Phi(\Gamma)-\widetilde\Phi(\Gamma')=\widetilde\Phi(\Gamma\sqcup\overline{\Gamma'})-\widetilde\Phi(\Gamma'\sqcup\overline{\Gamma'})=0$. Therefore $\widetilde\Phi$ descends to the free abelian group on closed oriented PL $d$-manifolds, where it respects the relation $M\sqcup M'=M+M'$ by construction and vanishes on boundaries by Lemma \ref{lem:daughters}, hence to $\Omega^{\mathrm{SPL}}_d$. It is surjective because every element of $R(\Z[d\text{-GEMs}])$ is the residue of a formal combination of GEMs.

For the product, let $\Gamma$ and $\Gamma'$ be GEMs of dimensions $d$ and $d'$ (Figure \ref{fig:product}). The product $K(\Gamma)\times K(\Gamma')$ is the cell complex with cells the products $\sigma\times\rho$ of a cell of $K(\Gamma)$ and a cell of $K(\Gamma')$, the closed cells being the embedded products of two simplices, with boundary $\partial\sigma\times\rho\cup\sigma\times\partial\rho$. Its barycentric subdivision has a vertex at the barycenter of each cell and a simplex for each chain of cells ordered by inclusion, and is a simplicial complex. With each vertex colored by the dimension of the cell it subdivides, it is a $(d+d'+1)$-colored triangulation of $M_\Gamma\times M_{\Gamma'}$, and its dual graph $\Gamma\times\Gamma'$, with the bipartition determined by the product orientation, is a $(d+d')$-GEM of $M_\Gamma\times M_{\Gamma'}$ (Lemma \ref{lem:dual}). The vertex subdividing $\sigma\times\rho$, with $\sigma$ an $i$-cell and $\rho$ a $j$-cell, has color $i+j$, and its link is the join of the barycentric subdivision of $\partial(\sigma\times\rho)$ with the subcomplex spanned by the barycenters of the cells containing $\sigma\times\rho$. The latter is a copy of the barycentric subdivision of $\lk(\sigma\times\rho)=\lk\sigma*\lk\rho$, the cells containing $\sigma\times\rho$ corresponding to the cells of the link, where $\lk\sigma=K(\gamma)$ for the component $\gamma$ of $\rs{\Gamma}{\hat J}$ corresponding to $\sigma$, $J$ being the set of vertex colors of $\sigma$ (Section \ref{sec:boolean}), and similarly $\lk\rho=K(\gamma')$. The first factor of the join, with its coloring, depends only on $i$ and $j$, which are determined by the sets of colors of $\gamma$ and $\gamma'$, since $\partial(\sigma\times\rho)$ is the boundary of the product of an $i$-simplex and a $j$-simplex. In the second factor the barycenter of a cell $\tau\times\theta\supseteq\sigma\times\rho$ has the color $\dim\tau+\dim\theta$, which is determined by the corresponding cell of $\lk\sigma*\lk\rho$ together with $i$ and $j$. The oriented colored link of this vertex is therefore determined by the pair $(\gamma,\gamma')$ of residue components, and the number of such vertices is $\#(\gamma,\Gamma)\,\#(\gamma',\Gamma')$. Hence the multiset of colored vertex links of $\Gamma\times\Gamma'$ is a bilinear function of $(R(\Gamma),R(\Gamma'))$, and by (\ref{vertex-links-determine}) so is the full residue: $R(\Gamma\times\Gamma')=\beta(R(\Gamma),R(\Gamma'))$ for a bilinear map $\beta$. If $S$ is a spherical $d$-GEM, then $S\times\Gamma'$ is a GEM of $S^d\times M_{\Gamma'}=\partial(D^{d+1}\times M_{\Gamma'})$, so that $\beta(R(S),R(\Gamma'))\in R(\Z[\text{spherical }(d+d')\text{-GEMs}])$ by the first part, and similarly in the second argument. Therefore $\beta$ descends to $\Qd\otimes\mathcal{Q}_{d'}\rightarrow\mathcal{Q}_{d+d'}$, and (\ref{Q-product}) is well defined and bilinear. Since $\Gamma\times\Gamma'$ is a GEM of $M_\Gamma\times M_{\Gamma'}$, we have $\Phi([M]\cdot[M'])=\Phi[M]\cdot\Phi[M']$, and the associativity and graded commutativity of the product on $\mathcal{Q}_*$ follow from those of $\Omega^{\mathrm{SPL}}_*$ by the surjectivity of $\Phi$.
\end{proof}

\begin{proof}[Proof of the converse part of Proposition \ref{prop:equivalence}]
Assume Conjecture TS, and let $F\subseteq\Z[d\text{-GEMs}]$ be the subgroup generated by the connected $d$-GEMs $\Gamma$ with $Z(M_\Gamma)\neq 0$. Since the components of a GEM with non-vanishing partition function have non-vanishing partition functions, the condition on $\Gamma$ in Proposition \ref{prop:equivalence} is that $NR(\Gamma)\in R(F)$ for some $N\geq 1$.

For a connected $\Gamma$ with $Z(M_\Gamma)\neq 0$, the left-hand side of (\ref{conj-TS}) for the collection consisting of $\Gamma$ twice, with the exponents $1$ and $-1$, is only defined if $\CZ(\Gamma;|\psi\rangle)\neq 0$, so that Conjecture TS includes the statement that $\CZ(\Gamma;|\psi\rangle)\neq 0$ for $L$ large enough. Now fix $L$. For a connected $\Gamma\in F$ set $u(\Gamma):=\CZ(\Gamma;|\psi\rangle)/Z(M_\Gamma)$ if $\CZ(\Gamma;|\psi\rangle)\neq 0$ and $u(\Gamma):=1$ otherwise, and extend $u$ to a homomorphism $u:F\rightarrow\C^*$. The image $R(F)$ is a subgroup of the free abelian group $\Z[\text{spherical }(<d)\text{-GEMs}]$ and therefore free, so that there is a homomorphism $s:R(F)\rightarrow F$ with $R\circ s=\mathrm{id}$. Since $\C^*$ is divisible, the homomorphism $u\circ s$ extends from the subgroup $R(F)$ to a homomorphism $A:\Z[\text{spherical }(<d)\text{-GEMs}]\rightarrow\C^*$, and we set $A(\gamma;|\psi\rangle):=A(\gamma)$. The numbers $A(\gamma)$ depend on $L$, on the state and on the choices of $s$ and of the extension, but not on $\Gamma$.

For a connected $\Gamma\in F$, the element $\Gamma-s(R(\Gamma))$ is a finite combination $\sum_in_i\Gamma_i$ of GEMs with non-vanishing partition functions and $\sum_in_iR(\Gamma_i)=0$. For $L$ large enough none of the multi-invariants of these GEMs vanishes, and Conjecture TS gives $u(\Gamma-s(R(\Gamma)))=1+O(e^{-\mathrm{const}\cdot L})$, that is
\begin{equation}
    \CZ(\Gamma;|\psi\rangle)=A(R(\Gamma))\,\big(Z(M_\Gamma)+O(e^{-\mathrm{const}\cdot L})\big),
    \label{ELD-from-TS}
\end{equation}
which is (\ref{conj-ELD}) with $U(\Gamma;|\psi\rangle)=A(R(\Gamma))=\prod_\gamma A(\gamma)^{\#(\gamma,\Gamma)}$. By the multiplicativity (\ref{multiplicativity-graphs}), the same holds for disconnected GEMs with non-vanishing partition functions.

Now let $\Gamma$ be a $d$-GEM with $Z(M_\Gamma)=0$ and $NR(\Gamma)\in R(F)$ for some $N\geq 1$. Thus there exist connected $\Gamma_i\in F$ such that $NR(\Gamma)-\sum_i n_iR(\Gamma_i)=0$ for some integers $n_i$. For $L$ large enough the multi-invariants of the $\Gamma_i$ do not vanish, and
\begin{multline}
     \CZ(\Gamma;|\psi\rangle)^N=O(e^{-\mathrm{const}\cdot L})\prod_i\CZ(\Gamma_i;|\psi\rangle)^{n_i}\\
     =\prod_i A(R(\Gamma_i))^{n_i}\,O(e^{-\mathrm{const}\cdot L})=A(R(\Gamma))^N\,O(e^{-\mathrm{const}\cdot L}).
\end{multline}
In the first equality we used Conjecture TS for the collection $\Gamma,\,\Gamma_i$ with the exponents $N$ and $-n_i$, and the non-vanishing of $Z(M_{\Gamma_i})$. In the second equality we used (\ref{ELD-from-TS}) for the $\Gamma_i$, and in the last equality we used again that $NR(\Gamma)=\sum_{i}n_iR(\Gamma_i)$. Taking an $N$-th root gives $\CZ(\Gamma;|\psi\rangle)=A(R(\Gamma))\,O(e^{-\mathrm{const}\cdot L/N})$, which is (\ref{conj-ELD}) for $\Gamma$. Conjecture TS constrains the remaining GEMs, those with $Z(M_\Gamma)=0$ and $NR(\Gamma)\notin R(F)$ for all $N\geq 1$, only through combinations in which they enter with positive exponents.

Finally, let $\Gamma$ be a $d$-GEM with $N[M_\Gamma]=0\in\Omega^{\mathrm{SPL}}_d$ for some $N\geq 1$. By Theorem \ref{thm:bordism}, $N[\Gamma]=\Phi(N[M_\Gamma])=0\in\Qd$, that is, $NR(\Gamma)=R(\eta)$ for a combination $\eta$ of spherical $d$-GEMs. Since $Z(S^d)>0$ by (\ref{Z-sphere-positive}), the spherical $d$-GEMs belong to $F$, and therefore $NR(\Gamma)\in R(F)$. Moreover, by the same argument the condition of Proposition~\ref{prop:equivalence} holds for every $\Gamma$ as soon as $Z(M_i)\neq 0$ for some manifolds $M_i$ with classes generating $\Omega^{\mathrm{SPL}}_d\otimes\Q$. For example, for $d=4$, where $\Omega^{\mathrm{SPL}}_4\cong\Z$ is detected by the signature, every $4$-GEM satisfies the condition if $Z(M)\neq 0$ for a single $M$ with $\sigma(M)\neq 0$. For $d\not\equiv 0\mod 4$ there is no condition at all: the group $\Omega^{\mathrm{SPL}}_d\otimes\Q\cong\Omega^{\mathrm{SO}}_d\otimes\Q$ vanishes \cite{Thom1954}, so that every class is torsion, and the two conjectures are always equivalent.
\end{proof}

\medskip\noindent
The remarks after Corollary \ref{cor:extract} are justified as follows. Let $\Gamma$ be a $d$-GEM with $[M_\Gamma]=0\in\Omega^{\mathrm{SPL}}_d$ and let $W$ be a compact oriented PL manifold with $\partial W=M_\Gamma$. Theorem \ref{thm:filling} and Lemma \ref{lem:daughters} applied to $W$ give the spherical GEMs $T$ and $H_J$, $J\neq\emptyset$, and the combination $\eta=S(\Gamma)$ of (\ref{S-def}) with $R(\eta)=R(\Gamma)$, which is the construction of Remark 4, and (\ref{extract-S}) identifies the right-hand side of (\ref{extract-null}) with the signal (\ref{conj-A-signal}). The statements about $N_\Gamma$ in Remark 2 follow from Proposition \ref{prop:exponent} and Lemma \ref{lem:parity}: for even $d$ the power is $\chi(M_\Gamma)/2$. For odd $d$, adding multiples of the $(d+2)$-cube relation $\zeta$ to $\eta$ shifts $N_\Gamma$ by even integers, and for $d\equiv 3\mod 4$ the relation $\zeta'$ shifts it by one, while for $d\equiv 1\mod 4$ the parity of $N_\Gamma$ is $\kQ(M_\Gamma)$ by Lemma \ref{lem:parity}. Remark 3 is Proposition \ref{prop:exponent}(ii) combined with Conjecture TS, i.e.\ (\ref{cube-signal}).

\subsection{The kernel: characteristic numbers are local}
\label{sec:kernel}

\begin{proposition}
\label{prop:kernel}
Let $d=4k$ and let $P$ be a Pontryagin number of degree $d$. There is a $\Q$-linear functional $\lambda_P$ on $\Z[\text{spherical }(<d)\text{-GEMs}]$ such that $P[M_\Gamma]=\lambda_P(R(\Gamma))$ for every bipartite $d$-GEM $\Gamma$. Consequently, $\Phi\otimes\Q$ is an isomorphism, the kernel of $\Phi$ is torsion, and thus $\Phi$ itself is an isomorphism whenever $\Omega^{\mathrm{SPL}}_d$ is torsion-free. For $d=4$, $\Phi:\Omega^{\mathrm{SPL}}_4=\Omega^{\mathrm{SO}}_4\cong\Z\rightarrow\mathcal{Q}_4$ is an isomorphism. In particular, the residue of a GEM of $\CP^2$ is not the residue of any combination of spherical GEMs.
\end{proposition}

\begin{proof}
The existence of local combinatorial formulas for rational Pontryagin classes goes back to Levitt and Rourke \cite{LevittRourke1978}. In the explicit form due to Gaifullin \cite{Gaifullin2004,Gaifullin2005}, for every $P$ there is a function $f_P$ on the isomorphism classes of oriented combinatorial $(d-1)$-spheres, odd under orientation reversal, such that for every oriented combinatorial $d$-manifold $X$ (a simplicial complex all the vertex links of which are PL spheres)
\begin{equation}
    P[X]=\sum_{v\in \mathrm{Vertices}(X)}f_P(\lk_Xv),
    \label{gaifullin}
\end{equation}
with the links oriented by $X$. We apply this to $X=\sd K(\Gamma)$, the barycentric subdivision of the colored complex of $\Gamma$, which is a simplicial complex triangulating $M_\Gamma$ \cite{Gaifullin2008}. The vertices of $X$ are the barycenters $b_\tau$ of the cells $\tau$ of $K(\Gamma)$. If $\dim\tau\geq 1$, the link of $b_\tau$ is isomorphic to the join $\sd(\partial\tau)*\sd(\lk_{K(\Gamma)}\tau)$, and a transposition of two vertices of the simplex $\tau$ induces an orientation reversing automorphism of it. Since $f_P$ is odd, such vertices contribute zero to (\ref{gaifullin}). If $\tau=v$ is a vertex of $K(\Gamma)$ of color $\alpha$, corresponding to a connected component $Q$ of $\rs{\Gamma}{\hat\alpha}$, then $\lk_Xv=\sd K(Q)$, with the orientation differing from the one induced by the orientation of $Q$ as a bipartite graph by a sign $\epsilon_\alpha$ depending only on $\alpha$ and $d$. Hence
\begin{equation}
    P[M_\Gamma]=\sum_{\alpha\in\Dd}\epsilon_\alpha\sum_{Q\in\pi_0(\rs{\Gamma}{\hat\alpha})}f_P\big(\sd K(Q)\big),
\end{equation}
which is a linear function $\lambda_P$ of the part of $R(\Gamma)$ with $|I|=d$. We extend $\lambda_P$ by zero on the generator with the color set $\emptyset$. For a spherical $T$ one has $\lambda_P(R(T))=P[S^d]=0$.

Now if $\sum_im_iR(\Gamma_i)=R(\eta)$ with $\eta$ a combination of spherical GEMs, then $\sum_im_iP[M_{\Gamma_i}]=\lambda_P(R(\eta))=0$ for all Pontryagin numbers $P$. Rational Pontryagin numbers detect $\Omega^{\mathrm{SO}}_*\otimes\Q$ \cite{Thom1954}, and $\Omega^{\mathrm{SO}}_*\otimes\Q\rightarrow\Omega^{\mathrm{SPL}}_*\otimes\Q$ is an isomorphism since the homotopy groups of $PL/O$ are finite \cite{KirbySiebenmann1977}.\footnote{Here $PL/O$ is the homotopy fiber of the map $BO\rightarrow BPL$ between the classifying spaces for smooth and PL bundles. It measures the difference between the two categories. Its homotopy group $\pi_i(PL/O)$ classifies the smooth structures on the PL sphere $S^i$ up to isotopy, and coincides for $i\geq 7$ with the Kervaire-Milnor group of exotic $i$-spheres. All these groups are finite, and $\pi_i(PL/O)=0$ for $i\leq 6$.} Therefore $\sum_im_i[M_{\Gamma_i}]$ is a torsion element of $\Omega^{\mathrm{SPL}}_d$, i.e. $\Phi\otimes\Q$ is injective, and it is surjective by Theorem \ref{thm:bordism}. For $d=4$, $\Omega^{\mathrm{SPL}}_4=\Omega^{\mathrm{SO}}_4\cong\Z$, detected by $\sigma=\frac13p_1$, so that $\Phi$ is an isomorphism and $\Phi[\CP^2]\neq 0$.
\end{proof}

\begin{proof}[Proof of Theorem \ref{thm:structure}]
The map $\Phi$ is well defined and surjective by Theorem \ref{thm:bordism}, which also contains (iv), (i) is Proposition \ref{prop:kernel}, and (iii) is Proposition \ref{prop:smooth-torsion} below. For (ii), recall from Section \ref{sec:gem} that $\Omega^{\mathrm{SPL}}_d=\Omega^{\mathrm{SO}}_d$ for $d\leq 7$. Hence $\Omega^{\mathrm{SPL}}_d=0$ for $d\leq 3$, so that $\Qd=0$ by surjectivity, while for $d=4$ the statement is part of Proposition \ref{prop:kernel}. The claims about the composite $\Omega^{\mathrm{SO}}_d\hookrightarrow\Omega^{\mathrm{SPL}}_d\rightarrow\Qd$ made after the theorem follow from the smoothing theory recalled in Section \ref{sec:gem}: the obstructions to smoothing a PL manifold $V$ lie in $H^{i+1}(V;\pi_i(PL/O))$, and $\pi_i(PL/O)=0$ for $i\leq 6$, so that every PL manifold of dimension at most $7$ admits a smooth structure. Therefore $\Omega^{\mathrm{SO}}_d\hookrightarrow\Omega^{\mathrm{SPL}}_d$ is surjective for $d\leq 7$, hence an isomorphism for $d\leq 7$ since it is injective, and it is an isomorphism after tensoring with $\Q$ for all $d$ since the groups $\pi_i(PL/O)$ are finite.
\end{proof}

\medskip\noindent
The following is the strong form of Theorem \ref{thm:structure}(iii). It says that the kernel of $\Phi$ is invisible to smooth manifolds: the torsion of $\Omega^{\mathrm{SO}}_d$ is generated by manifolds which, like the Dold manifold $P(1,2)$ in the explicit example at the beginning of the proof, have ``untwisted'' partners with identical residues.

\begin{proposition}
\label{prop:smooth-torsion}
Let $M$ be a closed oriented smooth manifold with torsion class in $\Omega^{\mathrm{SO}}_d$. Then $[\Gamma]=0\in\Qd$ for every GEM $\Gamma$ of $M$. Consequently, the kernel of the composite $\Omega^{\mathrm{SO}}_d\hookrightarrow\Omega^{\mathrm{SPL}}_d\rightarrow\Qd$ is exactly the torsion subgroup of $\Omega^{\mathrm{SO}}_d$, the induced map $\Omega^{\mathrm{SO}}_d/\mathrm{torsion}\rightarrow\Qd$ is injective with image of finite index, and $\Phi$ induces a surjection
\begin{equation}
    \mathrm{Tor}\,\Omega^{\mathrm{SPL}}_d\big/\mathrm{im}\,\mathrm{Tor}\,\Omega^{\mathrm{SO}}_d\;\twoheadrightarrow\;\mathrm{Tor}\,\Qd,
    \label{torsion-surjection}
\end{equation}
where $\mathrm{Tor}$ denotes the torsion subgroup and $\mathrm{im}\,\mathrm{Tor}\,\Omega^{\mathrm{SO}}_d\subseteq\mathrm{Tor}\,\Omega^{\mathrm{SPL}}_d$ is the subgroup of torsion classes represented by smooth manifolds. For $d=5$, explicitly, the Dold manifold $P(1,2)$, i.e. the mapping torus of complex conjugation on $\CP^2$, and $\CP^2\times S^1$ have GEMs with identical residues. Since $P(1,2)$ generates $\Omega^{\mathrm{SPL}}_5=\Omega^{\mathrm{SO}}_5\cong\Z_2$ \cite{Dold1956,Wall1960}, it follows that $\mathcal{Q}_5=0$.
\end{proposition}

\begin{proof}
We use Wall's determination of the torsion of $\Omega^{\mathrm{SO}}_*$ \cite{Wall1960} in the explicit form of \cite[Corollary 6.8]{EbertWiemeler2024}: every torsion class of $\Omega^{\mathrm{SO}}_d$ is an integer combination of classes $[E]$, each of which is itself torsion, of total spaces of smooth fiber bundles $E\rightarrow B$ over closed oriented manifolds $B$, with fiber $F=\CP^{2q}$, $q\geq 1$, and structure group $\Z_2$ acting on the fiber by complex conjugation.\footnote{Explicitly, the generators are products of even-dimensional complex projective spaces, of the Dold manifolds $P(2^r-1,2^rs)=S^{2^r-1}\times_{\Z_2}\CP^{2^rs}$, which are $\CP^{2^rs}$-bundles over $\RP^{2^r-1}$, and of the manifolds obtained by Wall's boundary construction from products of the Wall manifolds $Q(2^r-1,2^rs)$, the mapping tori of the involutions of $P(2^r-1,2^rs)$ induced by a reflection of $S^{2^r-1}$. The boundary construction restricts a product $Q(m_1,n_1)\times\cdots\times Q(m_p,n_p)\rightarrow(S^1)^p$ to the kernel of the multiplication map $(S^1)^p\rightarrow S^1$. Since $Q(m,n)$ is a $\CP^n$-bundle over $Q(m,0)$ with structure group $\Z_2$, compatibly with the map to $S^1$, the result is a $\CP^{n_1}$-bundle with structure group $\Z_2$ over the restriction of $Q(m_1,0)\times Q(m_2,n_2)\times\cdots\times Q(m_p,n_p)$ to the same kernel. The base of each resulting bundle is orientable since the total space is, the fiber being simply connected and the structure group preserving its orientation.} Since $\Phi$ is a homomorphism defined on bordism classes (Theorem \ref{thm:bordism}), it suffices to prove that $\Phi[E]=0$ for each such bundle with $[E]$ torsion. We argue by induction on $d$. There is nothing to prove for $d\leq 4$, where $\Omega^{\mathrm{SO}}_d$ is torsion-free.

\textit{Quotients.} We will twice use the following observation (Figure \ref{fig:quotient}). Let $\widetilde\Gamma$ be a $d$-GEM, $X:=K(\widetilde\Gamma)$, and let $g$ be a free simplicial involution of $X$ preserving the vertex colors and the orientation. Then the quotient graph $\widetilde\Gamma/g$, the vertices and edges of which are the $g$-orbits of the vertices and edges of $\widetilde\Gamma$, is a $d$-GEM of $X/g$ with
\begin{equation}
    R(\widetilde\Gamma/g)=\tfrac12R(\widetilde\Gamma).
    \label{residue-quotient}
\end{equation}
Indeed, since $g$ preserves colors and the vertices of a cell have distinct colors, no two points of a closed cell of $X$ are identified, so $X/g$ is a properly colored simplicial cell complex, the cells of which are the orbits of the cells of $X$, and it is a closed oriented PL manifold. No cell is glued to its own image along a face, since such a face would be $g$-invariant and its barycenter fixed. Therefore Lemma \ref{lem:dual} applies to $X/g$, and its dual graph, which is $\widetilde\Gamma/g$, is a bipartite $d$-GEM of $X/g$. For the residues, recall from Section \ref{sec:boolean} that the connected components of $\rs{\widetilde\Gamma}{I}$ correspond to the cells $\tau$ of $X$ with the complementary set of vertex colors, the component $C_\tau$ being formed by the cells containing $\tau$. No cell contains both $\tau$ and $g\tau$, which have the same colors, so $C_\tau$ and $C_{g\tau}$ are disjoint, interchanged by $g$, and mapped injectively to $\widetilde\Gamma/g$. The components of $\rs{(\widetilde\Gamma/g)}{I}$ are therefore the images of the $C_\tau$, one for each orbit $\{\tau,g\tau\}$ and isomorphic to $C_\tau$, which proves (\ref{residue-quotient}). Note that only the freeness of $g$ is used: the closed stars of a vertex and of its image may well intersect, in which case $X/g$ is not a simplicial complex, but it is a simplicial cell complex, which is all that GEMs require.

\begin{figure}[htp]
\centering
\begin{tikzpicture}[scale=1.0]
\input{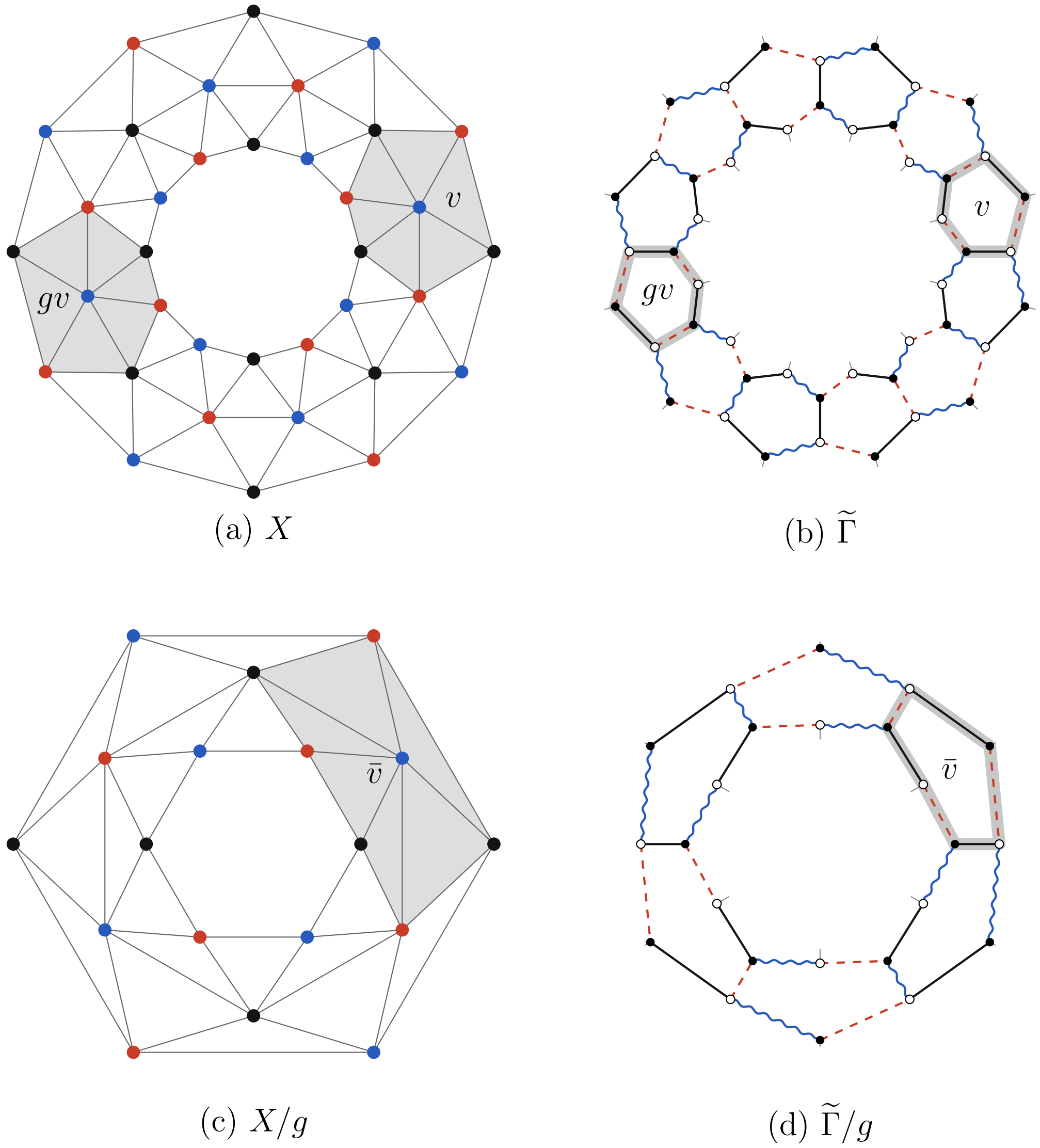}
\end{tikzpicture}
\caption{The quotient of a GEM by a free color-preserving involution, see (\ref{residue-quotient}). (a) A $3$-colored triangulated surface $X$, here an annulus, with the free involution $g$ given by the half-turn, a vertex $v$ and its image $gv$. The cells containing $v$ and those containing $gv$ (shaded) are disjoint, since no cell contains two vertices of the same color. (b) The dual graph $\widetilde\Gamma$, with the dual edges crossing the boundary of the annulus drawn as stubs. The components of $\rs{\widetilde\Gamma}{\hat\alpha}$, $\alpha$ the color of $v$, formed by the cells containing $v$ and $gv$ (highlighted) are two disjoint isomorphic cycles interchanged by $g$. (c) The quotient $X/g$ with the image $\bar v$ of $v$. (d) The quotient graph $\widetilde\Gamma/g$ contains one copy of the cycle. The same holds for every proper subset of colors, so that the residue of $\widetilde\Gamma/g$ is half of that of $\widetilde\Gamma$.}
\label{fig:quotient}
\end{figure}

\textit{The Dold manifold.} For $d=5$ the torsion is all of $\Omega^{\mathrm{SO}}_5\cong\Z_2$, generated by $P(1,2)$, for which $\int_{P(1,2)}w_2w_3\neq 0$. This case is covered by the general argument below, but the following explicit construction, used in Section \ref{sec:dold-example}, shows what it produces. Take a triangulation of $\CP^2$ on which complex conjugation acts simplicially (a smooth action of a finite group on a smooth manifold admits an equivariant triangulation \cite{Illman1978}), take its product with a cycle with an even number $2n$ of vertices as a triangulation of $S^1$ (a simplicial cell complex, $n=1$ being allowed), take the barycentric subdivision of the resulting cell complex and color its vertices by the dimension of the cells they subdivide. The result is a $5$-GEM $\widetilde\Gamma$ of $\CP^2\times S^1$ with two free color- and orientation-preserving involutions, the rotation of the cycle by $n$ steps and the same rotation combined with complex conjugation. By the observation above, the two quotient graphs, which we call $\Gamma_P$ and $\Gamma_D$, are GEMs of $\CP^2\times S^1$ and of the mapping torus of complex conjugation, which is the Dold manifold $P(1,2)$, and both have residue $\frac12R(\widetilde\Gamma)$. The construction is summarized in Figure \ref{fig:dold}. Since $\CP^2\times S^1=\partial(\CP^2\times D^2)$, both residues lie in $R(\Z[\text{spherical }5\text{-GEMs}])$ by Theorem \ref{thm:bordism}, so that $\Phi[P(1,2)]=0$, and $\mathcal{Q}_5=0$ since $\Phi$ is surjective.\footnote{An explicit pair of GEMs of $\CP^2\times S^1$ and of $P(1,2)$ with identical residues and of moderate size can be built from K\"uhnel's $9$-vertex triangulation of $\CP^2$ \cite{KuehnelBanchoff1983}, on which complex conjugation acts simplicially \cite[\S 9, \S 11]{MorinYoshida1991}. All nine involutions of its automorphism group of order $54$ are conjugate, so that complex conjugation can be taken to be $\mathfrak{c}=(12)(46)(89)$ in the labeling of \cite{KuehnelBanchoff1983}, with fixed point set a $6$-vertex $\RP^2$. The barycentric subdivision colored by dimension is a $4$-GEM of $\CP^2$ with $4320$ vertices on which $\mathfrak{c}$ acts freely, and canceling dipoles in $\mathfrak{c}$-orbits reduces it to a $\mathfrak{c}$-invariant $4$-GEM with $664$ vertices. Cutting $S^1$ into two intervals, triangulating each of the two copies of the product with $[0,1]$ by the standard triangulation of the prisms over the $4$-simplices, with the vertices ordered by color, the colors shifted by one along the first copy and shifted back along the second, and inserting $\mathfrak{c}$ across one of the two seams for the mapping torus, gives GEMs $\Gamma_P$ of $\CP^2\times S^1$ and $\Gamma_D$ of $P(1,2)$ with $6640$ vertices each. Their residues were verified by computer to be identical.}

\begin{figure}[ht]
\centering
\begin{tikzpicture}[scale=0.9,font=\small,
    cp/.style={draw,thick,fill=black!6,rounded corners=1pt,minimum width=1.05cm,minimum height=0.7cm,inner sep=1pt},
    cpbar/.style={cp,fill=c1!15},
    ar/.style={-{Stealth[length=5pt]},thick},
    inv/.style={-{Stealth[length=4pt]},c2,thick,dashed}]
  \def\Rc{2.2}
  \foreach \k in {0,...,5}{
    \pgfmathsetmacro{\ang}{90-60*\k}
    \node[cp] (c\k) at (\ang:\Rc) {$\CP^2$};
  }
  \foreach \k/\l in {0/1,1/2,2/3,3/4,4/5,5/0}{ \draw[thick] (c\k) -- (c\l); }
  \node at (0,3.15) {\small $X=\CP^2\times S^1$, triangulated with a cycle of $2n$ vertices};
  \draw[inv] (c0.south) to[bend left=18] node[right,pos=0.15,black]{\scriptsize $a$} (c3.north);
  \draw[inv,c3] (c1.west) to[bend left=18] node[below,pos=0.12,black]{\scriptsize $b$} (c4.east);
  \node[anchor=west,align=left,font=\scriptsize] at (3.1,0.35) {$a$: rotation by half a period};
  \node[anchor=west,align=left,font=\scriptsize] at (3.1,-0.25) {$b$: $a$ composed with $z\mapsto\bar z$ on $\CP^2$};
  \begin{scope}[shift={(-4.6,-5.2)}]
    \foreach \k in {0,1,2}{ \pgfmathsetmacro{\ang}{90-120*\k} \node[cp] (l\k) at (\ang:1.3) {$\CP^2$}; }
    \draw[thick] (l0) -- (l1) -- (l2) -- (l0);
    \node at (0,-2.1) {$X/a=\CP^2\times S^1$};
    \node at (0,-2.6) {\scriptsize $=\partial(\CP^2\times D^2)$};
  \end{scope}
  \begin{scope}[shift={(4.6,-5.2)}]
    \foreach \k in {0,1,2}{ \pgfmathsetmacro{\ang}{90-120*\k} \node[cp] (r\k) at (\ang:1.3) {$\CP^2$}; }
    \draw[thick] (r0) -- (r1) -- (r2);
    \draw[thick,c1] (r2) -- node[left=4pt,black,align=center]{\scriptsize glue by\\[-2pt]\scriptsize $z\mapsto\bar z$} (r0);
    \node at (0,-2.1) {$X/b=P(1,2)$};
    \node at (0,-2.6) {\scriptsize mapping torus of conjugation};
  \end{scope}
  \draw[ar] (-1.4,-2.6) -- node[above left]{\scriptsize $/a$} (-3.2,-3.9);
  \draw[ar] ( 1.4,-2.6) -- node[above right]{\scriptsize $/b$} ( 3.2,-3.9);
  \node[align=center,font=\scriptsize] at (0,-9.0) {both quotients are GEMs with the same multiset of colored vertex links,\\ each link occurring half as often as in $X$};
\end{tikzpicture}
\caption{The construction in the proof of Proposition \ref{prop:smooth-torsion} for $d=5$. The double cover $X=\CP^2\times S^1$ is triangulated as the product of a conjugation-invariant triangulation of $\CP^2$ with a cycle of $2n$ vertices (here $n=3$), barycentrically subdivided and colored by dimension. The rotation $a$ by half a period and its composition $b$ with complex conjugation are free color-preserving involutions. Their quotients are $\CP^2\times S^1$, which bounds, and the Dold manifold $P(1,2)$, which does not, yet the two quotient GEMs have identical residues.}
\label{fig:dold}
\end{figure}
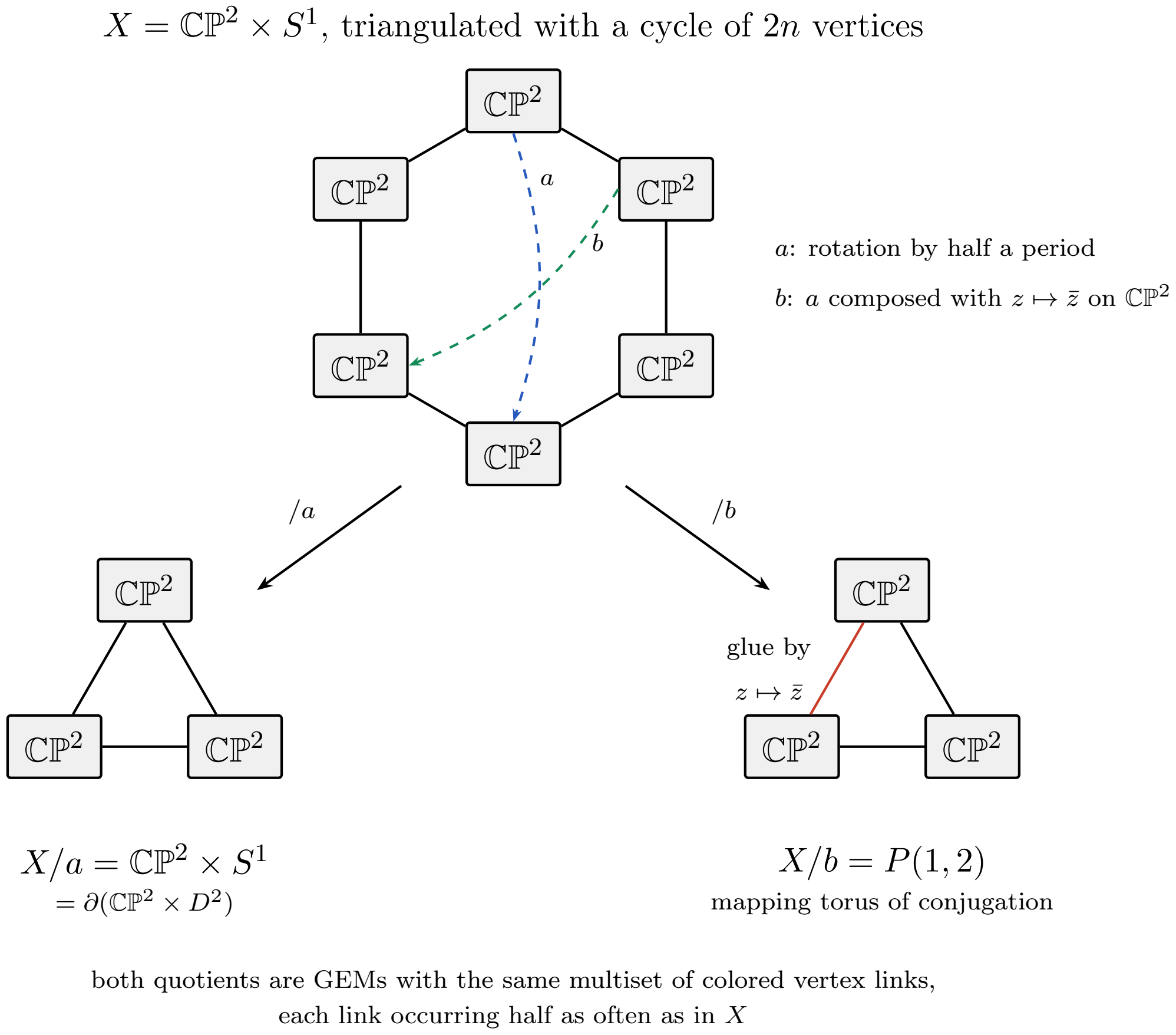
\textit{Untwisting.} Let $\widetilde B\rightarrow B$ be the double cover of the base associated with the bundle, with deck transformation $\mathfrak{t}$, and let $\mathfrak{c}$ be complex conjugation on $F$, so that $E=X/(\mathfrak{t}\times\mathfrak{c})$ and $B\times F=X/(\mathfrak{t}\times 1)$ for $X:=\widetilde B\times F$. Both involutions of $X$ are free and orientation preserving ($\mathfrak{c}$ preserves the orientation of $\CP^{2q}$), and they generate a smooth action of $\Z_2\times\Z_2$. By \cite{Illman1978}, $X$ has a triangulation on which this group acts simplicially, and its barycentric subdivision, with each vertex colored by the dimension of the simplex it subdivides, is a $(d+1)$-colored triangulation of $X$ on which both involutions act simplicially, preserving the colors and the orientation. Its dual graph $\widetilde\Gamma$ is a $d$-GEM of $X$ (Lemma \ref{lem:dual}), and by the observation above the quotients $\Gamma_E:=\widetilde\Gamma/(\mathfrak{t}\times\mathfrak{c})$ and $\Gamma_{B\times F}:=\widetilde\Gamma/(\mathfrak{t}\times 1)$ are GEMs of $E$ and of $B\times F$ with $R(\Gamma_E)=\frac12R(\widetilde\Gamma)=R(\Gamma_{B\times F})$, i.e. $\Phi[E]=\Phi[B\times F]$.

\textit{The base vanishes in $\mathcal{Q}_b$.} Pontryagin numbers are multiplied by the degree under finite covers, and $X$ is a double cover of both $E$ and $B\times F$. Hence $[B\times F]=[B][F]$ is a torsion class, as $[E]$ is. Since $\Omega^{\mathrm{SO}}_*\otimes\Q$ is a polynomial ring \cite{Thom1954} in which $[\CP^{2q}]\neq 0$, the class $[B]\in\Omega^{\mathrm{SO}}_b$, $b=d-4q<d$, is torsion. If $b=0$, then $B$ is a finite set of points with total sign zero, $B\times F$ is null-bordant, and $\Phi[E]=\Phi[B\times F]=0$. Otherwise the induction hypothesis gives $\Phi[B]=0$, hence $\Phi[B\times F]=\Phi[B]\cdot\Phi[F]=0$ since $\Phi$ is a ring homomorphism (Theorem \ref{thm:bordism}), and thus $\Phi[E]=0$ by the untwisting step. This completes the induction.
For the consequences, if $\Phi[M]=0$ for a smooth $M$, then all rational Pontryagin numbers of $M$ vanish by Proposition \ref{prop:kernel} and $[M]$ is torsion \cite{Thom1954}, so that the kernel of the composite is exactly the torsion subgroup. Its image has finite index since $\Phi\otimes\Q$ is an isomorphism and $\Qd$ is finitely generated. Finally, a torsion class of $\Omega^{\mathrm{SPL}}_d$ in the image of $\Omega^{\mathrm{SO}}_d$ comes from a torsion class of $\Omega^{\mathrm{SO}}_d$, again because the rational Pontryagin numbers detect $\Omega^{\mathrm{SO}}_d\otimes\Q$. Hence the kernel of $\Phi$, which consists of torsion classes by Proposition \ref{prop:kernel}, contains all torsion classes represented by smooth manifolds. Moreover, since $\Phi$ is surjective (Theorem \ref{thm:bordism}) with torsion kernel (Proposition \ref{prop:kernel}), the torsion subgroup of $\Qd$ is the image of the torsion subgroup of $\Omega^{\mathrm{SPL}}_d$: if $\Phi(y)$ is torsion, then $Ny\in\ker\Phi$ for some $N\geq 1$, so $Ny$ and hence $y$ are torsion. Therefore $\mathrm{Tor}\,\Qd=\mathrm{Tor}\,\Omega^{\mathrm{SPL}}_d/\ker\Phi$, and since $\ker\Phi\supseteq\mathrm{im}\,\mathrm{Tor}\,\Omega^{\mathrm{SO}}_d$, this gives (\ref{torsion-surjection}).
\end{proof}

\medskip\noindent
The argument does not extend to all of $\Omega^{\mathrm{SPL}}_d$, since no description by explicit generators is available for its torsion, which in high dimensions contains classes not represented by smooth manifolds, including classes of odd order \cite[Chapter 14]{MadsenMilgram1979}, whereas $\Omega^{\mathrm{SO}}_*$ has only $2$-torsion \cite{Wall1960}. By Proposition \ref{prop:smooth-torsion}, these are the only classes on which $\Phi$ could fail to be injective modulo torsion.

\medskip\noindent
Finally, the entries of Table \ref{tab:dims} follow from Proposition \ref{prop:extract}, Theorem \ref{thm:structure} and Corollary \ref{cor:extract}. For $d=4$, $\Phi$ is an isomorphism detected by the signature, so $\sum_in_i[\Gamma_i]=0$ if and only if $\sum_in_i\sigma(M_{\Gamma_i})=0$. In particular $[\Gamma]+[\overline\Gamma]=0$ always and $|Z(M)|^2/Z(S^4)^{\chi(M)}$ is extractable for every $M$. For $d=5$ one has $\mathcal{Q}_5=0$. For odd $d$ the power of $Z(S^d)$ is removed using Proposition \ref{prop:exponent}(ii) and (\ref{Z-sphere-positive}). For $d=4k$ in general, $[\Gamma]=0$ implies the vanishing of the rational Pontryagin numbers by Theorem \ref{thm:structure}(i), and conversely a class in $\Omega^{\mathrm{SPL}}_{4k}$ with vanishing rational Pontryagin numbers is torsion \cite{Thom1954,Wall1960}, hence maps to a torsion element of $\Qd$, which is zero if the class is represented by a smooth manifold (Theorem \ref{thm:structure}(iii)) or if $\Qd$ is torsion-free.

\section{Example: an invertible phase in \texorpdfstring{$4+1$}{4+1} dimensions}
\label{sec:dold-example}

In this section we illustrate the results above, and in particular Theorem \ref{thm:structure}(iii), in an exactly solvable example with $d=5$. The theory is the non-trivial bosonic invertible phase without symmetry in $4+1$ dimensions \cite{Kapustin:2014tfa,Freed:2016rqq}, with partition function
\begin{equation}
    Z(M)=(-1)^{\nu(M)},\qquad \nu(M):=\int_M w_2(TM)\cup w_3(TM)\in\Z_2,
    \label{Z-w2w3}
\end{equation}
the generator of $\mathrm{Hom}(\Omega^{\mathrm{SO}}_5,U(1))\cong\Z_2$. It is realized by the exactly solvable lattice models of \cite{Chen:2021xks,Fidkowski:2021unr}, the boundary of which is the anomalous $\Z_2$ gauge theory with fermionic particles and fermionic loops. For the ground state of \cite{Chen:2021xks} the multi-invariants can be computed exactly for every $5$-GEM: Conjecture ELD holds exactly, without any large-region limit, with the universal factor (\ref{Z-w2w3}) and a non-universal factor of residue form. In particular, for the GEMs $\Gamma_D$ of the Dold manifold and $\Gamma_P$ of $\CP^2\times S^1$ with identical residues constructed in the proof of Proposition \ref{prop:smooth-torsion}, the two multi-invariants are equal in absolute value and opposite in sign. The torsion bordism invariant $\nu$ does not admit a local formula, and accordingly it is extracted exactly by the scheme of Proposition \ref{prop:extract}. The proofs are collected in Appendix \ref{app:dold}.

\subsection{The state and the regions}
\label{sec:ch-state}

Let $X$ be a triangulated closed $4$-manifold with a branching structure (a total order on the vertex set of each simplex, such that the order on a subsimplex is the restriction of the order on the simplex, for instance the orders induced on all simplices by a single total order of all the vertices of $X$), with a qubit on every $2$-simplex and every $3$-simplex, so that a basis of the Hilbert space is labeled by pairs of cochains $A\in C^2(X;\Z_2)$, $B\in C^3(X;\Z_2)$: $|A,B\rangle$. The ground state of the local commuting-projector model of \cite{Chen:2021xks} for the phase (\ref{Z-w2w3}) is
\begin{equation}
\begin{gathered}
    |\Psi\rangle\;\propto\;\sum_{a\in C^1(X;\Z_2)}\;\sum_{b\in C^2(X;\Z_2)}(-1)^{\int_X\phi_4(a,b)}\,|\delta a,\delta b\rangle,\\[2pt]
    \phi_4(a,b)=b\cup b+b\cupi{1}\delta b+a\cup\delta b+a\cup(\delta a\cupi{1}\delta a),
\end{gathered}
    \label{Psi-def}
\end{equation}
where $\cup_i$ are Steenrod's cup-$i$ products defined by the branching structure (the explicit formulas are recalled in Appendix \ref{app:ch-conventions}). The $4$-cochain $\phi_4$ satisfies the exact identity
\begin{equation}
    \delta\phi_4(a,b)=\phi_5(\delta a,\delta b),\qquad \phi_5(A,B)=B\cupi{1}B+A\cup B+A\cup(A\cupi{1}A),
    \label{phi5}
\end{equation}
and $\phi_5$ is the action of the $\Z_2$ two-form/three-form gauge theory of \cite{Chen:2021xks}: on a closed oriented $5$-manifold $M$ the integral $\int_M\phi_5(A,B)$ depends only on the classes $[A]\in H^2(M;\Z_2)$, $[B]\in H^3(M;\Z_2)$ (Appendix \ref{app:ch-proof}), and
\begin{equation}
    \sum_{[A]\in H^2(M;\Z_2)}\;\sum_{[B]\in H^3(M;\Z_2)}(-1)^{\int_M\phi_5(A,B)}=|H^3(M;\Z_2)|\,(-1)^{\nu(M)}.
    \label{gauged}
\end{equation}
Indeed, by the Wu formula $\int_MB\cupi{1}B=\int_M\Sq^2B=\int_Mw_2\cup B$, so that the sum over $[B]$ vanishes unless $[A]=w_2$ by Poincar\'e duality, and then $\int_MA\cup(A\cupi{1}A)=\int_Mw_2\cup\Sq^1w_2=\int_Mw_2w_3$.

For $X\cong S^4$ the sign in (\ref{Psi-def}) depends only on $(A,B)=(\delta a,\delta b)$ (the cocycles $A$, $B$, in the notation of \cite{Chen:2021xks}, are not to be confused with the factors $A(\gamma)$ of Conjecture ELD) (Appendix \ref{app:ch-conventions}), and we fix the overall constant by
\begin{equation}
    |\Psi\rangle=\sum_{\substack{A\in Z^2(X;\Z_2)\\ B\in Z^3(X;\Z_2)}}(-1)^{F(A,B)}\,|A,B\rangle,\qquad F(A,B):=\int_X\phi_4(a,b),
    \label{Psi-F}
\end{equation}
with any $a,b$ such that $\delta a=A$, $\delta b=B$. The state is not normalized, which is immaterial: the normalization would only contribute the factor $A(\bullet\sqcup\circ)^r$ of (\ref{conj-ELD}) to the multi-invariants (Section \ref{sec:residues}), which cancels in all ratios. Note that $F$ is a cubic, not quadratic, function of $(A,B)$ over $\Z_2$, because of the last term of $\phi_4$.

\begin{figure}[ht]
\centering
\begin{tikzpicture}[scale=0.95,font=\scriptsize]
  \begin{scope}
  \fill[black!4] (0.000,0.000) -- (1.000,0.000) -- (2.000,0.000) -- (3.000,0.000) -- (2.500,0.866) -- (2.000,1.732) -- (1.500,2.598) -- (1.000,1.732) -- (0.500,0.866) -- cycle;
  \draw[black!45,thin] (1.500,0.866) -- (0.000,0.000);
  \draw[black!45,thin] (1.500,0.866) -- (1.000,0.000);
  \draw[black!45,thin] (1.500,0.866) -- (2.000,0.000);
  \draw[black!45,thin] (1.500,0.866) -- (3.000,0.000);
  \draw[black!45,thin] (1.500,0.866) -- (2.500,0.866);
  \draw[black!45,thin] (1.500,0.866) -- (2.000,1.732);
  \draw[black!45,thin] (1.500,0.866) -- (1.500,2.598);
  \draw[black!45,thin] (1.500,0.866) -- (1.000,1.732);
  \draw[black!45,thin] (1.500,0.866) -- (0.500,0.866);
  \draw[c0,very thick] (0.000,0.000) -- (3.000,0.000);
  \draw[c1,very thick] (3.000,0.000) -- (1.500,2.598);
  \draw[c2,very thick] (1.500,2.598) -- (0.000,0.000);
  \node[circle,inner sep=1.15pt,fill=c0] at (1.000,0.000) {};
  \node[circle,inner sep=1.15pt,fill=c0] at (2.000,0.000) {};
  \node[circle,inner sep=1.15pt,fill=c1] at (2.500,0.866) {};
  \node[circle,inner sep=1.15pt,fill=c1] at (2.000,1.732) {};
  \node[circle,inner sep=1.15pt,fill=c2] at (1.000,1.732) {};
  \node[circle,inner sep=1.15pt,fill=c2] at (0.500,0.866) {};
  \node[circle,inner sep=1.15pt,fill=c0] at (0.320,0.000) {};
  \node[circle,inner sep=1.15pt,fill=c2] at (0.160,0.277) {};
  \draw[black,thin,fill=white] (0.000,0.000) circle (0.07);
  \node[circle,inner sep=1.15pt,fill=c0] at (2.680,0.000) {};
  \node[circle,inner sep=1.15pt,fill=c1] at (2.840,0.277) {};
  \draw[black,thin,fill=white] (3.000,0.000) circle (0.07);
  \node[circle,inner sep=1.15pt,fill=c1] at (1.660,2.321) {};
  \node[circle,inner sep=1.15pt,fill=c2] at (1.340,2.321) {};
  \draw[black,thin,fill=white] (1.500,2.598) circle (0.07);
  \node[black!70] at (1.500,0.330) {$W$};
  \node[c0,below] at (1.500,-0.060) {$0$};
  \node[c1,above right,inner sep=1.5pt] at (2.250,1.299) {$1$};
  \node[c2,above left,inner sep=1.5pt] at (0.750,1.299) {$2$};
  \node[above,inner sep=4pt] at (1.500,2.598) {$|x\rangle\mapsto|x,x\rangle$};
  \node[gray] at (-0.550,1.299) {(a)};
  \end{scope}
  \begin{scope}[shift={(5.200,1.299)}]
  \fill[black!4] (0.000,0.000) -- (1.000,0.000) -- (2.000,0.000) -- (3.000,0.000) -- (2.500,0.866) -- (2.000,1.732) -- (1.500,2.598) -- (1.000,1.732) -- (0.500,0.866) -- cycle;
  \draw[black!45,thin] (1.500,0.866) -- (0.000,0.000);
  \draw[black!45,thin] (1.500,0.866) -- (1.000,0.000);
  \draw[black!45,thin] (1.500,0.866) -- (2.000,0.000);
  \draw[black!45,thin] (1.500,0.866) -- (3.000,0.000);
  \draw[black!45,thin] (1.500,0.866) -- (2.500,0.866);
  \draw[black!45,thin] (1.500,0.866) -- (2.000,1.732);
  \draw[black!45,thin] (1.500,0.866) -- (1.500,2.598);
  \draw[black!45,thin] (1.500,0.866) -- (1.000,1.732);
  \draw[black!45,thin] (1.500,0.866) -- (0.500,0.866);
  \fill[black!4] (0.000,0.000) -- (1.000,0.000) -- (2.000,0.000) -- (3.000,0.000) -- (2.500,-0.866) -- (2.000,-1.732) -- (1.500,-2.598) -- (1.000,-1.732) -- (0.500,-0.866) -- cycle;
  \draw[black!45,thin] (1.500,-0.866) -- (0.000,0.000);
  \draw[black!45,thin] (1.500,-0.866) -- (1.000,0.000);
  \draw[black!45,thin] (1.500,-0.866) -- (2.000,0.000);
  \draw[black!45,thin] (1.500,-0.866) -- (3.000,0.000);
  \draw[black!45,thin] (1.500,-0.866) -- (2.500,-0.866);
  \draw[black!45,thin] (1.500,-0.866) -- (2.000,-1.732);
  \draw[black!45,thin] (1.500,-0.866) -- (1.500,-2.598);
  \draw[black!45,thin] (1.500,-0.866) -- (1.000,-1.732);
  \draw[black!45,thin] (1.500,-0.866) -- (0.500,-0.866);
  \draw[c0,very thick] (0.000,0.000) -- (3.000,0.000);
  \draw[c1,very thick] (3.000,0.000) -- (1.500,2.598);
  \draw[c2,very thick] (1.500,2.598) -- (0.000,0.000);
  \draw[c1,very thick] (3.000,0.000) -- (1.500,-2.598);
  \draw[c2,very thick] (1.500,-2.598) -- (0.000,0.000);
  \node[circle,inner sep=1.15pt,fill=c0] at (1.000,0.000) {};
  \node[circle,inner sep=1.15pt,fill=c0] at (2.000,0.000) {};
  \node[circle,inner sep=1.15pt,fill=c1] at (2.500,0.866) {};
  \node[circle,inner sep=1.15pt,fill=c1] at (2.000,1.732) {};
  \node[circle,inner sep=1.15pt,fill=c2] at (1.000,1.732) {};
  \node[circle,inner sep=1.15pt,fill=c2] at (0.500,0.866) {};
  \node[circle,inner sep=1.15pt,fill=c1] at (2.500,-0.866) {};
  \node[circle,inner sep=1.15pt,fill=c1] at (2.000,-1.732) {};
  \node[circle,inner sep=1.15pt,fill=c2] at (1.000,-1.732) {};
  \node[circle,inner sep=1.15pt,fill=c2] at (0.500,-0.866) {};
  \node[circle,inner sep=1.15pt,fill=c0] at (0.320,0.000) {};
  \node[circle,inner sep=1.15pt,fill=c2] at (0.160,0.277) {};
  \node[circle,inner sep=1.15pt,fill=c2] at (0.160,-0.277) {};
  \draw[black,thin,fill=white] (0.000,0.000) circle (0.07);
  \node[circle,inner sep=1.15pt,fill=c0] at (2.680,0.000) {};
  \node[circle,inner sep=1.15pt,fill=c1] at (2.840,0.277) {};
  \node[circle,inner sep=1.15pt,fill=c1] at (2.840,-0.277) {};
  \draw[black,thin,fill=white] (3.000,0.000) circle (0.07);
  \node[circle,inner sep=1.15pt,fill=c1] at (1.660,2.321) {};
  \node[circle,inner sep=1.15pt,fill=c2] at (1.340,2.321) {};
  \draw[black,thin,fill=white] (1.500,2.598) circle (0.07);
  \node[circle,inner sep=1.15pt,fill=c1] at (1.660,-2.321) {};
  \node[circle,inner sep=1.15pt,fill=c2] at (1.340,-2.321) {};
  \draw[black,thin,fill=white] (1.500,-2.598) circle (0.07);
  \node[bl] (kv) at (1.500,0.866) {};
  \node[wh] (bv) at (1.500,-0.866) {};
  \draw[cc0] (kv) -- (bv);
  \draw[cc1] (kv) -- (2.236,1.291);
  \draw[cc2] (kv) -- (0.764,1.291);
  \draw[cc1] (bv) -- (2.236,-1.291);
  \draw[cc2] (bv) -- (0.764,-1.291);
  \node[above,inner sep=4pt] at (1.500,0.866) {$i$};
  \node[below,inner sep=4pt] at (1.500,-0.866) {$\sigma_0(i)$};
  \node[c0,right,inner sep=3pt] at (3.000,0.000) {$0$};
  \node[gray] at (-0.550,0.000) {(b)};
  \end{scope}
\end{tikzpicture}
\caption{A two-dimensional analog of the lattice of this section and of the gluing in the proof of Proposition \ref{prop:ch} (Appendix \ref{app:ch-proof}). (a) The space $S^1=\partial\Delta^2$, with the parties $0,1,2$ its edges, is triangulated by $X$, here with each party cut into three segments, and $W$ (gray) is a triangulation of the disc with $\partial W=X$. In this analog the qubits (dots) sit on the vertices of $X$. A vertex of $\Delta^2$ lies in two parties and carries one qubit in each of them, drawn displaced into the two parties, and the state is embedded by copying. (b) The copies of $W$ for a black vertex $i$ and a white vertex $\sigma_0(i)$ of $\Gamma$, the latter a mirror image, glued along the party $0$ by the edge of color $0$ of $\Gamma$ between them, as in Figure \ref{fig:gems}. The qubits of the party $0$ of the two replicas are identified, and the two copies of a corner qubit within each replica agree. Therefore the corner, a cell of $K(\Gamma)$ of type $\{0,2\}$ (left) or $\{0,1\}$ (right), carries a single value on the whole connected component of $\rs{\Gamma}{\{0,2\}}$, respectively $\rs{\Gamma}{\{0,1\}}$, containing this edge.}
\label{fig:ch-lattice}
\end{figure}
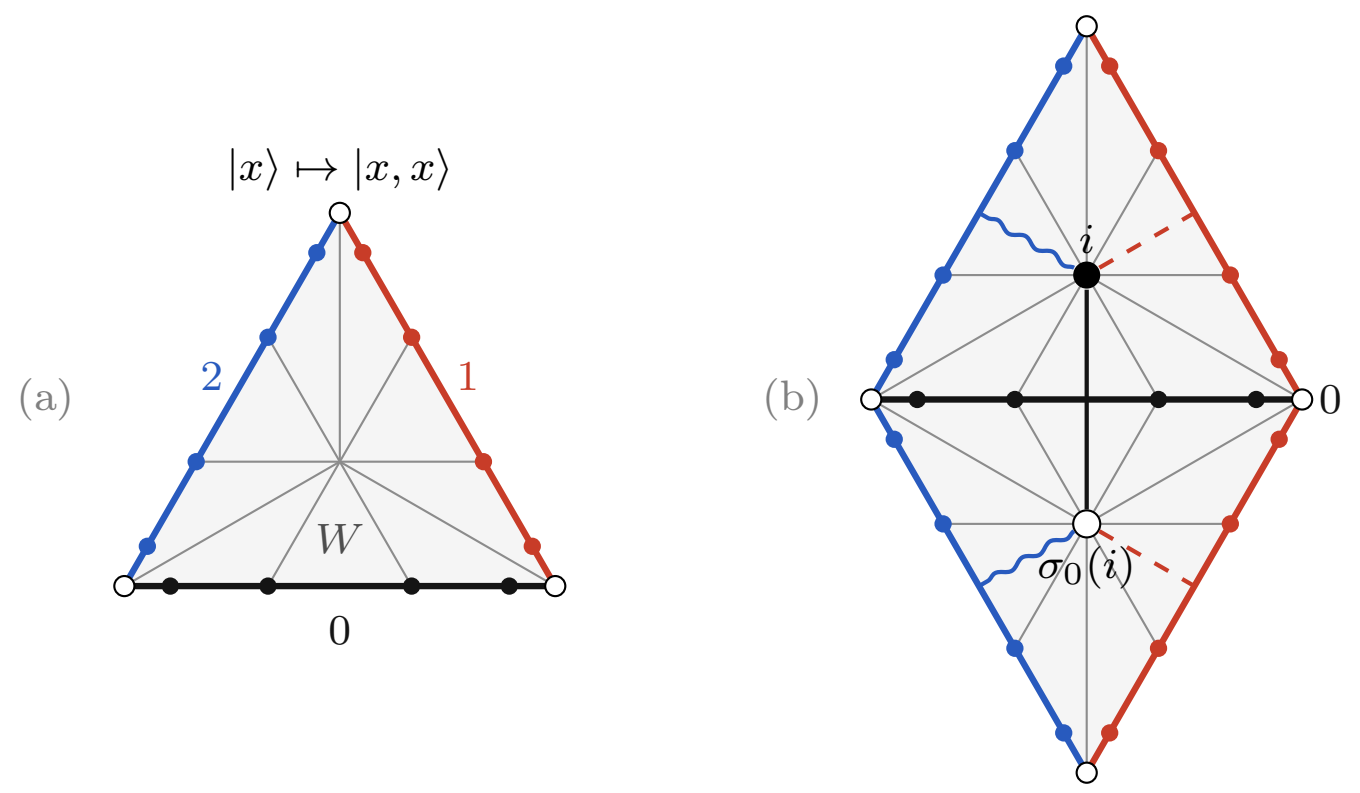

The space $S^4$ is decomposed into the six parties as in Section \ref{sec:multi-invariants}: $S^4=\partial\Delta^5$ with vertices labeled by the colors $\{0..5\}$, the party $\alpha$ being the $4$-dimensional face $\Delta_\alpha$ opposite to the vertex $\alpha$. The lattice is an arbitrary triangulation $X$ of $\partial\Delta^5$ refining its simplicial structure, with an arbitrary branching structure. For a non-empty $I\subseteq\{0..5\}$, the subsimplex of $\Delta^5$ contained in exactly the parties $\Delta_\alpha$, $\alpha\in I$, is the simplex spanned by $\hat I$, of dimension $5-|I|$. We denote by $\kappa_I$ the number of $2$-simplices of $X$ with interior contained in the interior of that subsimplex (so that for $|I|=3$, where the subsimplex is itself a triangle, $\kappa_I$ is the number of triangles of $X$ subdividing it), so that $\kappa_I=0$ for $|I|\geq 4$, and $\sum_I\kappa_I=c_2(X)$, the number of triangles of $X$. A simplex of $X$ lying in the closure of several parties is represented by one qubit in the Hilbert space of \textit{each} of them (Figure \ref{fig:ch-lattice}(a)), and the state is embedded isometrically by copying, $|x\rangle\mapsto|x,\ldots,x\rangle$ (CNOT gates with the original qubit $|x\rangle$ as the control and auxiliary qubits initialized in $|0\rangle$ as the targets, acting near the interfaces, so that the embedded state is the exact ground state of a local commuting Hamiltonian in the same phase). This is the extended Hilbert space mentioned in Section \ref{sec:multi-invariants}, the standard choice for gauge-theory-like models \cite{levin2006detecting,DelZotto:2026dgp}. We keep the notation $|\Psi\rangle$ for the embedded state.

\begin{proposition}
\label{prop:ch}
Let $\Gamma$ be a $5$-GEM with $r$ replicas. Then, exactly,
\begin{equation}
    \CZ(\Gamma;|\Psi\rangle)=(-1)^{\nu(M_\Gamma)}\;2^{\,\sum_{1\leq|I|\leq 3}\kappa_I\,\#\pi_0(\rs{\Gamma}{I})}.
    \label{ch-main}
\end{equation}
\end{proposition}

\noindent
For instance, for $X=\partial\Delta^5$ itself, $\kappa_I=1$ for $|I|=3$ and $\kappa_I=0$ otherwise, and
\begin{equation}
    \CZ(\Gamma;|\Psi\rangle)=(-1)^{\nu(M_\Gamma)}\,2^{\,c_2(K(\Gamma))},\qquad c_2(K(\Gamma)):=\sum_{|I|=3}\#\pi_0(\rs{\Gamma}{I})=\sum_{|J|=3}c_J(\Gamma),
    \label{ch-bare}
\end{equation}
$c_2(K(\Gamma))$ being the number of connected $3$-colored subgraphs of $\Gamma$, i.e. of triangles of $K(\Gamma)$. In general, the exponent in (\ref{ch-main}) has three kinds of contributions. The terms with $|I|=1$ add up to $r\sum_{|I|=1}\kappa_I$, since $\#\pi_0(\rs{\Gamma}{I})=r$ for a single color, and depend on $\Gamma$ only through $r$. The terms with $|I|=2$ come from the two-party interfaces, which are three-dimensional, and the terms with $|I|=3$ from the three-party junctions, which are two-dimensional. Under a refinement of $X$ with lattice spacing $\epsilon$ at fixed size $L$ of the regions, $\kappa_I$ grows like $(L/\epsilon)^3$ for $|I|=2$ and like $(L/\epsilon)^2$ for $|I|=3$, while no triangle fits into a junction of four or more regions. Each factor $2^{\#\pi_0(\rs{\Gamma}{I})}$ is precisely the multi-invariant of an unnormalized Bell pair ($|I|=2$) or three-party Greenberger-Horne-Zeilinger (GHZ) state ($|I|=3$), $|0\ldots0\rangle+|1\ldots1\rangle$ shared by the parties $I$, cf.\ (\ref{lower-partite-factorization}): the non-universal factor is that of one such state per interface or junction triangle, although the state (\ref{Psi-def}) itself does not factorize.

\begin{corollary}
\label{cor:ch}
Conjecture ELD holds exactly for $|\Psi\rangle$, on every lattice $X$ as above, with $Z(M)=(-1)^{\nu(M)}$ and with $A(\gamma)=2^{\kappa_I}$ for $\gamma$ with color set $I$, $1\leq|I|\leq 3$, and $A(\gamma)=1$ otherwise. Consequently, every signal of Corollary \ref{cor:S} and Proposition \ref{prop:extract} evaluates exactly to the corresponding product of partition functions (\ref{Z-w2w3}). In particular, for two GEMs with the same residue the ratio of the multi-invariants is the ratio of the partition functions.
\end{corollary}

\subsection{The Dold manifold}
\label{sec:dold-signal}

Let $\Gamma_D$ and $\Gamma_P$ be GEMs of $P(1,2)$ and of $\CP^2\times S^1$ with the same residue, as constructed in the proof of Proposition \ref{prop:smooth-torsion}. They have the same replica number and the same numbers $\#\pi_0(\rs{\Gamma}{I})$ for all $I$, so that by Proposition \ref{prop:ch}
\begin{equation}
    \frac{\CZ(\Gamma_D;|\Psi\rangle)}{\CZ(\Gamma_P;|\Psi\rangle)}=\frac{Z(P(1,2))}{Z(\CP^2\times S^1)}=(-1)^{\nu(P(1,2))}=-1
    \label{dold-ratio}
\end{equation}
exactly, on every lattice $X$. Since $P(1,2)$ does not bound, Corollary \ref{cor:extract} does not apply to it, but $\mathcal{Q}_5=0$ by Theorem \ref{thm:structure}(iii), so that Proposition \ref{prop:extract}(i) does, and (\ref{dold-ratio}) is its explicit form: $Z(P(1,2))$ is the ratio (\ref{dold-ratio}) times $Z(\CP^2\times S^1)$, and the latter is in turn a substitution-graph signal built from the null-bordism $\CP^2\times D^2$ (Corollary \ref{cor:S}), equal to $1$ here.

It is instructive to see where the sign comes from. The mod $2$ cohomology ring of the Dold manifold is $H^*(P(1,2);\Z_2)=\Z_2[u,v]/(u^2,v^3)$ with $|u|=1$, $|v|=2$ and $\int uv^2=1$, and its total Stiefel-Whitney class is $w=(1+u)(1+u+v)^3=1+v+uv+v^2$ \cite{Dold1956}, so that $w_2=v$, $w_3=uv$ and $\Sq^1v=uv$. For $\CP^2\times S^1$ the ring is the same, with $u$ the class of the circle, again $w_2=v$, and $\Sq^1v=0$. In both cases $H^2$ and $H^3$ have two elements each, $[A]=xv$ and $[B]=y\,uv$ with $x,y\in\{0,1\}$, and, since $x^2=x$ and $y^2=y$, the exponent in (\ref{gauged}) is $\int\phi_5=y\int\Sq^2(uv)+xy\int v\cup uv+x\int v\cup\Sq^1v=x\int v\cup\Sq^1v+y\int w_2\cup uv+xy\int v\cup uv$, i.e.
\begin{equation}
    \frac{1}{|H^3|}\sum_{x,y}(-1)^{\int\phi_5}=\begin{cases}
    \frac12\sum_{x,y\in\{0,1\}}(-1)^{x+y+xy}=\frac{1-1-1-1}{2}=-1 & \text{for }P(1,2),\\[4pt]
    \frac12\sum_{x,y\in\{0,1\}}(-1)^{y+xy}=\frac{1+1-1+1}{2}=+1 & \text{for }\CP^2\times S^1.
    \end{cases}
    \label{four-sectors}
\end{equation}
The difference is the single term $\int v\cup\Sq^1v$. The mod $2$ cohomology rings of the two manifolds are isomorphic, but they are different modules over the Steenrod algebra: for $\CP^2\times S^1$ the class $v$ is the reduction of an integral class, so that $\Sq^1v=0$, whereas for $P(1,2)$ going once around the circle with complex conjugation obstructs the integral lift of $v$, and $\Sq^1v=uv\neq 0$. It is this difference that the sign detects. This is a general feature of mapping tori. If $f$ is an orientation preserving self-map of $\CP^2$ acting as $-1$ on $H^2(\CP^2;\Z)$, the Wang sequence gives $H^2(M_f;\Z)=0$ and $H^2(M_f;\Z_2)=\Z_2$, generated by the class restricting to the generator of the fiber, which is $w_2(M_f)$. Since it does not lift to an integral class, its Bockstein is the non-zero element of $H^3(M_f;\Z)=\mathrm{coker}(f^*-1)=\Z_2$, which is not divisible by $2$, so that its reduction $w_3(M_f)=\Sq^1w_2(M_f)$ is non-zero and $\nu(M_f)=1$. For $f=\mathrm{id}$, $w_2$ lifts and $w_3=0$.

The sign in (\ref{dold-ratio}) is a genuinely non-local property of the state: no product over the vertex links of $\Gamma$, or over any of its proper-color subgraphs, can produce it, since the same product would appear for $\Gamma_P$. This is consistent with the fact that the state (\ref{Psi-def}) is not a finite-depth circuit applied to a product state \cite{Chen:2021xks,Fidkowski:2021unr}. The same computation should apply to the model of \cite{Fidkowski:2021unr} and, with symmetry-refined multi-invariants, to the $\Z_2$-symmetric beyond-cohomology phase of \cite{Chen:2021xks}.

One further remark is in order, concerning the parity of the power of $Z(S^5)$ discussed after Lemma \ref{lem:parity}. Since $\Gamma_D$ and $\Gamma_P$ have the same residue, every combination $\eta$ of spherical GEMs with $R(\eta)=R(\Gamma_P)$ also satisfies $R(\eta)=R(\Gamma_D)$, and for $\eta$ obtained from a filling of $\CP^2\times S^1$ the coefficient sum is odd by Lemma \ref{lem:parity}, $\kQ(\CP^2\times S^1)=1$. On the other hand $\kQ(P(1,2))=0$. The difference is again the invariant $\nu$: by a theorem of \cite{LusztigMilnorPeterson1969}, for a closed oriented manifold $M$ of dimension $4k+1$ the rational and the mod $2$ semicharacteristics satisfy $\kQ(M)-k_{\Z_2}(M)=\nu(M)$, where $k_{\Z_2}(M):=\sum_{i=0}^{2k}\dim H_i(M;\Z_2)\mod 2$, and the mod $2$ semicharacteristics of $P(1,2)$ and $\CP^2\times S^1$ agree, since their mod $2$ cohomologies are isomorphic. Thus $\kQ(M_\Gamma)$ is not a function of the residue, and the parity of the power of $Z(S^5)$ is not given by $\kQ(M_\Gamma)$ beyond null-bordant manifolds.


\section{Discussion and future directions}
\label{sec:discussion}

Let us comment on the physical meaning of the results and on some natural directions for future work.

\paragraph{Local formulas and invertible phases.}
Parts (i) and (iii) of Theorem \ref{thm:structure} have a common origin. A ratio of multi-invariants of the type considered in this paper is insensitive to any factor of the form $U(R(\Gamma))$, whatever its origin. The Euler characteristic and the rational Pontryagin numbers are of this form: by (\ref{chi-linear}) and Proposition \ref{prop:kernel} they are sums over the vertices of the triangulation of quantities determined by the colored links. The partition function of an invertible TQFT \cite{Kapustin:2014tfa,Freed:2016rqq} with a continuous character of the bordism group, $e^{i\theta\sigma(M)}$ in $d=4$, and likewise the Euler counterterm $e^{\lambda\chi(M)}$ in any even $d$, which is not a bordism invariant but is a local formula by (\ref{chi-linear}), is therefore indistinguishable from an ultraviolet contribution, and the corresponding ambiguity in $Z(M)$ cannot be resolved by these schemes. This is the content of the second part of Proposition \ref{prop:extract} and of Table \ref{tab:dims}. Conversely, everything which is not of this form can be extracted, and for null-bordant manifolds this is everything. By Theorem \ref{thm:structure}, the group of bordism invariants admitting a local formula in this sense, i.e. of characters of $\Omega^{\mathrm{SPL}}_d$ factoring through $R$, is the character group of $\Qd$. It contains all the rational Pontryagin numbers, and the expectation $\Qd\cong\Omega^{\mathrm{SPL}}_d/\mathrm{torsion}$ is equivalent to the statement that no torsion bordism invariant admits such a local formula. Local formulas of this kind include those of \cite{Gaifullin2008}: a function of the isomorphism class of the oriented link of a vertex, with values in an arbitrary abelian group, such that its sum over the vertices of every oriented combinatorial manifold is a bordism invariant. Every such formula restricts to a character of $\Qd$ (apply it to the barycentric subdivision of $K(\Gamma)$ as in the proof of Proposition \ref{prop:kernel}). By Theorem \ref{thm:structure}(iii), a character of $\Omega^{\mathrm{SO}}_d$ admits a local formula, of either kind, if and only if it is trivial on the torsion subgroup, i.e. if and only if it is a function of the rational Pontryagin numbers. For smooth manifolds this answers the question, left open in \cite{Gaifullin2008}, of which bordism invariants admit local formulas with non-rational coefficients. What remains open, there and here, is the torsion of $\Omega^{\mathrm{SPL}}_d$ that is not represented by smooth manifolds. In particular, the invariant $\int w_2w_3$ in $d=5$, or any Stiefel-Whitney number that is non-trivial on the torsion subgroup, admits no local formula, and the corresponding invertible phases, unlike the ones above, are all detectable by multi-invariants. Section \ref{sec:dold-example} uses this explicitly for $\int w_2w_3$ and the Chen-Hsin state.

This is also the natural place to comment on which invertible phases can be detected at all. The classification of invertible topological phases by characters of the bordism group \cite{Kapustin:2014tfa,Freed:2016rqq} is formulated for smooth manifolds, whereas a lattice system defines its partition function, and its multi-invariants, on triangulated, i.e. PL, manifolds. The same classification then applies with $\Omega^{\mathrm{SPL}}_d$ in place of $\Omega^{\mathrm{SO}}_d$. The natural map $\Omega^{\mathrm{SO}}_d\hookrightarrow\Omega^{\mathrm{SPL}}_d$ is injective and an isomorphism for $d\leq 7$ (Section \ref{sec:gem}). From $d=8$ on it has a cokernel. For instance, coning off the boundary of the $E_8$-plumbing of eight copies of the tangent disk bundle of $S^4$, which is PL homeomorphic to $S^7$, gives a non-smoothable closed PL $8$-manifold with signature $8$ and $p_1=0$, hence with the fractional Pontryagin number $p_2=360/7$ \cite{KirbySiebenmann1977}, and in higher dimensions there are torsion classes with no smooth representative \cite[Chapter 14]{MadsenMilgram1979}. Accordingly, every invertible phase of smooth manifolds extends, non-uniquely, to PL manifolds, and there are invertible phases visible only on non-smoothable spacetimes. Assuming Conjecture ELD, an invertible phase is undetectable by the multi-invariants precisely when its partition function, as a character of $\Omega^{\mathrm{SPL}}_d$, factors through $\Phi$, i.e. is the pull-back of a character of $\Qd$. The situation is summarized in the diagram
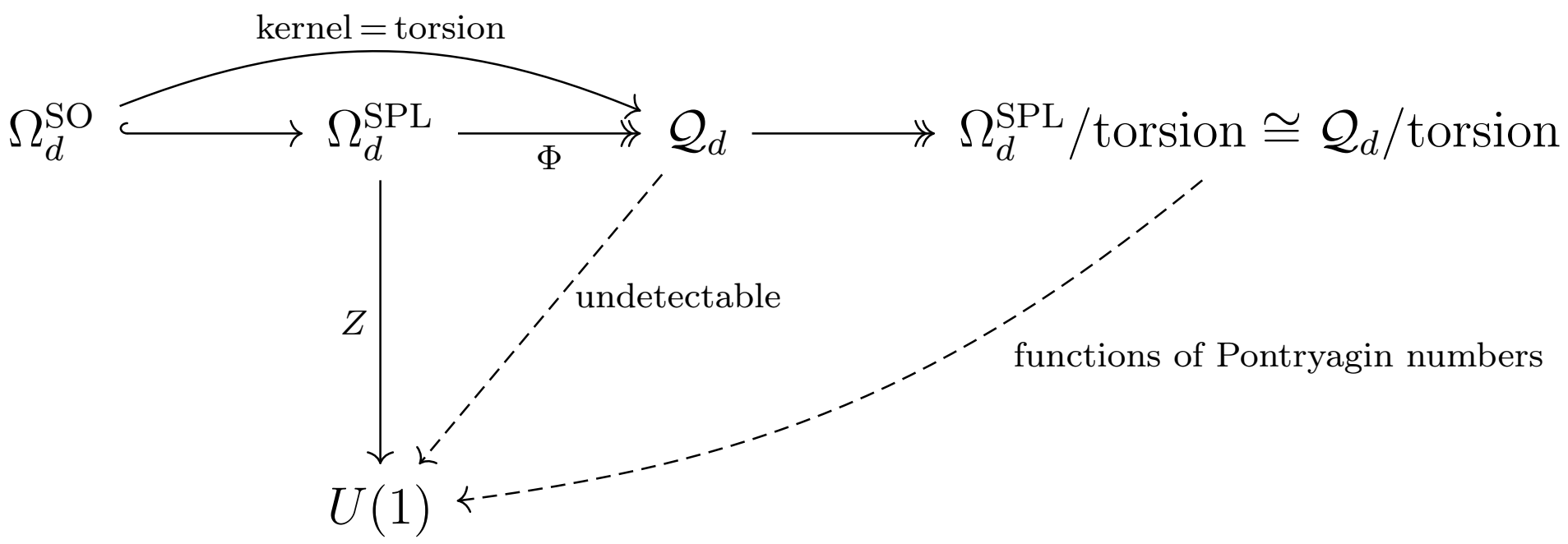
\begin{equation}
\begin{tikzcd}[column sep=large,row sep=2.3cm]
\Omega^{\mathrm{SO}}_d \arrow[r,hook] \arrow[rr,bend left=22,"\text{kernel}\,=\,\text{torsion}"] & \Omega^{\mathrm{SPL}}_d \arrow[r,two heads,"\Phi"'] \arrow[d,"Z"'] & \Qd \arrow[r,two heads] \arrow[dl,dashed,"\text{undetectable}"{anchor=south west,pos=0.5,xshift=5pt}] & \Omega^{\mathrm{SPL}}_d/\mathrm{torsion}\cong\Qd/\mathrm{torsion} \arrow[dll,dashed,bend left=16,"\text{functions of Pontryagin numbers}"{anchor=north west,pos=0.3,xshift=2pt}] \\
& U(1) & &
\end{tikzcd}
\label{detection-diagram}
\end{equation}
in which the top row consists of the natural maps, the curved arrow is the statement of Theorem \ref{thm:structure}(iii), and the dashed arrows are the factorizations of the partition function $Z$ which may or may not exist: $Z$ is undetectable if and only if the first one does. By Theorem \ref{thm:structure}(iii) this forces its restriction to smooth manifolds to be a function of the rational Pontryagin numbers, so that every phase with a partition function which is a non-trivial torsion invariant of smooth manifolds is detected by some pair of GEMs with identical residues, as in Section \ref{sec:dold-example}. A phase which is trivial on the whole torsion subgroup of $\Omega^{\mathrm{SPL}}_d$ factors through $\Omega^{\mathrm{SPL}}_d/\mathrm{torsion}\cong\Qd/\mathrm{torsion}$, the isomorphism being induced by $\Phi$ (the second dashed arrow in (\ref{detection-diagram})), and is therefore undetectable. This includes the theta terms with the fractional periods allowed on PL manifolds. The remaining phases are non-trivial on the torsion of $\Omega^{\mathrm{SPL}}_d$ but, by Theorem \ref{thm:structure}(iii), trivial on that of $\Omega^{\mathrm{SO}}_d$, so that they differ from such functions only on non-smoothable manifolds. They are undetectable exactly when they are pulled back from a character of $\Qd$ which is non-trivial on its torsion subgroup, and such characters exist if and only if $\Qd$ has torsion, which is the question left open above. Whether a lattice system can realize such a phase at all is a separate open question.

\paragraph{Odd dimensions and the sphere.}
For odd $d$ there is no Euler counterterm, and Proposition \ref{prop:exponent}(ii) provides $Z(S^d)^2$ as a $(d+1)$-partite signal with $2^{d+1}$ terms and $2^{d+1}$ replicas. In $d=3$ this recovers the total quantum dimension, $Z(S^3)^2=\CD^{-2}$, from a single multi-invariant signal rather than from a combination of entanglement entropies \cite{Kitaev:2005dm,levin2006detecting}. The sign of $Z(S^d)$ is fixed by unitarity. Whether $Z(S^d)$ itself, rather than its square, is a ratio of multi-invariants with integer exponents is settled affirmatively for $d\equiv 3\mod 4$ and is open for $d\equiv 1\mod 4$, $d\geq 5$ (Lemma \ref{lem:parity}). Curiously, the parity of the power of $Z(S^d)$ produced by a null-bordism is the rational Kervaire semicharacteristic of $M_\Gamma$.

\paragraph{Reversing the implication.}
Corollary \ref{cor:extract} (NBS) derives the extraction scheme for null-bordant manifolds from Conjecture ELD, and Proposition \ref{prop:equivalence} shows that Conjecture ELD is equivalent to the extraction scheme for all signals built out of GEMs, Conjecture TS, for $d\not\equiv 0\mod 4$ and, in general, for non-vanishing partition functions. One can ask whether the much smaller family of signals of Remark 4 after Corollary \ref{cor:extract} already suffices: if for a given state (\ref{extract-S}) holds for every GEM and every null-bordism, do the non-universal factors of its multi-invariants necessarily have the form (\ref{conj-ELD})? The question has a purely combinatorial core. Extend $\CZ$ and $Z$ multiplicatively to $\Z[d\text{-GEMs}]$, as in (\ref{multiplicativity-graphs}), and assume for simplicity that $Z(M)\neq 0$ for all $M$. The hypothesis says that the ratio $\Gamma\mapsto\CZ(\Gamma;|\psi\rangle)/Z(M_\Gamma)$ tends to $1$ on the subgroup $\mathcal{B}$ generated by the elements $\Gamma-S(\Gamma)$ of (\ref{S-def}), over all bipartite $(d+2)$-colored graphs $\hG$ satisfying (i) and (ii) of Theorem \ref{thm:filling}, while Conjecture ELD restricted to null-bordant GEMs says that it tends to $1$ on the kernel of $R$ restricted to the subgroup spanned by null-bordant GEMs. The first subgroup is contained in the second, and every null-bordant GEM is congruent to a combination of spherical GEMs modulo the first. The two subgroups coincide, therefore, if and only if every relation $\sum_in_iR(\Gamma_i)=0$ among the residues of spherical $d$-GEMs is a consequence of the relations $R(\Gamma)=R(S(\Gamma))$ produced by admissible graphs, such as the $(d+2)$-cube relation of Proposition \ref{prop:exponent}(ii). We do not know whether this is the case. It is a question about the kernel of $R$ on spherical GEMs, complementary to Theorem \ref{thm:structure}, which describes the image of $R$ modulo spherical GEMs. A positive answer would show that, as far as null-bordant manifolds are concerned, the residue form of the non-universal factors is not an additional assumption, but is equivalent to the consistency of the extraction scheme for all choices of the null-bordism. GEMs of manifolds which are not null-bordant are not constrained by (\ref{extract-S}) at all, and for them Conjecture ELD is a genuinely stronger statement. For instance, it is what predicts the ratio (\ref{dold-ratio}) for the pair $\Gamma_D$, $\Gamma_P$ of Section \ref{sec:dold-example}.

\paragraph{Locality of the ground state.}
The converse part of Proposition \ref{prop:extract} assumes that the non-universal homomorphism $U$ can be arbitrary. For ground states of local Hamiltonians this is not evidently the case: for example, in the Levin-Wen models the factors $A(\gamma;|\psi\rangle)$ are trivial for $\gamma$ with $|I|=3$ \cite{DelZotto:2026dgp}, for the matrix product ground states of $(1+1)$-dimensional symmetry-protected phases the analogous factors are moments of the bipartite entanglement spectrum, one for each vertex of the triangulation \cite{SPT2d:2026}, and for the Chen-Hsin state of Section \ref{sec:dold-example} they are exactly those of a collection of Bell pairs and GHZ states located at the triangles of the interfaces and junctions. It would be interesting to understand what constraints locality of $H$ imposes on $U$, and whether they allow one to extract more than what is listed in Table \ref{tab:dims}. In the opposite direction, one can ask whether Conjecture TS, or its restriction to null-bordant manifolds in Corollary \ref{cor:extract}, can serve as a bottom-up characterization of a topological phase, as speculated in \cite{DelZotto:2026dgp}.

\paragraph{Finite-depth circuits and quantum cellular automata.}
A sharper form of the last question is suggested by the origin of the form (\ref{conj-ELD}) of the non-universal factors. By (\ref{lower-partite-factorization}), a factor of this form is exactly what a state shared by a proper subset of the parties contributes to a multi-invariant. Call two states \textit{residue-equivalent} if for every $d$-GEM $\Gamma$ their multi-invariants differ, in the limit $L\rightarrow\infty$, by a factor $U(R(\Gamma))$ with $U$ a homomorphism as in (\ref{conj-ELD}). On the other hand, two states belong to the same gapped phase if they are related by adding ancillary degrees of freedom in a product state and applying a quantum circuit of finite depth, bounded independently of $L$ \cite{Chen:2010gda}. One can ask whether, under appropriate assumptions on the states, the two equivalence relations coincide. That states in the same phase are residue-equivalent is expected from locality. A finite-depth circuit is the product of a unitary of the form $U_0\otimes\ldots\otimes U_d$, which does not affect the multi-invariants, and of the gates in the forward light cone of the boundaries of the regions, a unitary supported within a distance of the order of the depth from the boundaries. The latter changes only the short-range correlations across the interfaces and junctions, which is what (\ref{conj-ELD}) is designed to account for, although this is not a proof, since a unitary supported near the boundaries does not act on the parties separately. The converse, that residue-equivalent states are in the same phase, is the non-trivial direction. It would mean that the multi-invariants modulo factors of residue form are complete invariants of gapped phases. For invertible phases it is consistent with the results of this paper. Assuming Conjecture ELD, two invertible states are residue-equivalent precisely when their partition functions differ by a character of $\Omega^{\mathrm{SPL}}_d$ factoring through $\Phi$. By Theorem \ref{thm:structure}(iii), such a character restricted to smooth manifolds is a function of the rational Pontryagin numbers, i.e. a theta term with a continuous parameter, and the corresponding invertible theories are deformations of the trivial one, in agreement with the classification of \cite{Freed:2016rqq}, in which only the torsion characters correspond to distinct deformation classes. In this setting the question can be phrased in terms of quantum cellular automata (QCA), i.e. locality-preserving unitaries, which need not be finite-depth circuits \cite{Freedman:2019ncn,Haah:2018jdf}. Applying a QCA $\alpha$ to a product state $|0\rangle$ gives an invertible state, and it is expected that QCAs modulo finite-depth circuits and ancillas are classified by invertible phases. In particular, the state of Section \ref{sec:dold-example} is argued in \cite{Chen:2021xks,Fidkowski:2021unr} to be produced by a non-trivial QCA. The ratio of the multi-invariants of $\alpha|0\rangle$ and of $|0\rangle$, modulo factors of residue form, is then an invariant of $\alpha$ modulo circuits, conjecturally the partition function of the corresponding invertible phase, and the non-trivial direction above asserts that it is a complete one: $\alpha|0\rangle$ is residue-equivalent to $|0\rangle$ only if $\alpha$ is a finite-depth circuit, up to ancillas.

\paragraph{Other structures.}
Beyond oriented TQFTs, one can consider theories that require a spin structure \cite{Gaiotto:2015zta}, a background gauge field, or other tangential structures. The corresponding generalization replaces $\Omega^{\mathrm{SPL}}_d$ by the appropriate bordism group and $R$ by a residue keeping track of the structure. The identity of Lemma \ref{lem:boolean} and the filling construction of Theorem \ref{thm:filling} are insensitive to this, and we expect the same dichotomy between bordism invariants with and without local formulas. The case of a background gauge field for a finite group $G$, relevant for symmetry-protected phases, comes with a caveat. The multi-invariants are then refined by symmetry twists along the edges of $\Gamma$, which define a flat $G$-bundle on $M_\Gamma$, and the natural residue records the twists restricted to each link. The partition function of a group-cohomology symmetry-protected topological (SPT) phase is, however, itself a product over the vertices of the triangulation of a quantity determined by the twisted link, namely the state sum of the Dijkgraaf-Witten cocycle \cite{Dijkgraaf:1989pz}, i.e. it is of the same form as the non-universal factor $U$ in (\ref{conj-ELD}). A scheme designed to cancel arbitrary non-universal factors of this form would therefore cancel the SPT partition function as well. What distinguishes the two is gauge covariance: the non-universal factors are expected to be gauge invariant link by link, hence independent of the background, since a flat bundle restricted to a link is gauge-trivial, whereas the factors making up the SPT partition function are gauge invariant only in their product. Under this additional assumption, which is verified for matrix product states in \cite{SPT2d:2026}, the partition function of a $G$-symmetric invertible theory on $(M_\Gamma,f:M_\Gamma \rightarrow BG)$ is extracted from the ratio of the multi-invariant of $\Gamma$ with the twists to the one without, with no null-bordance condition. The same caveat and the same ratio scheme apply to spin and pin structures: the signs entering the fermionic permutation operators define a combinatorial spin structure on $M_\Gamma$, the role of the gauge transformations is played by $(-1)^F$ acting on individual replicas, and the partition functions of the Arf theory \cite{Kapustin:2014dxa,Gaiotto:2015zta} and of the Gu-Wen phases \cite{Gu:2012ib} are again products of local, individually gauge-variant weights. 

\section*{Acknowledgments}

The authors would like to thank Sriram Akella, Markus Dierigl, Po-Shen Hsin, Kantaro Ohmori, Rajath Radhakrishnan, Ingo Runkel and Yarden Sheffer for useful discussions and correspondence. The work of MDZ is supported by the European Union (ERC, HIGH, 101171852). Views and opinions expressed are, however, those of the author(s) only and do not necessarily reflect those of the European Union or the European Research Council. Neither the European Union nor the granting authority can be held responsible for them. MDZ also acknowledges support by the VR project grant No. 2023-05590, the VR Centre for Geometry and Physics (VR grant No. 2022-06593), and the Simons Foundation International Grant for the Simons Collaboration on Global Categorical Symmetries. AG acknowledges the support of the Department of Atomic Energy, Government of India, under Project Identification No. RTI 4012. His work is also supported in part by the Infosys Endowment for the Study of the Quantum Structure of Spacetime.

\medskip\noindent
\textit{Use of AI.} AI assistants, Claude Fable 5.1 (Anthropic) and GPT-6 Astra (OpenAI) used through Claude Code and Codex, respectively, were employed in the preparation of this paper, for drafting and editing the text and the figures, for writing the code used in the computer verifications mentioned in the text, for checking computations and for verifying the internal consistency of the manuscript. Moreover, the relevance of bordism to the problem of the extraction of TQFT partition functions, together with the useful results of \cite{Gaifullin2004,Gaifullin2005,Gaifullin2008}, was brought to the attention of the authors by AI. All results and their proofs were checked by the authors, who take full responsibility for the content.

\appendix

\section{Colored fillings with prescribed boundary}
\label{app:filling}

In this appendix we prove Theorem \ref{thm:filling}. The construction is classical in crystallization theory \cite{CavicchioliGagliardi1980,Gagliardi1983}, where, for crystallizations $\Gamma$, $K_W$ below is a ``crystallization with boundary''. We present a self-contained argument based on the relative coloring theorem of \cite{IzmestievKleeNovik2017}, which applies to arbitrary GEMs and keeps track of all the colors and orientations explicitly, and comment on the relation with the classical results at the end of the appendix.

We begin with three lemmas. Recall that a $(d+2)$-colored graph $\hG$ is called admissible if every connected component of every $\rs{\hG}{\hat\alpha}$ is a $d$-GEM.

\begin{lemma}
\label{lem:pseudo}
Let $\hG$ be a bipartite $(d+2)$-colored graph with colors $\{0..d,*\}$ and $\hW:=K(\hG)$.
\begin{enumerate}
    \item[(i)] For every $J\subseteq\{0..d,*\}$, the cells of $\hW$ with the set of vertex colors $J$ are in one-to-one correspondence with the connected components of $\rs{\hG}{\hat J}$, the link of the cell corresponding to a component $Q$ being $K(Q)$. In particular, the link of a vertex of color $\alpha$ is $K(Q)$ for a component $Q$ of $\rs{\hG}{\hat\alpha}$.
    \item[(ii)] If $\hG$ is admissible, then $\hW$ is a closed oriented $(d+1)$-dimensional pseudomanifold, the only possible non-manifold points of which are the vertices. That is, the link of every cell of dimension $\geq 1$ is a PL sphere.
\end{enumerate}
\end{lemma}

Part (i) is the general fact about colored complexes recalled in Section \ref{sec:boolean}, valid for any colored graph and holding by construction of $K(\hG)$: a cell with vertex colors $J$ is the common subsimplex of the top-dimensional cells that are connected to each other through the faces opposite to the vertices of colors not in $J$, i.e. through the edges of $\rs{\hG}{\hat J}$. Only (ii) requires proof.

\begin{proof}
For a cell $\tau$ of a simplicial cell complex and a vertex $v\in\tau$ one has $\lk_{\hW}\tau=\lk_{\lk_{\hW}v}(\tau\setminus v)$: the link of $\tau$ can be computed in two steps, by first passing to the link of the vertex $v$, in which the face $\tau\setminus v$ of $\tau$ opposite to $v$ appears as a cell, and then taking the link of that cell there. By (i) and admissibility, $\lk_{\hW}v=K(Q)$ is a closed PL $d$-manifold, so the link of any cell of positive dimension is the link of a cell in a PL manifold, hence a PL sphere. For the orientation, orient the simplex of a black vertex by the color order $(0,\ldots,d,*)$ and that of a white vertex oppositely. Two simplices sharing a face have opposite colors, so the induced orientations on the face cancel.
\end{proof}

The construction $\hG\mapsto K(\hG)$ can be inverted.

\begin{lemma}
\label{lem:dual}
Let $X$ be a $(d+1)$-dimensional simplicial cell complex with vertices properly colored by $d+2$ colors, such that every face lies in exactly two top-dimensional cells and every cell of codimension $\geq 2$ has a connected link. Let $\hG$ be the dual graph of $X$, with a vertex for each top-dimensional cell and an edge for each face, colored by the color of the vertex opposite to that face in either of the two cells containing it. Then $\hG$ is a $(d+2)$-regular properly edge-colored graph with $K(\hG)=X$, and $\hG$ is bipartite if $X$ is orientable.
\end{lemma}

\begin{proof}
Since the vertex coloring is proper, the two vertices opposite to a face in the two cells containing it have the same color (the one missing from the face), so the edge coloring is well defined, and each top-dimensional cell has exactly one face of each color, so $\hG$ is $(d+2)$-regular and properly edge-colored. The natural map from $K(\hG)$ to $X$ is bijective on top-dimensional cells and on faces by construction. A cell of $X$ with the set of colors $J$ is the common subsimplex of the top-dimensional cells containing it, and these form a single connected component of the subgraph with the complementary colors precisely because the link of the cell is connected. Therefore the map is bijective on all cells. Orientability provides the bipartition as in the proof of Lemma \ref{lem:pseudo}(ii).
\end{proof}

\begin{lemma}[\cite{GrasselliMulazzani2020}, Theorem 4.1]
\label{lem:susp}
Let $\Lambda$ be an $(m+1)$-colored graph and $\Lambda'$ the $(m+2)$-colored graph obtained by adding a copy of the edges of some color $\alpha$, with a new color. Then $M_{\Lambda'}$ is the suspension of $M_\Lambda$, componentwise. In particular, if $\Lambda$ is a disjoint union of spherical $m$-GEMs, then $\Lambda'$ is a disjoint union of spherical $(m+1)$-GEMs.
\end{lemma}

\begin{figure}[!t]
\centering
\begin{tikzpicture}[scale=0.80,font=\scriptsize,
    dbl/.style={circle,fill=black,inner sep=1.3pt}, dwh/.style={circle,draw=black,fill=white,inner sep=1.3pt}]
  \begin{scope}
  \draw[black!60,thick] (1.299,-0.750) -- (-0.000,-1.500);
  \draw[black!60,thick] (1.299,-0.750) -- (1.299,0.750);
  \draw[black!60,thick] (-0.000,-1.500) -- (-1.299,-0.750);
  \draw[black!60,thick] (-1.299,0.750) -- (0.000,1.500);
  \draw[black!60,thick] (-1.299,-0.750) -- (-1.299,0.750);
  \draw[black,very thick] (1.299,-0.750) -- (0.000,1.500);
  \draw[black,very thick] (-1.299,-0.750) -- (0.000,1.500);
  \draw[black!60,thin] (-0.000,-1.500) -- (0.000,1.500);
  \draw[black!60,thick] (1.299,0.750) -- (0.000,1.500);
  \fill[c0] (0.000,1.500) circle (2.2pt);
  \fill[c1] (-1.299,0.750) circle (2.2pt);
  \fill[c0] (-1.299,-0.750) circle (2.2pt);
  \fill[c1] (-0.000,-1.500) circle (2.2pt);
  \fill[c0] (1.299,-0.750) circle (2.2pt);
  \fill[c1] (1.299,0.750) circle (2.2pt);
  \node[gray] at (0,-2.050) {(a)};
  \node[black!70,above,inner sep=3pt] at (0.000,1.500) {$K$};
  \node[black!70] at (0.55,-0.35) {$T_1$};
  \end{scope}
  \begin{scope}[shift={(4.200,0)}]
  \draw[black!60,thin] (-0.546,0.555) -- (-0.000,-1.500);
  \draw[black!60,thin] (0.546,0.555) -- (-0.000,-1.500);
  \draw[black!60,thin] (0.546,0.555) -- (1.299,0.750);
  \draw[black!60,thick] (1.299,-0.750) -- (-0.000,-1.500);
  \draw[black!60,thick] (1.299,-0.750) -- (1.299,0.750);
  \draw[black!60,thick] (-0.000,-1.500) -- (-1.299,-0.750);
  \draw[black!60,thin] (-0.546,0.555) -- (0.000,1.500);
  \draw[black!60,thick] (-1.299,0.750) -- (0.000,1.500);
  \draw[black!60,thin] (0.546,0.555) -- (0.000,1.500);
  \draw[black!60,thin] (0.546,0.555) -- (1.299,-0.750);
  \draw[black!60,thick] (-1.299,-0.750) -- (-1.299,0.750);
  \draw[black!60,thin] (-0.546,0.555) -- (-1.299,-0.750);
  \draw[black!60,thin] (-0.546,0.555) -- (-1.299,0.750);
  \draw[black!60,thin] (-0.000,-1.500) -- (0.000,1.500);
  \draw[black!60,thick] (1.299,0.750) -- (0.000,1.500);
  \fill[c0] (0.000,1.500) circle (2.2pt);
  \fill[c1] (-1.299,0.750) circle (2.2pt);
  \fill[c0] (-1.299,-0.750) circle (2.2pt);
  \fill[c1] (-0.000,-1.500) circle (2.2pt);
  \fill[c0] (1.299,-0.750) circle (2.2pt);
  \fill[c1] (1.299,0.750) circle (2.2pt);
  \fill[c4] (-0.546,0.555) circle (2.2pt);
  \fill[c4] (0.546,0.555) circle (2.2pt);
  \node[gray] at (0,-2.050) {(b)};
  \node[black!70] at (0.55,-0.35) {$K_W$};
  \end{scope}
  \begin{scope}[shift={(11.518,0)},scale=1.85]
  \draw[black!35,thin] (-0.546,0.555) -- (-0.000,-1.500);
  \draw[black!35,thin] (0.546,0.555) -- (-0.000,-1.500);
  \draw[black!35,thin] (0.546,0.555) -- (1.299,0.750);
  \draw[black!35,thick] (1.299,-0.750) -- (-0.000,-1.500);
  \draw[black!35,thick] (1.299,-0.750) -- (1.299,0.750);
  \draw[black!35,thick] (-0.000,-1.500) -- (-1.299,-0.750);
  \draw[black!35,thin] (-0.546,0.555) -- (0.000,1.500);
  \draw[black!35,thick] (-1.299,0.750) -- (0.000,1.500);
  \draw[black!35,thin] (0.546,0.555) -- (0.000,1.500);
  \draw[black!35,thin] (0.546,0.555) -- (1.299,-0.750);
  \draw[black!35,thick] (-1.299,-0.750) -- (-1.299,0.750);
  \draw[black!35,thin] (-0.546,0.555) -- (-1.299,-0.750);
  \draw[black!35,thin] (-0.546,0.555) -- (-1.299,0.750);
  \draw[black!35,thin] (-0.000,-1.500) -- (0.000,1.500);
  \draw[black!35,thick] (1.299,0.750) -- (0.000,1.500);
  \draw[black!35,thin] (0.000,1.500) -- (0.000,2.550);
  \draw[black!35,thin] (-1.299,0.750) -- (-2.208,1.275);
  \draw[black!35,thin] (-1.299,-0.750) -- (-2.208,-1.275);
  \draw[black!35,thin] (-0.000,-1.500) -- (-0.000,-2.550);
  \draw[black!35,thin] (1.299,-0.750) -- (2.208,-1.275);
  \draw[black!35,thin] (1.299,0.750) -- (2.208,1.275);
  \fill[c0] (0.000,1.500) circle (1.6pt);
  \fill[c1] (-1.299,0.750) circle (1.6pt);
  \fill[c0] (-1.299,-0.750) circle (1.6pt);
  \fill[c1] (-0.000,-1.500) circle (1.6pt);
  \fill[c0] (1.299,-0.750) circle (1.6pt);
  \fill[c1] (1.299,0.750) circle (1.6pt);
  \fill[c4] (-0.546,0.555) circle (1.6pt);
  \fill[c4] (0.546,0.555) circle (1.6pt);
  \draw[cc0] (-0.615,0.935) -- (-1.048,0.185);
  \draw[cc1] (-0.615,0.935) -- (-0.182,0.185);
  \draw[cc4] (-0.615,0.935) -- (-1.025,1.775);
  \draw[cc1] (-1.048,0.185) -- (-0.615,-0.565);
  \draw[cc4] (-1.048,0.185) -- (-2.050,0.000);
  \draw[cc0] (-0.182,0.185) -- (-0.615,-0.565);
  \draw[cc4] (-0.182,0.185) -- (0.182,0.185);
  \draw[cc4] (-0.615,-0.565) -- (-1.025,-1.775);
  \draw[cc0] (0.182,0.185) -- (0.615,-0.565);
  \draw[cc1] (0.182,0.185) -- (0.615,0.935);
  \draw[cc1] (0.615,-0.565) -- (1.048,0.185);
  \draw[cc4] (0.615,-0.565) -- (1.025,-1.775);
  \draw[cc0] (0.615,0.935) -- (1.048,0.185);
  \draw[cc4] (0.615,0.935) -- (1.025,1.775);
  \draw[cc4] (1.048,0.185) -- (2.050,-0.000);
  \draw[cc0] (-1.025,1.775) -- (-2.050,0.000);
  \draw[cc1] (-1.025,1.775) -- (1.025,1.775);
  \draw[cc1] (-2.050,0.000) -- (-1.025,-1.775);
  \draw[cc0] (-1.025,-1.775) -- (1.025,-1.775);
  \draw[cc1] (1.025,-1.775) -- (2.050,-0.000);
  \draw[cc0] (2.050,-0.000) -- (1.025,1.775);
  \node[dbl] at (-0.615,0.935) {};
  \node[dwh] at (-1.048,0.185) {};
  \node[dwh] at (-0.182,0.185) {};
  \node[dbl] at (-0.615,-0.565) {};
  \node[dbl] at (0.182,0.185) {};
  \node[dwh] at (0.615,-0.565) {};
  \node[dwh] at (0.615,0.935) {};
  \node[dbl] at (1.048,0.185) {};
  \node[dwh] at (-1.025,1.775) {};
  \node[dbl] at (-2.050,0.000) {};
  \node[dwh] at (-1.025,-1.775) {};
  \node[dbl] at (1.025,-1.775) {};
  \node[dwh] at (2.050,-0.000) {};
  \node[dbl] at (1.025,1.775) {};
  \node[gray] at (0,-2.900) {(c)};
  \node[c4] at (2.300,2.200) {$p=\infty$};
  \end{scope}
\end{tikzpicture}
\caption{The construction of the proof of Theorem \ref{thm:filling} for $d=1$: $W$ is a disc and $\Gamma$ the $6$-cycle, so that $K=K(\Gamma)$ is a hexagon with vertices colored alternately by $0$ and $1$. (a) Step 1: a triangulation $T_1$ of $W$ with $\partial T_1=K$. Two of its diagonals (thick) join vertices of the same color, so no proper vertex coloring of $T_1$ by three colors extends that of $K$. (b) Step 2: the stellar subdivision $K_W$ of $T_1$ at these two diagonals, with the two new vertices of color $*$. It is properly colored, and every triangle has one vertex of each color. (c) Step 3: the cone over $K$ from a vertex $p$ of color $*$, placed at infinity so that the six rays are the edges to $p$, closes $K_W$ to a $2$-sphere $\hW$, and $\hG$ is its dual graph, with the edge colors $0$ (black, solid), $1$ (red, dashed) and $*$ (purple, dotted), the same colors as those of the vertices of $\hW$, and the black and white vertices given by the orientation. The components of $\rs{\hG}{\hat *}$ are the cycles around the three vertices of color $*$: the outer $6$-cycle around $p$ is $\Gamma$ and the two $4$-cycles around the interior vertices form the graph $T$ of Theorem \ref{thm:filling}(i). For $d=1$ every $2$-colored cycle is a GEM of $S^1$, so that (ii) of the theorem is automatic, and the links of Step 4 are the cycles around the vertices of colors $0$ and $1$.}
\label{fig:filling}
\end{figure}
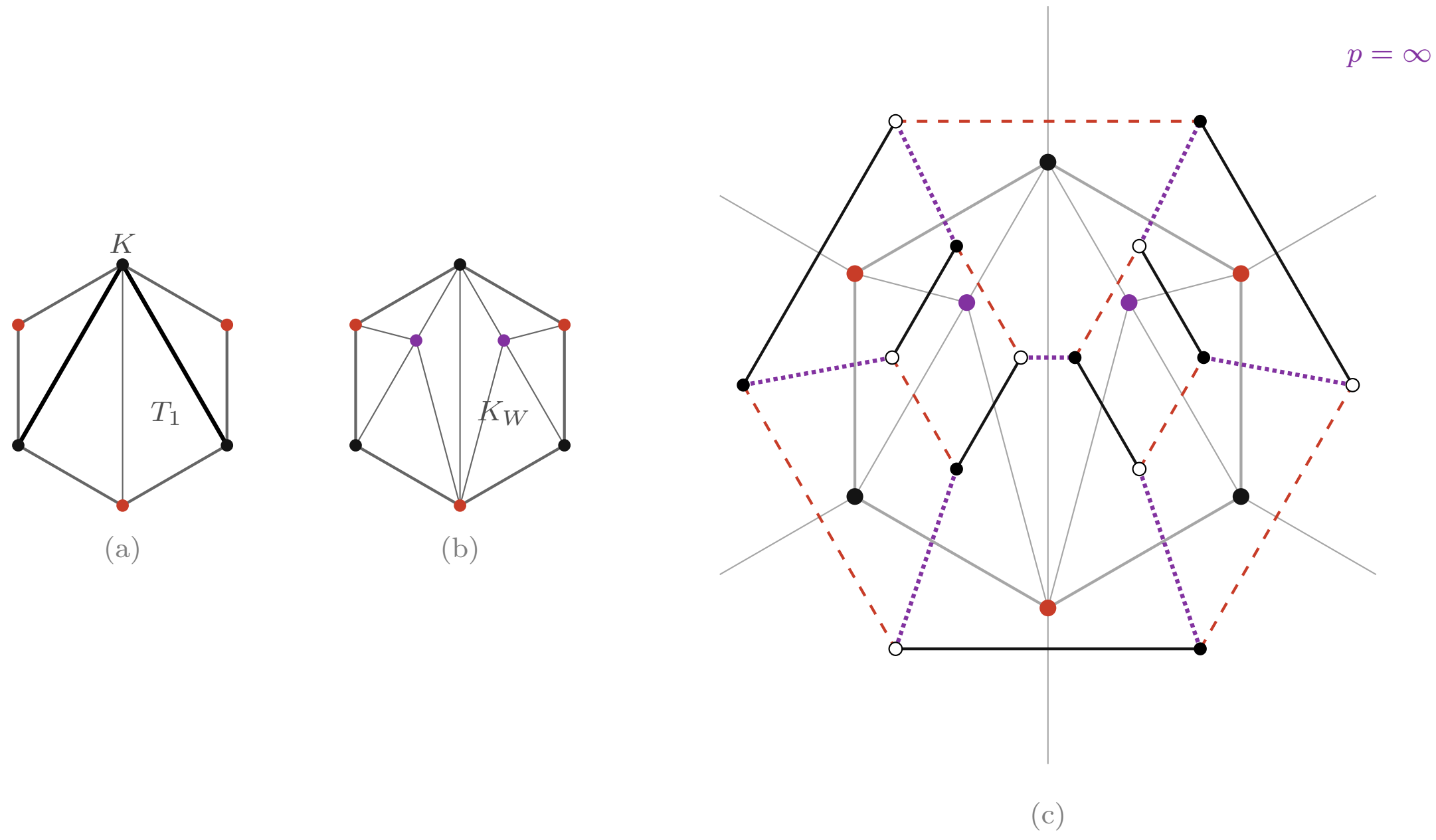

\begin{proof}[Proof of Theorem \ref{thm:filling}]
The construction is illustrated in Figure \ref{fig:filling} for $d=1$.

\textit{Step 1: a triangulation of $W$ with the prescribed boundary.} Let $\Gamma_1,\ldots,\Gamma_k$ be the connected components of $\Gamma$ and $K:=K(\Gamma)=\bigsqcup_iK(\Gamma_i)$, a simplicial cell complex with $|K|\cong\partial W$ (orientation preserving). Choose a PL triangulation $T_W$ of $W$. Both $\partial T_W$ and the barycentric subdivision $\sd K$ are simplicial complexes triangulating $\partial W$, so they have a common subdivision $C$ \cite{RourkeSanderson1972}. Attach two collars, $W\cup(\partial W\times[0,1])\cup(\partial W\times[1,2])\cong W$. Triangulate $\partial W\times[0,1]$ with the bottom $\partial T_W$ and the top $C$, and $\partial W\times[1,2]$ with the bottom $C$ and the top $K$, in both cases inductively over the cells $\tau$ of the coarser end: the prism $\tau\times[a,b]$ has the boundary consisting of $\tau$ at one end, its subdivision at the other end, and the already triangulated $\partial\tau\times[a,b]$, and we cone it off from an interior point. Every cell produced in this way is a cone over a simplex, hence a simplex, and no cell of $K$ is subdivided. The result $T_1$ is a simplicial cell complex (a \textit{simplicial poset} in the terminology of \cite{IzmestievKleeNovik2017}) with $|T_1|\cong W$ and $\partial T_1=K$.

\textit{Step 2: coloring.} The complex $K$ carries a proper vertex coloring by $\Dd$ coming from the GEMs. By Theorem 3.1 of \cite{IzmestievKleeNovik2017}, applied to the relative complex $(T_1,K)$, there is a stellar subdivision $T_2$ of $T_1$, obtained by starring cells not in $K$, together with a proper vertex coloring of $T_2$ by $d+2$ colors extending the given coloring of $K$. (Their theorem is stated for simplicial complexes, but Section 4.2 of \cite{IzmestievKleeNovik2017} points out that the statement and the proof extend verbatim to simplicial posets, which is what is needed here since $K(\Gamma)$ is in general not a simplicial complex.) We call the new color $*$. Since $T_2$ is $(d+1)$-dimensional and properly colored by $d+2$ colors, every $(d+1)$-cell has exactly one vertex of each color. Stellar subdivisions preserve the PL manifold structure, so $K_W:=T_2$ is a triangulation of $W$ with $\partial K_W=K$.

\textit{Step 3: coning.} Let $\hW:=K_W\cup\bigsqcup_i\Cone_{p_i}(K(\Gamma_i))$, with new vertices $p_i$ of color $*$, which is (iii) of the theorem. The cone cells $p_i*\tau$ are simplices with the vertex colors $(\text{colors of }\tau)\cup\{*\}$, so $\hW$ is a $(d+1)$-dimensional simplicial cell complex properly colored by $d+2$ colors. Every face of $\hW$ lies in exactly two $(d+1)$-cells: the interior faces of $K_W$ by the manifold property, and a boundary face $\tau\subset K$ in the cell of $K_W$ containing it and in $p_i*\tau$. Let $\hG$ be the dual graph of $\hW$, with the edge coloring of Lemma \ref{lem:dual}. The complex $\hW$ is orientable, and its cells of codimension $\geq 2$ have connected links (Step 4), so by Lemma \ref{lem:dual} $\hG$ is a $(d+2)$-regular, properly edge-colored, bipartite graph with $K(\hG)=\hW$.

\textit{Step 4: the links.} A vertex $v$ of $\hW$ is of one of three kinds. (a) $v=p_i$: then $\lk p_i=K(\Gamma_i)$. (b) $v$ is an interior vertex of $K_W$: then $\lk v$ is a PL $d$-sphere. (c) $v\in K(\Gamma_i)$, of color $\alpha\in\Dd$: then $\lk_{\hW}v=\lk_{K_W}v\cup\Cone_{p_i}(\lk_Kv)$, a $d$-ball glued to the cone over the $(d-1)$-sphere $\lk_Kv$ along that sphere, which is a PL $d$-sphere. By Lemma \ref{lem:pseudo}(i) the connected components of $\rs{\hG}{\hat\alpha}$ are the dual graphs of the links of the vertices of color $\alpha$. Hence the components of $\rs{\hG}{\hat\alpha}$ for $\alpha\in\Dd$ are spherical $d$-GEMs, which is part (ii) of the theorem, and the components of $\rs{\hG}{\hat *}$ are the graphs $\Gamma_i$ or their mirrors, together with the links of the interior vertices of color $*$, which are spheres.

\textit{Step 5: the bipartition of the cone components.} Orient $\hW$ so that $K_W$ carries the given orientation of $W$. The cone component over $K(\Gamma_i)$ is $\Gamma_i$ or $\overline{\Gamma_i}$ as a bipartite colored graph. Its vertices are the cone cells $p_i*\tau$, where $\tau$ runs over the $d$-cells of $K(\Gamma_i)$, i.e. over the vertices of $\Gamma_i$. The vertex $p_i*\tau$ is black if and only if the simplex $p_i*\tau$, with the vertices ordered as $(0,\ldots,d,*)$, is positively oriented in $\hW$. Since $\tau$ is the face opposite to the vertex $p_i$ of color $*$, the last one in this order, this holds if and only if $\tau$, with the vertices in the color order, carries $(-1)^{d+1}$ times the boundary orientation of the cone, i.e. $(-1)^d$ times the orientation of $\partial W$ at $\tau$. Therefore the cone component is $\Gamma_i$ for all $i$ if $d$ is even and $\overline{\Gamma_i}$ for all $i$ if $d$ is odd. In the latter case we replace $\hG$ by its mirror, which mirrors $T$ and all the substitution graphs, and hence preserves (i) and (ii).
\end{proof}

We conclude with an observation which shows that the conclusion of Theorem \ref{thm:filling} is stable under dipole moves applied to $\Gamma$ (Appendix \ref{app:glossary}), and which provides an alternative proof of the theorem via the classical results \cite{CavicchioliGagliardi1980,Gagliardi1983}, as explained after the proof.

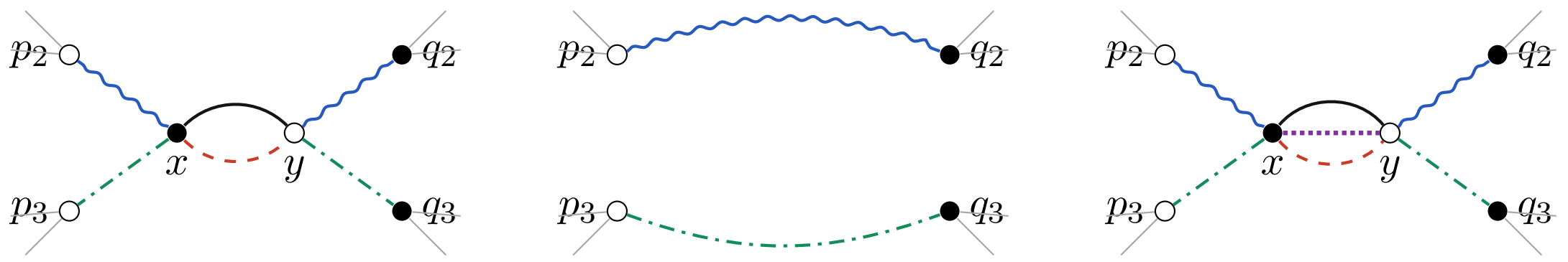
\begin{figure}[ht]
\centering
\begin{tikzpicture}[scale=0.9,font=\small,
  stub/.style={gray!70,thin}]
  \newcommand{\dipOuter}{
    \node[wh] (p2) at (-1.7,0.8) {}; \node[wh] (p3) at (-1.7,-0.8) {};
    \node[bl] (q2) at (1.7,0.8) {};  \node[bl] (q3) at (1.7,-0.8) {};
    \node[left=1pt] at (p2) {$p_2$}; \node[left=1pt] at (p3) {$p_3$};
    \node[right=1pt] at (q2) {$q_2$}; \node[right=1pt] at (q3) {$q_3$};
    \draw[stub] (p2) -- ++(-0.45,0.45); \draw[stub] (p2) -- ++(-0.6,0.05);
    \draw[stub] (p3) -- ++(-0.45,-0.45); \draw[stub] (p3) -- ++(-0.6,-0.05);
    \draw[stub] (q2) -- ++(0.45,0.45); \draw[stub] (q2) -- ++(0.6,0.05);
    \draw[stub] (q3) -- ++(0.45,-0.45); \draw[stub] (q3) -- ++(0.6,-0.05);
  }
  \begin{scope}[shift={(-5.6,0)}]
    \dipOuter
    \node[bl] (X) at (-0.6,0) {}; \node[wh] (Y) at (0.6,0) {};
    \node[below=2pt] at (X) {$x$}; \node[below=2pt] at (Y) {$y$};
    \draw[cc0] (X) to[bend left=45] (Y); \draw[cc1] (X) to[bend right=45] (Y);
    \draw[cc2] (X) -- (p2); \draw[cc3] (X) -- (p3);
    \draw[cc2] (Y) -- (q2); \draw[cc3] (Y) -- (q3);
    \node at (0,-1.6) {(a) $\Gamma$};
  \end{scope}
  \begin{scope}[shift={(0,0)}]
    \dipOuter
    \draw[cc2] (p2) to[bend left=20] (q2); \draw[cc3] (p3) to[bend right=20] (q3);
    \node at (0,-1.6) {(b) $\Gamma'$, and $\hG'$ near $\Gamma'$};
  \end{scope}
  \begin{scope}[shift={(5.6,0)}]
    \dipOuter
    \node[bl] (X) at (-0.6,0) {}; \node[wh] (Y) at (0.6,0) {};
    \node[below=2pt] at (X) {$x$}; \node[below=2pt] at (Y) {$y$};
    \draw[cc0] (X) to[bend left=50] (Y); \draw[cc1] (X) to[bend right=50] (Y); \draw[cc4] (X) -- (Y);
    \draw[cc2] (X) -- (p2); \draw[cc3] (X) -- (p3);
    \draw[cc2] (Y) -- (q2); \draw[cc3] (Y) -- (q3);
    \node at (0,-1.6) {(c) $\hG$};
  \end{scope}
\end{tikzpicture}
\caption{The dipole move of Lemma \ref{lem:dipole}, for $d=3$ and a $2$-dipole $\theta=\{x,y\}$ of colors $C=\{0,1\}$ (color $0$ solid, $1$ dashed, $2$ wavy, $3$ dash-dotted, $*$ dotted, cf.\ Figure \ref{fig:lens}). (a) The GEM $\Gamma$ near $\theta$. (b) Cancelling $\theta$ replaces the edges $(x,p_\alpha)$ and $(y,q_\alpha)$ of the colors $\alpha=2,3$ by the single edges $(p_\alpha,q_\alpha)$ of $\Gamma'$. The filling graph $\hG'$ of $\Gamma'$ contains this picture, with further edges of color $*$ at $p_\alpha$ and $q_\alpha$ not shown. (c) The graph $\hG$: the vertices $x$ and $y$ are re-inserted into $\hG'$ as a dipole of the colors $C\cup\{*\}=\{0,1,*\}$, with the edges of the colors $2,3$ of (a).}
\label{fig:dipole-lemma}
\end{figure}

\begin{lemma}
\label{lem:dipole}
Let $\Gamma$ be a $d$-GEM, let $\theta=\{x,y\}$ be a dipole of $\Gamma$ of type $h$ with colors $C\subseteq\Dd$, and let $\Gamma'$ be the GEM obtained from $\Gamma$ by cancelling it. Suppose that $\hG'$ is a bipartite $(d+2)$-colored graph satisfying (i) and (ii) of Theorem \ref{thm:filling} for $\Gamma'$, with $\rs{\hG'}{\hat *}=\Gamma'\sqcup T$. Let $\hG$ be the graph obtained from $\hG'$ by adding the two vertices $x$ and $y$ back, joined to each other by edges of the colors $C\cup\{*\}$ and to the other vertices by the same edges of the colors $\Dd\setminus C$ as in $\Gamma$. Then $\hG$ satisfies (i) and (ii) for $\Gamma$, with the same $T$ (Figure \ref{fig:dipole-lemma}).
\end{lemma}

\begin{proof}
Cancelling $\theta$ replaces, for every color $\alpha\in\Dd\setminus C$, the two $\alpha$-edges $(x,p_\alpha)$ and $(y,q_\alpha)$ of $\Gamma$ by the single $\alpha$-edge $(p_\alpha,q_\alpha)$ of $\Gamma'$, and the construction of $\hG$ reverses this inside $\hG'$. Therefore $\rs{\hG}{\hat *}=\Gamma\sqcup T$, which is (i) of the theorem, and $\hG$ is bipartite, since $x$ and $y$ are adjacent while $p_\alpha$ and $q_\alpha$ lie on opposite sides of the bipartition of $\Gamma'$. It remains to show (ii), i.e. that for every $\alpha\in\Dd$ all the connected components of $\rs{\hG}{I}$, $I:=\hat\alpha$, are spherical GEMs. If $I$ contains a color outside $C\cup\{*\}$, i.e. if $C\neq\Dd\setminus\{\alpha\}$, then $\{x,y\}$ is a dipole of $\rs{\hG}{I}$, with the colors $I\cap(C\cup\{*\})$. Removing the edges of those colors produces connected components of $\rs{\Gamma}{I\setminus(C\cup\{*\})}$ containing $x$ and $y$, which are distinct because $\theta$ is a dipole of $\Gamma$, and are spherical GEMs because $I\setminus(C\cup\{*\})$ is a proper subset of $\Dd$ (Section \ref{sec:residues}). Cancelling the dipole gives back $\rs{\hG'}{I}$ and does not change the represented space \cite{FerriGagliardi1982,FerriGagliardiGrasselli1986}, so the connected components of $\rs{\hG}{I}$ are those of $\rs{\hG'}{I}$, up to a dipole move in one of them, and thus are PL spheres. Otherwise $I=C\cup\{*\}$, and the connected components of $\rs{\hG}{I}$ are those of $\rs{\hG'}{I}$ together with the two-vertex GEM $\{x,y\}$ of a sphere.
\end{proof}

Every GEM is reduced to a crystallization of the same manifold by cancelling $1$-dipoles, which are always proper \cite{FerriGagliardiGrasselli1986}, and the components of $\Gamma$ can be treated simultaneously. Lemma \ref{lem:dipole} therefore reduces parts (i) and (ii) of Theorem \ref{thm:filling} to the case in which every component of $\Gamma$ is a crystallization. In that case the existence of a colored triangulation $K_W$ of $W$ with $\partial K_W=K(\Gamma)$ is the classical theorem of \cite{CavicchioliGagliardi1980,Gagliardi1983}: in the language of crystallization theory, the dual graph of $K_W$ is a $(d+2)$-colored graph with boundary, regular with respect to the color $*$ and with boundary graph $\Gamma$, and $\hG$ is obtained from it by capping the boundary with a copy of $\Gamma$, which is Step 3 above. We have preferred the argument based on \cite{IzmestievKleeNovik2017}, which treats all GEMs at once and keeps the orientations explicit.

\section{Parity of the sphere exponent for \texorpdfstring{$d\equiv 1\mod 4$}{d = 1 mod 4}}
\label{app:parity}

In this appendix we prove Lemma \ref{lem:parity}. Let $d+1=2m$ with $m$ odd, let $M:=M_\Gamma$, and write $a_i:=\dim H_i(W;\Q)$, $b_i:=\dim H_i(M;\Q)$ and $\rho_i:=\mathrm{rank}\big(H_i(M;\Q)\rightarrow H_i(W;\Q)\big)$, with $\rho_{2m}:=0$. Exactness of the long exact sequence
\begin{equation}
    \ldots\longrightarrow H_i(M)\longrightarrow H_i(W)\longrightarrow H_i(W,M)\longrightarrow H_{i-1}(M)\longrightarrow\ldots
\end{equation}
and Lefschetz duality $\dim H_i(W,M;\Q)=a_{2m-i}$ give $b_i=\rho_i+\rho_{i+1}+a_{2m-1-i}-a_{i+1}$ for $0\leq i\leq 2m-1$. Summing over $0\leq i\leq m-1$,
\begin{equation}
    \sum_{i=0}^{m-1}b_i=\rho_0+\rho_m+2\sum_{i=1}^{m-1}\rho_i+\sum_{j=m}^{2m-1}a_j-\sum_{j=1}^{m}a_j\;\equiv\;\rho_0+\rho_m+a_m+\sum_{j=1}^{2m-1}a_j\mod 2.
\end{equation}
Since $\chi(W)\equiv\sum_{j=0}^{2m}a_j$, this gives
\begin{equation}
    \kQ(M)-\chi(W)\equiv (\rho_0-a_0-a_{2m})+(a_m-\rho_m)\mod 2.
\end{equation}
The first bracket is even: $a_0-\rho_0$ is the number of connected components of $W$ that do not meet $M$, i.e. the closed ones, and $a_{2m}$ is the number of closed components as well, since $H_{2m}(W';\Q)=0$ for a connected manifold $W'$ with non-empty boundary. The image of $H_m(M;\Q)$ in $H_m(W;\Q)$ is the radical of the intersection pairing on $H_m(W;\Q)$, so $a_m-\rho_m$ is the rank of this pairing. The pairing is antisymmetric because $m$ is odd, hence its rank is even. Together with (\ref{N-chi}) this proves (\ref{chi-semichar}).

For $d\equiv 3\mod 4$ the intersection pairing on $H_m(W;\Q)$ is symmetric and the argument fails, as it must by Proposition \ref{prop:exponent}(iii). For example, for $W=D^4$ and $W=D^4\#\CP^2$, both with boundary $S^3$, the Euler characteristics are $1$ and $2$.

\section{The substitution graphs of the lens space GEM}
\label{app:daughters}

\begin{figure}[H]
\centering
\input{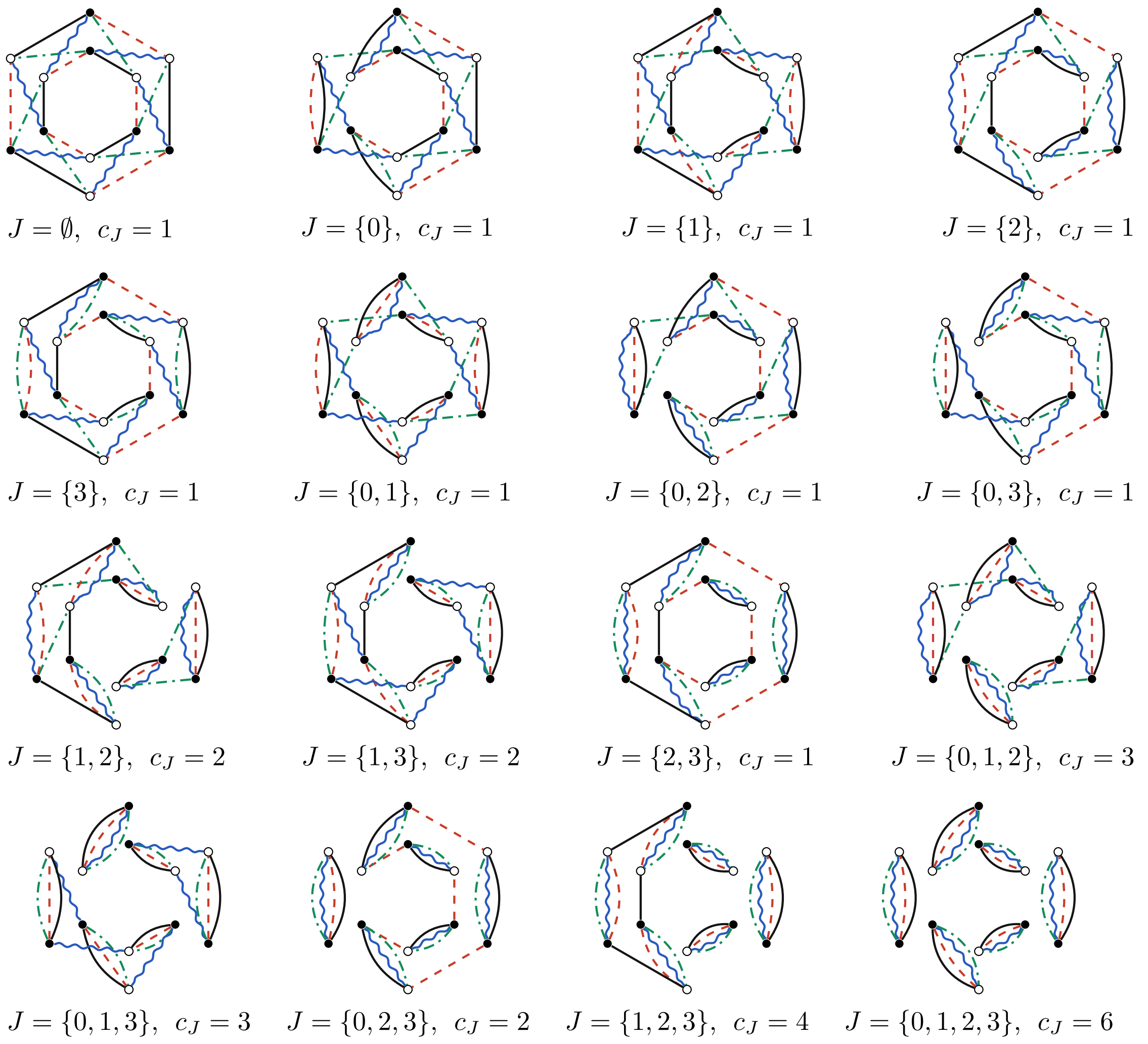}
\caption{The substitution graphs $H_J$ of the five permutations $(\sigma_0,\sigma_1,\sigma_2,\sigma_3,\sigma_*)$ of the $L(3,1)$ example, with the same vertex positions and edge styles as in Figure \ref{fig:lens}.}
\label{fig:daughters}
\end{figure}

Figure \ref{fig:daughters} shows all the substitution graphs $H_J$, $J\subseteq\{0..3\}$, of the $5$-colored graph defined by the permutations (\ref{L31-perms}) and (\ref{L31-star}) of Section \ref{sec:example}, in the layout of Figure \ref{fig:lens}. In $H_J$ the edges of the colors $\alpha\in J$ are those of $\sigma_*$, i.e. they join $\bullet i$ to $\circ\sigma_*(i)$ with $\sigma_*=(364)$. Parallel edges are drawn as arcs. Below each graph we indicate $J$ and the number $c_J$ of connected components. For $J\neq\emptyset$ every component is a GEM of $S^3$, and $H_\emptyset=\Gamma$ is the GEM of $L(3,1)$.

\section{The Chen-Hsin state: conventions and proof}
\label{app:dold}

\subsection{Cochain conventions and the wavefunction}
\label{app:ch-conventions}

All cochains are with $\Z_2$ coefficients and all signs are irrelevant. For a $p$-cochain $f$ and a $q$-cochain $g$ on a simplex with ordered vertices $(0,1,\ldots)$,
\begin{equation}
\begin{aligned}
    (f\cup g)(0,\ldots,p+q)&=f(0,\ldots,p)\,g(p,\ldots,p+q),\\
    (f\cupi{1}g)(0,\ldots,p+q-1)&=\sum_{j=0}^{p-1}f(0,\ldots,j,j+q,\ldots,p+q-1)\,g(j,\ldots,j+q),
\end{aligned}
    \label{cup-conventions}
\end{equation}
and similarly for the higher $\cup_i$ \cite{Chen:2021xks}. They satisfy the Leibniz rule $\delta(x\cupi{i}y)=\delta x\cupi{i}y+x\cupi{i}\delta y+x\cupi{i-1}y+y\cupi{i-1}x$, with $\cup_0=\cup$ and $\cup_{-1}=0$. On cocycles, $x\cupi{i}x$ represents $\Sq^{p-i}[x]$. The identity (\ref{phi5}) follows: $\delta(b\cup b+b\cupi{1}\delta b)=\delta b\cupi{1}\delta b$, $\delta(a\cup\delta b)=\delta a\cup\delta b$ and $\delta(a\cup(\delta a\cupi{1}\delta a))=\delta a\cup(\delta a\cupi{1}\delta a)$, since $\delta(\delta a\cupi{1}\delta a)=2\,\delta a\cup\delta a=0$.

The sign $(-1)^{\int_X\phi_4(a,b)}$ in (\ref{Psi-def}) depends only on $(\delta a,\delta b)$ for $X\cong S^4$. Let $W$ be a triangulation of the $5$-ball with $\partial W=X$ and with a branching structure extending the one of $X$, for instance the cone over $X$ with the apex ordered after all the vertices of $X$, and let $\tilde a$, $\tilde b$ be arbitrary extensions of $a$, $b$ to $W$. By Stokes' theorem and (\ref{phi5}),
\begin{equation}
    \int_X\phi_4(a,b)=\int_W\delta\phi_4(\tilde a,\tilde b)=\int_W\phi_5(\delta\tilde a,\delta\tilde b).
    \label{F-stokes}
\end{equation}
Since $X\cong S^4$, so that $H^i(X;\Z_2)=0$ for $1\leq i\leq 3$, any other potentials for the same $(\delta a,\delta b)$ are of the form $a+\delta\mu$, $b+\delta\xi$, and they extend to $\tilde a+\delta\tilde\mu$, $\tilde b+\delta\tilde\xi$ with the same coboundaries, so that the right-hand side of (\ref{F-stokes}) is unchanged. Consequently, the sum over potentials in (\ref{Psi-def}) is a constant multiple of (\ref{Psi-F}), every cocycle pair on $X$ being a coboundary pair, and (\ref{Psi-F}) is in turn, up to a constant, the sum over all cochains $(\tilde a,\tilde b)\in C^1(W;\Z_2)\times C^2(W;\Z_2)$ of $(-1)^{\int_W\phi_5(\delta\tilde a,\delta\tilde b)}\,|\delta\tilde a|_X,\delta\tilde b|_X\rangle$, or equivalently over all cocycle pairs $(\widetilde A,\widetilde B)\in Z^2(W;\Z_2)\times Z^3(W;\Z_2)$: in each case the map to $Z^2(X;\Z_2)\times Z^3(X;\Z_2)$ is linear and surjective, so that every $(A,B)$ arises the same number of times.

Explicitly, writing $A_{ijk}$ and $B_{ijkl}$ for the values on the triangles and tetrahedra, the exponent on the $5$-simplex $(012345)$ is
\begin{equation}
\begin{aligned}
    \phi_5(A,B)(012345)={}&B_{0345}B_{0123}+B_{0145}B_{1234}+B_{0125}B_{2345}\\
    &+A_{012}B_{2345}+A_{012}A_{245}A_{234}+A_{012}A_{235}A_{345}.
\end{aligned}
    \label{phi5-explicit}
\end{equation}
For $X=\partial\Delta^5$ one can take $W=\Delta^5$, and every cochain on $X$ is a cochain on $\Delta^5$, so that (\ref{F-stokes}) gives $F(A,B)=\phi_5(A,B)(012345)$: the state (\ref{Psi-F}) on $\partial\Delta^5$ is completely specified by the flatness constraints and these six monomials.

\subsection{Proof of Proposition \ref{prop:ch}}
\label{app:ch-proof}

Let $W$ be a triangulation of the $5$-ball with $\partial W=X$ as in Appendix \ref{app:ch-conventions}, and let $K^W_\Gamma$ be the complex obtained by gluing $2r$ copies of $W$, one for each vertex of $\Gamma$, along the parties as prescribed by $\Gamma$: for every edge of color $\alpha$ between $\bullet i$ and $\circ\sigma_\alpha(i)$, the two copies are glued along the party $\Delta_\alpha\subset X$ (Figure \ref{fig:ch-lattice}(b)). The copies of the white vertices are mirror images with the same vertex orders, so that $K^W_\Gamma$ is a triangulation of $M=M_\Gamma$ (as a simplicial cell complex) refining $K(\Gamma)$, with a branching structure extending the ones of the replicas. Let $K^{(4)}_\Gamma\subset K^W_\Gamma$ be the union of the $6r$ copies of the parties triangulated by $X$, i.e. $K(\Gamma)$ with its $4$-cells refined by $X$. This is the complex on which the qubits live. The multi-invariant (\ref{multi-invariant-def}) is the sum, over the configurations $(A_i,B_i)$ of the $r$ ket replicas and $(A'_i,B'_i)$ of the $r$ bra replicas, of the product of the coefficients in (\ref{Psi-F}), the configuration of the bra replica $\sigma_\alpha(i)$ on the party $\alpha$ being that of the ket replica $i$.

\textit{Step 1: agreement.} An edge of color $\alpha$ of $\Gamma$ identifies all the qubits of the party $\alpha$, i.e. the values on the entire closure of the face $\Delta_\alpha$, between the ket $i$ and the bra $\sigma_\alpha(i)$. Within each replica, the qubits representing the copies of a simplex shared by several parties carry the same value, or else the coefficient of the copied state vanishes. Therefore, for every cell of $K(\Gamma)$ of type $I$, i.e. a copy of the subsimplex of $\Delta^5$ contained in exactly the parties $I$, the value on a simplex of $X$ in its interior is the same in all the replicas of the connected component of $\rs{\Gamma}{I}$ corresponding to that cell (Section \ref{sec:residues}). The configurations with non-zero contribution are therefore exactly the pairs $(A,B)\in Z^2(K^{(4)}_\Gamma;\Z_2)\times Z^3(K^{(4)}_\Gamma;\Z_2)$, each contributing $\pm1$ (the coefficients are real, and the bra replicas are the mirror images of the ket replicas with the same vertex orders).

\textit{Step 2: the exponent is the gauge theory action on $M$.} By (\ref{F-stokes}), the contribution of $(A,B)$ is $(-1)^{E(A,B)}$ with
\begin{equation}
    E(A,B)=\sum_{v}\int_{W_v}\phi_5(\delta\tilde a_v,\delta\tilde b_v)=\int_M\phi_5(\widetilde A,\widetilde B),
\end{equation}
where $v$ runs over the vertices of $\Gamma$, $(\tilde a_v,\tilde b_v)$ are potentials on the copy $W_v$ of $W$ for the restriction of $(A,B)$ to $\partial W_v$, and $\widetilde A$, $\widetilde B$ are the cocycles on $K^W_\Gamma$ assembled from the $\delta\tilde a_v$, $\delta\tilde b_v$, which agree with $A$, $B$ on the glued parties. Every cocycle pair $(\widetilde A,\widetilde B)\in Z^2(K^W_\Gamma;\Z_2)\times Z^3(K^W_\Gamma;\Z_2)$ extending $(A,B)$ arises in this way, since on a ball every cocycle is a coboundary, and their number does not depend on $(A,B)$: on each copy the extensions of $A$ form a coset of $Z^2(W,X;\Z_2)$ and those of $B$ a coset of $Z^3(W,X;\Z_2)$, and since the relative cochain complex of $(W,X)$ is exact in degrees $\leq 3$, $\dim Z^2(W,X)+\dim Z^3(W,X)=c_2(W,X)$, the number of triangles of $W$ not in $X$. Hence
\begin{equation}
    \CZ(\Gamma;|\Psi\rangle)=2^{-2r\,c_2(W,X)}\sum_{(\widetilde A,\widetilde B)\in Z^2(K^W_\Gamma)\times Z^3(K^W_\Gamma)}(-1)^{\int_M\phi_5(\widetilde A,\widetilde B)}.
    \label{Z-overcount}
\end{equation}

\textit{Step 3: the action depends only on the classes.} On cocycles, $B\cupi{1}B$ and $A\cupi{1}A$ represent $\Sq^2[B]$ and $\Sq^1[A]$ (Appendix \ref{app:ch-conventions}), so that
\begin{equation}
    [\phi_5(A,B)]=\Sq^2[B]+[A]\cup[B]+[A]\cup\Sq^1[A]\;\in\;H^5(M;\Z_2),
\end{equation}
and $\int_M\phi_5(A,B)$ depends only on $([A],[B])$, cf.\ \cite{Chen:2021xks}. This is what is used in (\ref{gauged}).

\textit{Step 4: the sum.} Every pair of classes is represented by $|B^2(M)|\,|B^3(M)|$ cocycle pairs, so that by (\ref{gauged}) the sum in (\ref{Z-overcount}) equals
\begin{equation}
\begin{aligned}
    |B^2(M)|\,|B^3(M)|\,|H^3(M;\Z_2)|\,(-1)^{\nu(M)}&=|B^2(M)|\,|Z^3(M)|\,(-1)^{\nu(M)}\\
    &=2^{c_2(K^W_\Gamma)}\,(-1)^{\nu(M)},
\end{aligned}
\end{equation}
since $\dim B^2+\dim Z^3=\dim C^2$ is the number of triangles of $K^W_\Gamma$. Finally, the number of triangles of $K^{(4)}_\Gamma$ is
\begin{equation}
    c_2(K^W_\Gamma)-2r\,c_2(W,X)=\sum_{1\leq|I|\leq 3}\kappa_I\,\#\pi_0(\rs{\Gamma}{I}),
\end{equation}
since a triangle of $X$ interior to the subsimplex of type $I$ has one copy in $K^{(4)}_\Gamma$ for each cell of $K(\Gamma)$ of that type, and (\ref{Z-overcount}) gives (\ref{ch-main}). \qed

\medskip\noindent
The argument uses about $\Gamma$ only that $K(\Gamma)$ is a closed manifold, through the Wu formula and Poincar\'e duality. The copying of the shared qubits is also essential. The other natural way to decompose the Hilbert space into the parties is to assign each simplex of $X$ lying in the closure of several parties to one of them. In that case the agreement of Step 1 fails, the exponent is a genuinely cubic function of the configuration, and we have no evaluation.

Formula (\ref{ch-main}) was also verified by computer for the lattice $X$ obtained from $\partial\Delta^5$ by starring each of its six $4$-simplices, for which $\kappa_I=10$ for $|I|=1$, $\kappa_I=1$ for $|I|=3$ and $\kappa_I=0$ otherwise, by evaluating the sum over the cocycle configurations of Step 1 exactly, after checking that the exponent $E$ is a quadratic function on the configuration space. This was done for the two-vertex GEM of $S^5$, for GEMs of $S^4\times S^1$ and of $\CP^2\times S^1$, for the $64$ substitution graphs of the $7$-cube graph of Proposition \ref{prop:exponent}(ii), and for the graphs $\Gamma_P$ and $\Gamma_D$ with $6640$ vertices of the footnote in the proof of Proposition \ref{prop:smooth-torsion}, for which $\CZ(\Gamma_P;|\Psi\rangle)=-\CZ(\Gamma_D;|\Psi\rangle)=2^{209168}$.

\section{Terminology of combinatorial topology}
\label{app:glossary}

Tables \ref{tab:glossary} and \ref{tab:glossary-gems} summarize, for the reader not fluent in the language of combinatorial topology, the terms used in this paper and the role they play: Table \ref{tab:glossary} concerns complexes and manifolds, Table \ref{tab:glossary-gems} colored graphs. Standard references are \cite{RourkeSanderson1972} for PL topology and \cite{FerriGagliardiGrasselli1986,Lins1995} for GEMs and crystallizations.

\begin{table}[H]
\centering
\footnotesize
\begin{tabular}{@{}>{\raggedright\arraybackslash}p{2.5cm}p{12.1cm}@{}}
\toprule
Term & Meaning, and its role in this paper\\
\midrule
Simplicial complex & Simplices glued along common subsimplices in such a way that every simplex is determined by its set of vertices. A face of a simplex is a subsimplex of codimension one. In particular the intersection of two simplices is a single subsimplex, possibly empty, and two vertices are joined by at most one edge. Barycentric subdivisions (see below) are always simplicial complexes.\\
\addlinespace[4pt]
Simplicial cell complex (pseudocomplex, simplicial poset) & The same, except that two simplices may share several faces and two vertices may be joined by several edges. Each closed cell is still a genuine simplex. The complex $K(\Gamma)$ glued from a colored graph is of this type, e.g. the two-vertex GEM $\Gamma_{S^d}$ gives two simplices glued along their whole boundary. Links, stars, joins and subdivisions are defined as for simplicial complexes (next row).\\
\addlinespace[4pt]
Link, star, join, cone, suspension & The link of a simplex $\tau$ is the subcomplex formed by the simplices $\rho$ disjoint from $\tau$ such that $\rho$ and $\tau$ span a common simplex. For example, the link of a vertex of a triangulated surface is the cycle of edges opposite to it in the triangles containing it. The (closed) star of $\tau$ is the union of the simplices containing $\tau$, together with their subsimplices. The join $A*B$ of two disjoint complexes is the complex swept out by the segments connecting every point of $A$ to every point of $B$. The join of two simplices is a simplex, the join of a point with $B$ is the cone over $B$, and the join of $S^0$ (two points) with $B$ is the suspension of $B$. The star of a vertex is the cone from the vertex over its link. In a GEM the links of the simplices are encoded by the subgraphs $\rs{\Gamma}{I}$ (Section \ref{sec:residues}).\\
\addlinespace[4pt]
Subdivision, stellar and barycentric subdivision & A subdivision of a complex is a complex with the same realization, every simplex of which is contained in a simplex of the original. Starring a simplex $\sigma$ means adding a new vertex $v$ in the interior of $\sigma$ and replacing the closed star of $\sigma$ by the join of $v$ with its boundary, i.e. replacing every simplex $\rho\supseteq\sigma$ by the cones from $v$ over the subsimplices of $\rho$ not containing $\sigma$. A stellar subdivision is obtained by a sequence of starrings. The barycentric subdivision $\sd K$ is the stellar subdivision in which all simplices are starred in order of decreasing dimension: its vertices are the barycenters $b_\sigma$ of the simplices of $K$ and its simplices are the chains $\sigma_0\subset\sigma_1\subset\cdots\subset\sigma_k$. Coloring $b_\sigma$ by $\dim\sigma$ makes $\sd K$ a colored triangulation (next table), as used in Sections \ref{sec:bordism} and \ref{sec:kernel}. Two triangulations have a common subdivision if and only if they are related by a sequence of starrings and inverse starrings \cite{Alexander1930}.\\
\addlinespace[4pt]
Triangulation, PL manifold, PL structure & A triangulation of a space is a homeomorphism with (the realization of) a simplicial cell complex. A triangulated closed $d$-manifold is a PL (or combinatorial) manifold if the link of every vertex is a PL $(d-1)$-sphere, i.e. a sphere triangulated in a way that has a common subdivision with the boundary of a $d$-simplex. Two triangulations define the same PL structure if they have a common subdivision. Every smooth manifold has a canonical PL structure.\\
\bottomrule
\end{tabular}
\caption{Terminology of combinatorial topology used in this paper: complexes and manifolds.}
\label{tab:glossary}
\end{table}

\begin{table}[H]
\centering
\footnotesize
\begin{tabular}{@{}>{\raggedright\arraybackslash}p{2.5cm}p{12.1cm}@{}}
\toprule
Term & Meaning, and its role in this paper\\
\midrule
Pseudomanifold & A $d$-dimensional complex in which every $(d-1)$-cell lies in exactly two $d$-cells. The links of vertices need not be spheres, so singular points may occur. The complex $K(\hG)$ of an admissible $(d+2)$-colored graph $\hG$ of Section \ref{sec:daughters} is a pseudomanifold with singular points only at vertices (for the graphs of Theorem \ref{thm:filling}, only at vertices of color $*$).\\
\addlinespace[4pt]
Proper coloring & An edge coloring of a graph is proper if the edges meeting at any vertex have distinct colors. For a $(d+1)$-regular graph colored by $d+1$ colors this means that every vertex meets exactly one edge of each color, which is the case for all the graphs in this paper (Section \ref{sec:multi-invariants}). A vertex coloring of a complex is proper if the two endpoints of every edge have distinct colors. For a $d$-dimensional complex colored by $d+1$ colors this means that every $d$-simplex has one vertex of each color. The two notions are dual to each other: the dual graph of a properly vertex-colored triangulation is properly edge-colored, the edge crossing the face opposite to a vertex of color $\alpha$ receiving the color $\alpha$.\\
\addlinespace[4pt]
Colored triangulation, GEM & A triangulation is colored (or balanced) if its vertices are properly colored by $d+1$ colors (previous row), i.e. all colors are present in every $d$-simplex. Its dual graph is a $(d+1)$-colored graph, and conversely every $(d+1)$-colored graph $\Gamma$ defines a colored simplicial cell complex $K(\Gamma)$. A GEM (graph-encoded manifold) is a colored graph $\Gamma$ such that $|K(\Gamma)|$ is a PL manifold. Every closed PL manifold admits a GEM \cite{Pezzana1974,FerriGagliardiGrasselli1986}. The connected components of the subgraphs $\rs{\Gamma}{I}$ are called residues in the crystallization literature. Our residue $R(\Gamma)$ of Section \ref{sec:residues} is their formal sum.\\
\addlinespace[4pt]
Crystallization & A connected GEM $\Gamma$ such that $\rs{\Gamma}{\hat\alpha}$ is connected for every color $\alpha$, i.e. such that $K(\Gamma)$ has exactly $d+1$ vertices, one of each color: a GEM with the least possible number of vertices in its triangulation. Every closed connected PL manifold admits a crystallization \cite{Pezzana1974,FerriGagliardiGrasselli1986}, obtained from any GEM by cancelling $1$-dipoles (see below). The graph of Figure \ref{fig:lens}(a) is a crystallization of $L(3,1)$. In this paper GEMs are not required to be crystallizations.\\
\addlinespace[4pt]
Dipole moves & An $h$-dipole of $\Gamma$ is a pair of vertices joined by $h$ edges, of colors $C$, which lie in different connected components of $\rs{\Gamma}{\hat C}$. Cancelling it (deleting the two vertices and pasting the remaining edges colorwise) does not change $M_\Gamma$ if one of these two components is a spherical GEM, and adding a dipole is the inverse operation. For a GEM this condition is automatic, since the two components are proper-color subgraphs (Section \ref{sec:residues}), but it has to be checked for general colored graphs such as $\hG$ in Appendix \ref{app:filling}. Any two crystallizations of the same manifold are related by dipole moves \cite{FerriGagliardi1982}, and every GEM is reduced to a crystallization by cancelling $1$-dipoles. We use dipole moves in Section \ref{sec:example}, in Appendix \ref{app:filling} and in a footnote to the proof of Proposition \ref{prop:smooth-torsion}.\\
\bottomrule
\end{tabular}
\caption{Terminology of combinatorial topology used in this paper: pseudomanifolds and colored graphs.}
\label{tab:glossary-gems}
\end{table}

\bibliographystyle{JHEP}
\bibliography{refs}

\end{document}